\RequirePackage{fix-cm}
\documentclass[twocolumn,epjc3]{svjour3}  
\pdfoutput=1
\RequirePackage{graphicx}
\RequirePackage{cite}
\usepackage{xspace}
\usepackage{booktabs}
\usepackage{afterpage}
\usepackage{hepunits}
\usepackage{color}
\newcommand{\herwig}{\textsc{Herwig}}
\newcommand{\pythia}{\textsc{Pythia}}
\newcommand{\sherpa}{\textsc{Sherpa}}

\newcommand{\figref}[1]{Fig.~\ref{#1}}
\newcommand{\secref}[1]{Sec.~\ref{#1}}
\newcommand{\equref}[1]{Eq.~(\ref{#1})}

\newcommand{\ncluster}{\ensuremath{N_{\mathrm{cl}}}\xspace}
\newcommand{\ptmin}{p_{\perp}^{\rm min}}
\newcommand{\pdisrupt}{\ensuremath{p_{\rm disrupt}}\xspace}
\newcommand{\dNchgdetadphi}{\ensuremath {\langle \mathrm{d}^2N_\text{ch}/\mathrm{d}\eta\,\mathrm{d}\phi\rangle} \xspace}
\newcommand{\dpTsumdetadphi}{\ensuremath{\langle \mathrm{d}^2\sum p_t/\mathrm{d}\eta\,\mathrm{d}\phi \rangle}\xspace}
\newcommand{\ptlead}{\ensuremath{p_{\perp}^{\rm lead}}\xspace}
\newcommand{\ptsum}{\ensuremath{\sum p_{\perp}}\xspace}
\newcommand{\ptminnought}{\ensuremath{p_{\perp,0}^{\text{min}}}}
\newcommand{\lambdainit}{\ensuremath{\lambda_{\rm init}}\xspace}
\newcommand{\lambdafinal}{\ensuremath{\lambda_{\rm final}}\xspace}
\newcommand{\deltaif}{\ensuremath{ \Delta_{\rm if} }\xspace}
\newcommand{\mcut}{\ensuremath{m_{\mathrm{cut}}}\xspace}
\newcommand{\NC}{N_{\mathrm{c}}}

\newcommand{\preco}{\ensuremath{p_{\rm reco}}\xspace}
\newcommand{\SCRalpha}{\alpha\xspace}              
\newcommand{\SCRc}{c\xspace}                       
\newcommand{\SCRf}{f\xspace}                       
\newcommand{\SCRnsteps}{N_{\mathrm{steps}}\xspace} 
\newcommand{\pcr}{PCR\xspace}
\newcommand{\scr}{SCR\xspace}

\newcommand{\UEvii}{\textsc{ue7-2}\xspace}
\newcommand{\EEiii}{\textsc{ue-ee-3}\xspace}
\newcommand{\EEiiiCTEQ}{\textsc{ue-ee-3-cteq6l1}\xspace}
\newcommand{\EESCRCTEQ}{\textsc{ue-ee-scr-cteq6l1}\xspace}

\journalname{Eur. Phys. J. C}

\newlength{\colwidth}
\newlength{\subfigwidth}
\begin{document}

\newlength{\preprintwidth}
\settowidth{\preprintwidth}{\normalsize MAN/HEP/2012/03}

\title{Colour reconnections in Herwig++
\hfill
\parbox[b]{\preprintwidth}{
  \textmd{
    \normalsize
    \begin{flushright}
      MCnet-12-06\\
      KA-TP-17-2012\\
      MAN/HEP/2012/03
    \end{flushright}
  }
}
}

\author{Stefan Gieseke\thanksref{e1,addr1}
        \and
        Christian R\"ohr\thanksref{e2,addr1}
        \and
        Andrzej Si\'odmok\thanksref{e3,addr1,addr2}
}

\thankstext{e1}{stefan.gieseke@kit.edu}
\thankstext{e2}{christian.roehr@kit.edu}
\thankstext{e3}{andrzej.siodmok@manchester.ac.uk}

\institute{Institut f\"ur Theoretische Physik, Karlsruhe Institute of Technology
           (KIT), Karlsruhe, Germany \label{addr1}
           \and
           Consortium for Fundamental Physics,
           School of Physics and Astronomy, The University of Manchester,
           Manchester, U.K. \label{addr2}
}

\date{Received: date / Accepted: date}

\maketitle

\begin{abstract}
We describe the implementation details of the colour reconnection model in the
event generator \herwig{}++.  We study the impact on final-state observables in
detail and confirm the model idea from colour preconfinement on the basis of
studies within the cluster hadronization model. Moreover, we show that the
description of minimum bias and underlying event data at the LHC is improved
with this model and present results of a tune to available data.
\end{abstract}


\section{Introduction}

High-energy hadronic collisions at the Large Hadron Collider (LHC) require a
sound understanding of soft aspects of the collisions.  All hard collisions are
accompanied by the underlying event (UE) which adds hadronic activity in all
phase space regions.  The physics of the underlying event is similar to the
physics in minimum bias (MB) interactions and very important to understand to
quantify the impact of pile-up in high-luminosity runs at the LHC.  A wide range
of measurements at the Tevatron and the LHC gives us a good picture of MB
interactions and the UE \cite{Affolder:2001xt, Aaltonen:2010rm, Aad:2010rd,
Aad:2010fh, :2010ir, Aad:2011qe, Khachatryan:2010xs, Khachatryan:2010us,
Khachatryan:2011dx, Chatrchyan:2011id, Aamodt:2010ft, Aamodt:2010pp,
Aamodt:2010my}.  Data has also shown that a good part of the underlying event is
due to hard multiple partonic interactions (MPI).  By now, the three major Monte
Carlo event generators \herwig{} \cite{Bahr:2008pv}, \pythia{}
\cite{Sjostrand:2006za,Sjostrand:2007gs} and \sherpa{} \cite{Gleisberg:2008ta}
have an MPI model implemented to simulate the underlying event.  

Such a model of independent multiple partonic interactions was first implemented
in \pythia{} \cite{Sjostrand:1987su} where its relevance for a description of
hadron collider data was immediately shown.  On a similar physics basis, but
with some differences in the detailed modelling the \textsc{jimmy} add-on to the
old \herwig{} program, was introduced \cite{Butterworth:1996zw}.  In these
models, the average number of additional hard scatters is calculated from a few
input parameters and then for each hard event the additional number of hard
scatters is sampled.  The individual scatters in turn are modelled similarly to
the primary hard scatters from QCD $2\to 2$ interactions at leading order, with
parton shower and hadronization applied as usual.  The current underlying event
model in \sherpa{} \cite{Gleisberg:2008ta} is similar but will be replaced by a
new approach \cite{Khoze:2010by}.  The current model in \pythia{} differs from
the original development in some details and follows the idea of interleaved
partonic interactions and showering \cite{Sjostrand:2004pf,Sjostrand:2004ef}.

In the recent releases of \herwig{} an MPI model is also included
\cite{Bahr:2008dy}. It comes with two main parameters, the minimum transverse
momentum $\ptmin$ of the additional hard scatters and the parameter $\mu^2$,
that can be understood as the typical inverse proton radius squared and appears
in the spatial transverse overlap of the incoming hadrons.  Good agreement with
Tevatron data was found with this model.  Soft interactions were added to this
model in order to improve consistency with more general theoretical input as the
total cross section and the elastic slope parameter in high-energy hadronic
collisions \cite{Bahr:2009ek}.  The distribution of transverse momenta in the
non-perturbative region below $\ptmin$ was modelled similarly to the proposal in
\cite{Borozan:2002fk}.  Furthermore, it is assumed that the soft partons are
distributed differently from the hard partons inside the hadron.  The additional
parameters introduced here are fixed by requiring a description of the total
cross section and the slope parameter, so we are still left with only two
parameters.  Once again, a good description of Tevatron data on the UE was
found, now also where softer interactions play a role.  The model for soft
interactions smoothly extrapolates from the perturbative into the
non-perturbative region, similar to a model for intrinsic transverse momentum in
initial-state radiation \cite{Gieseke:2007ad}.

\begin{figure*}[htb]
  \includegraphics[width=\colwidth]{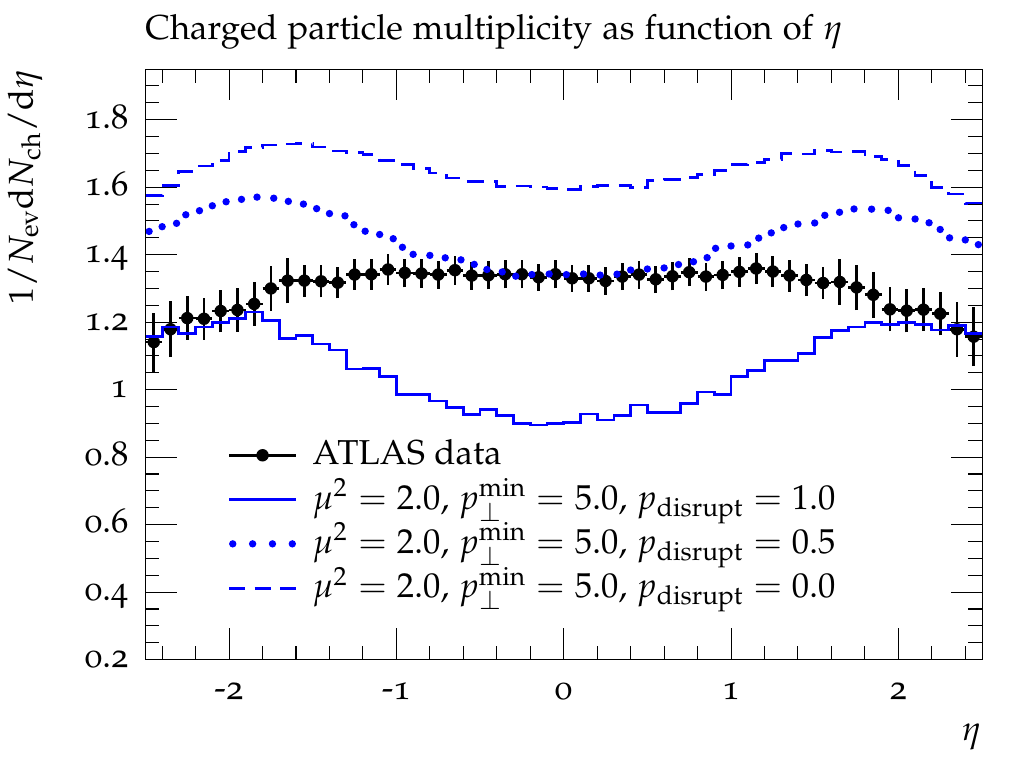}\hfill
  \includegraphics[width=\colwidth]{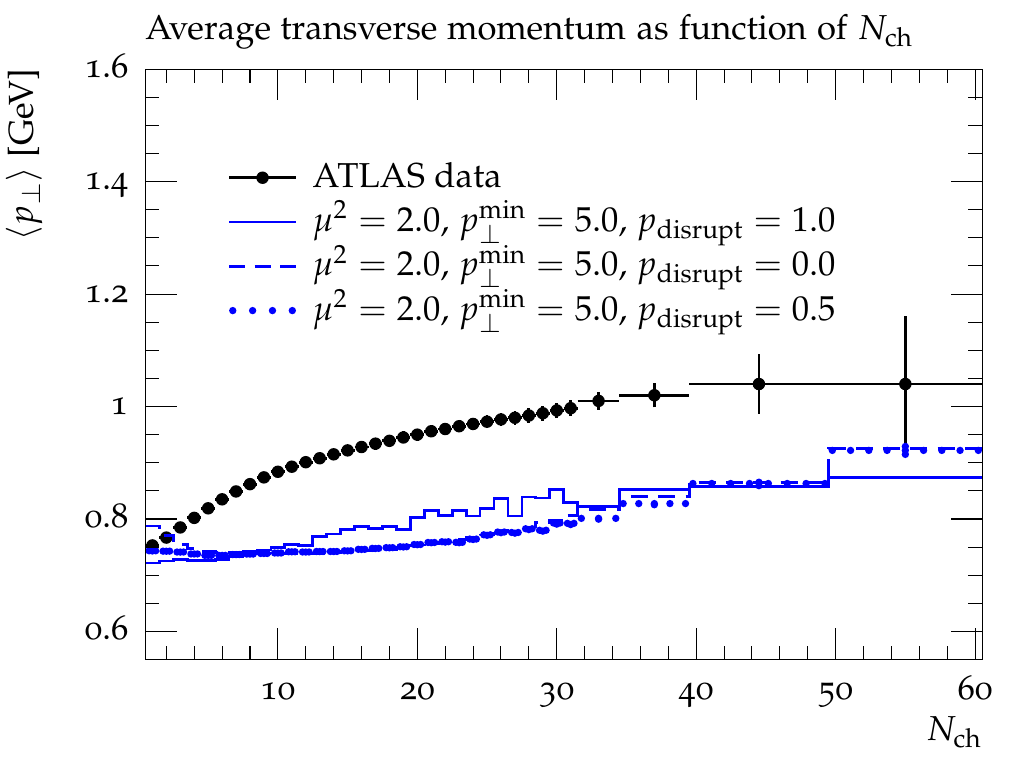}
  \caption{Comparison of \herwig{} 2.4.2 (without CR) to ATLAS minimum-bias
  distributions at $\sqrt{s}=\unit{0.9}{\TeV}$ with $N_{\mathrm{ch}} \ge 2$,
  $p_{\perp} > \unit{500}{\MeV}$ and $|\eta| < 2.5$. The \herwig{} results are
  obtained by using three different values for $\pdisrupt$: $0.0, 0.5$ and
  $1.0$.}
  \label{fig:ATLAS900_Nch_1_preco}
\end{figure*}

\begin{figure*}[htb]
  \includegraphics[width=\colwidth]{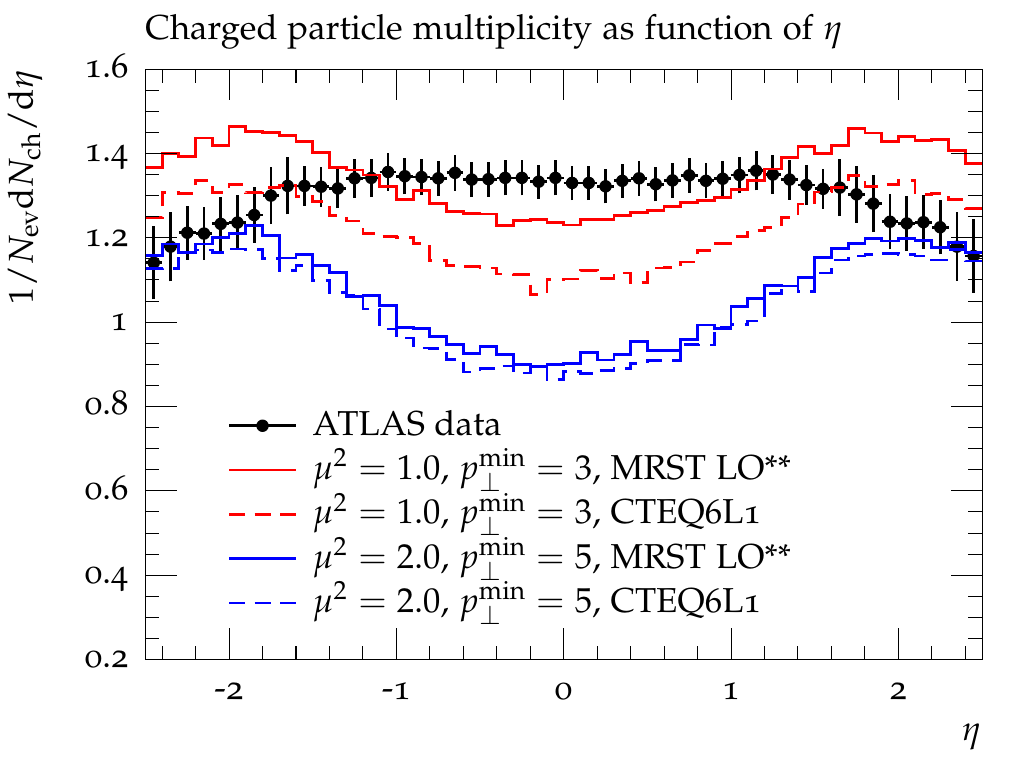}\hfill
  \includegraphics[width=\colwidth]{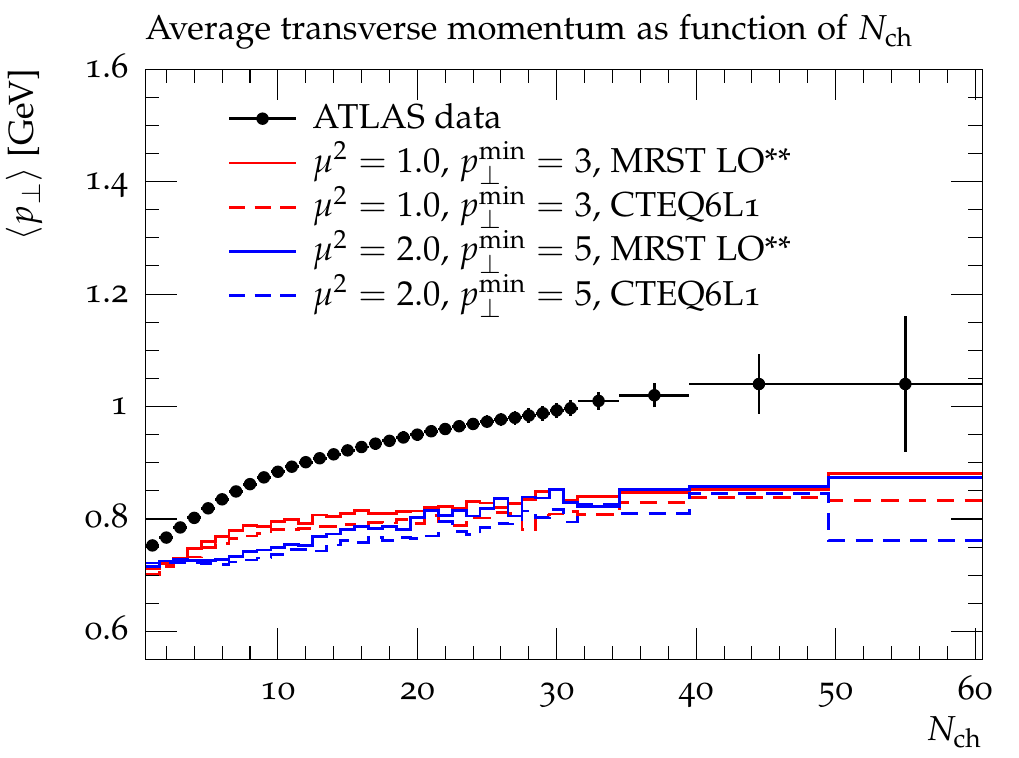}
  \caption{Dependence on the choice of the PDF set. The shown observables are
  the same as already introduced in Fig.~\ref{fig:ATLAS900_Nch_1_preco}. We show
  results from two parameter points of the MPI model. For each point, two
  different PDF sets are selected, CTEQ6L1 and MRST LO**. All settings give a
  satisfactory description of the Tevatron underlying-event data.}
  \label{fig:ATLAS900_Nch_1}
\end{figure*}

With the advent of new data from the LHC at \unit{900}{\GeV} \cite{Aad:2010rd}
we also considered new observables and found distinct disagreement with data,
e.g.\ in the pseudorapidity of charged particles.  It was clear that our
implementation was incomplete as we have not at all tried to modify the relative
colour structure of the multiple hard scatters.  In
Fig.~\ref{fig:ATLAS900_Nch_1_preco} we show the sensitivity to the parameter
\pdisrupt, which controls the colour structure of soft scatters and see a
partial refill of the central rapidity plateau.  This notable dependence on
\pdisrupt of soft scatters hints at the importance of colour correlations in a
more complete model.  Furthermore, we studied the dependence on other possible
sources, e.g.\ on the parton distribution functions (PDF), which are used to
extract the additional partons from the hadrons.  In
Fig.~\ref{fig:ATLAS900_Nch_1} we show the pseudorapidity of charged particles
and the average transverse momentum as a function of particle multiplicity,
$\langle p_{\perp}\rangle (N_{\rm ch})$, at that stage.  The lines represent
different settings of the parameter of soft colour disruption and two different
PDF sets: CTEQ6L1~\cite{Pumplin:2002vw} and MRST LO**~\cite{Sherstnev:2007nd}.
We stress that all settings gave a good description of the Tevatron UE data.
As discussed in more detail in \cite{Gieseke:2010zz,  Gieseke:2011xy,
Bartalini:2011jp}, even a dedicated tuning of the MPI model parameters did not
improve this description, which lead us to include a colour reconnection (CR)
model in order to improve the colour structure between various hard scatters in
the MPI model.  The starting point is the idea of colour preconfinement
\cite{Amati:1979fg}. While in a single hard interaction the colour structure is
given by (the leading part of) the colour matrices that appear in the Feynman
diagrams and also by the parton shower evolution, there is no such firm
prescription for the assignment of colour lines or colour connections
\emph{between} individual hard scatters.  Colour preconfinement leads us to the
\emph{assumption} that hard jets emerging from separate hard scatters should end up
colour-connected when they are produced nearby in momentum space.  As there is
no such correlation in the non-perturbative modelling of the multiple hard
interactions, we have to impose a model on it.  Studies of such a model were
carried out earlier in \cite{Sandhoff:2005jh,Skands:2007zg,Wicke:2008iz}.  In
this paper we describe the details of such a colour reconnection model and
confirm this physical picture with various analyses of the modelled hadronic
final state.  Finally, we present results of tuning this model to the currently
available data on MB interactions and the UE.

\section{Modelling colour reconnections}

The cluster hadronization model \cite{Webber:1983if} is based on planar diagram
theory \cite{'tHooft:1973jz}: The dominant colour structure of QCD diagrams in
the perturbation expansion in $1/\NC$ can be represented in a planar form using
colour lines, which is commonly known as the $\NC \to \infty$ limit. The
resulting colour topology in Monte Carlo events with partons in the final state
features open colour lines after the parton showers. Following a
non-perturbative isotropic decay of any left gluons in the parton jets to light
quark-antiquark pairs, the event finally consists of colour-connected partons in
colour triplet or anti-triplet states. These parton pairs form colour-singlet
clusters.

In dijet production via $e^+e^-$ annihilation the invariant mass spectrum of
these clusters is independent of the scale of the hard process
\cite{Webber:1983if, Gieseke:2003hm}. The mass distribution peaks at small
values, $\mathcal{O}(\unit{1}{\GeV})$, and quickly falls off at higher masses.
Descriptively speaking, the cluster constituents tend to be close in momentum
space.  This property of perturbative QCD is referred to as colour
preconfinement, as already stated above.  The invariant cluster mass largely
consists of the constituent rest masses, which gives rise to a pronounced peak
at the parton rest mass threshold.  Hence, clusters are interpreted as highly
excited pre-hadronic states. In the cluster hadronization model hadrons normally
arise from non-perturbative, isotropic cluster decays. The \herwig{}
implementation of this hadronization model is described in more detail in
Ref.~\cite{Bahr:2008pv}.

\begin{figure}[t]
  \centering
  \includegraphics[width=\linewidth,bb=8 2 340 130]{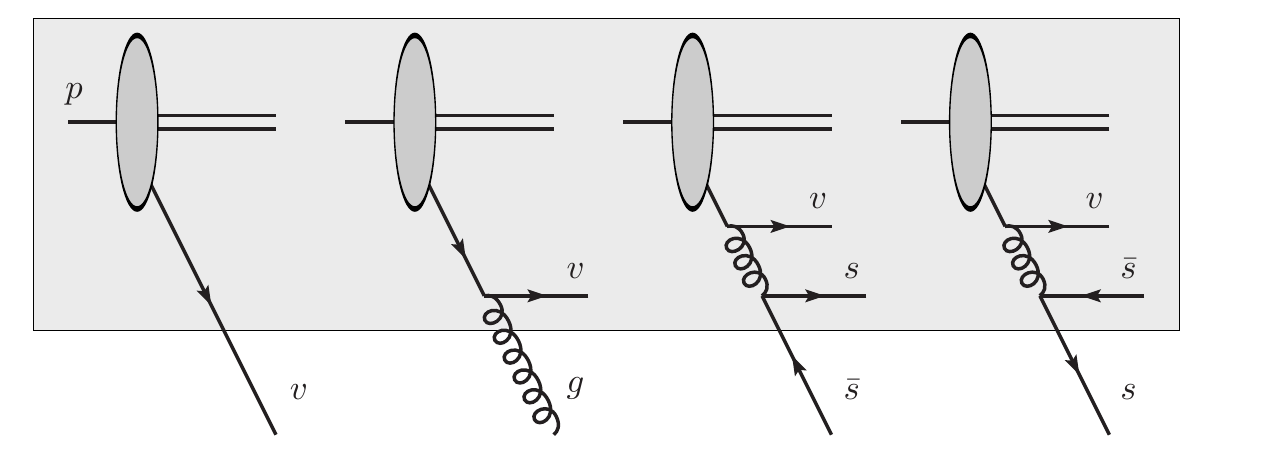}%
  \caption{For the hard subprocess a valence quark $v$ is extracted from the
  proton. Since the valence quark parton distribution functions dominate at
  large momentum fractions $x$ and small scales $Q^2$, the initial-state shower,
  which is generated backwards starting from the partonic scatter, commonly
  terminates on a valence quark. This situation is shown in the leftmost figure.
  If the perturbative evolution still terminates on a sea (anti)quark or a
  gluon, as indicated in the other figures, one or two additional
  non-perturbative splittings are performed to force the evolution to end with a
  valence quark. The grey-shaded area indicates this non-perturbative region,
  whereas the perturbative parton shower happens in the region below.}
  \label{fig:remnantextraction}
\end{figure}

The situation in hadron collisions is necessarily more complicated. In a typical
QCD $2\to 2$ scatter, there is QCD radiation from the initial-state parton
shower accompanied by jets emerging from outgoing partons. Due to colour charge
conservation, there are colour connections between the partonic subprocess and
the two hadron remnants. As sketched in \figref{fig:remnantextraction}, the
primary hard subprocess is modelled in \herwig{} as an interaction of two
valence (anti)quarks \cite{Bahr:2008pv}. Hence, in $pp$ ($p\bar{p}$) collisions
the hadron remnants are colour anti-triplets (triplets).  The typical length
scale of the valence parton extraction is the hadron size,
$\mathcal{O}(1\,\mathrm{fm})$, corresponding to energies where perturbation
theory is not applicable. Thus, perturbative QCD cannot be used to calculate or
assess the colour correlation between the partonic subprocess and the beam
remnants.

We face a similar situation if we consider multiple parton interactions in
single hadron collisions. The MPI model in \herwig{} equips the event with a
number of further QCD parton scatters, in addition to the primary partonic
subprocess. For each of these subprocesses a pair of gluons, initiating the
scatter, is extracted from the colliding hadrons. The chosen colour topology for
this extraction corresponds to the $\NC \to \infty$ limit.  As stated above,
this limit is justified in perturbative branchings. In non-perturbative regimes,
however, it is rather a QCD-motivated model than an assessable approximation.

\begin{figure*}[t]
  \sidecaption
  \resizebox{0.7\hsize}{!}{
  \begin{minipage}[t]{0.5\textwidth}
    \includegraphics[width=0.9\textwidth]{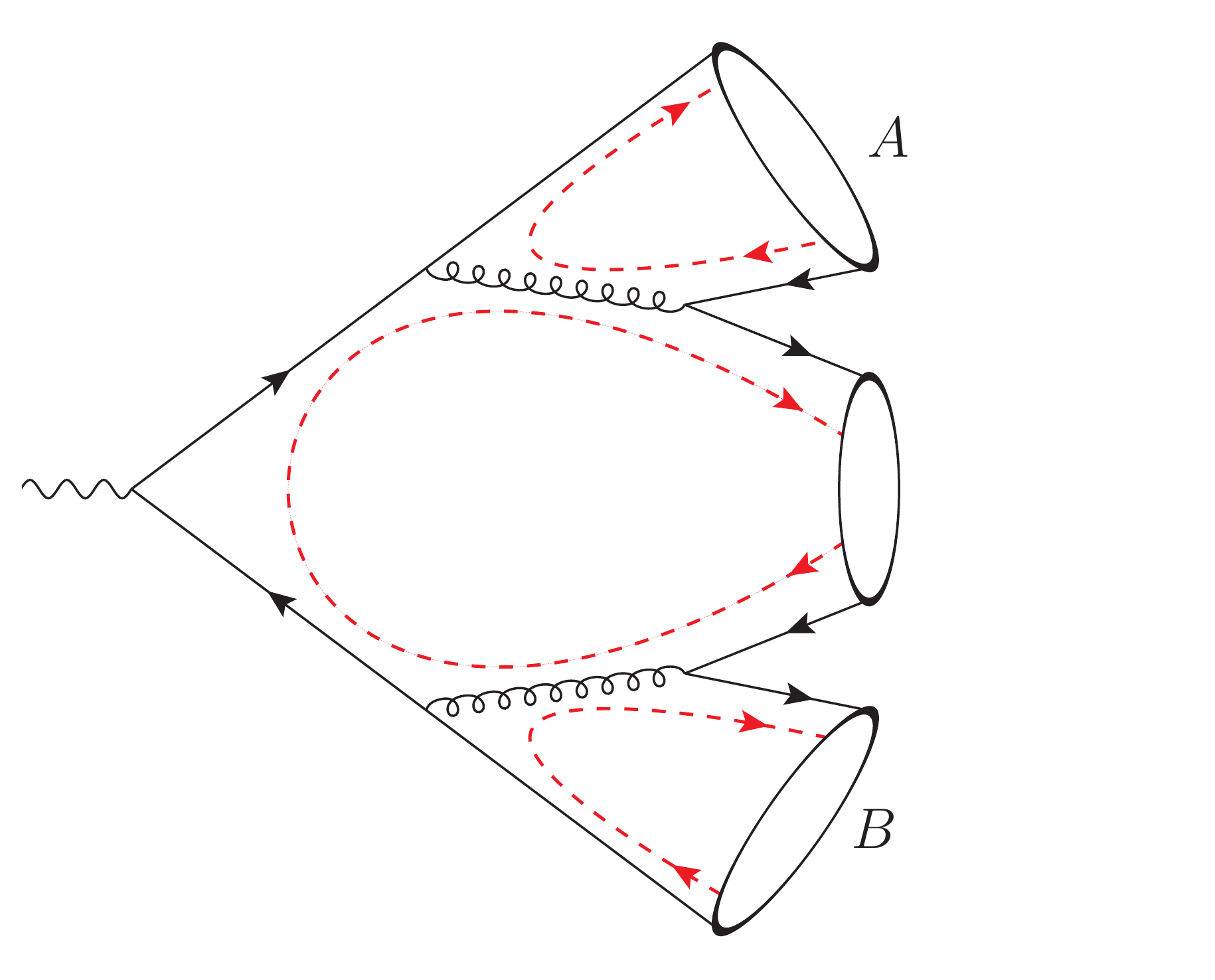}
  \end{minipage}
  \begin{minipage}[t]{0.5\textwidth}
    \includegraphics[width=0.9\textwidth]{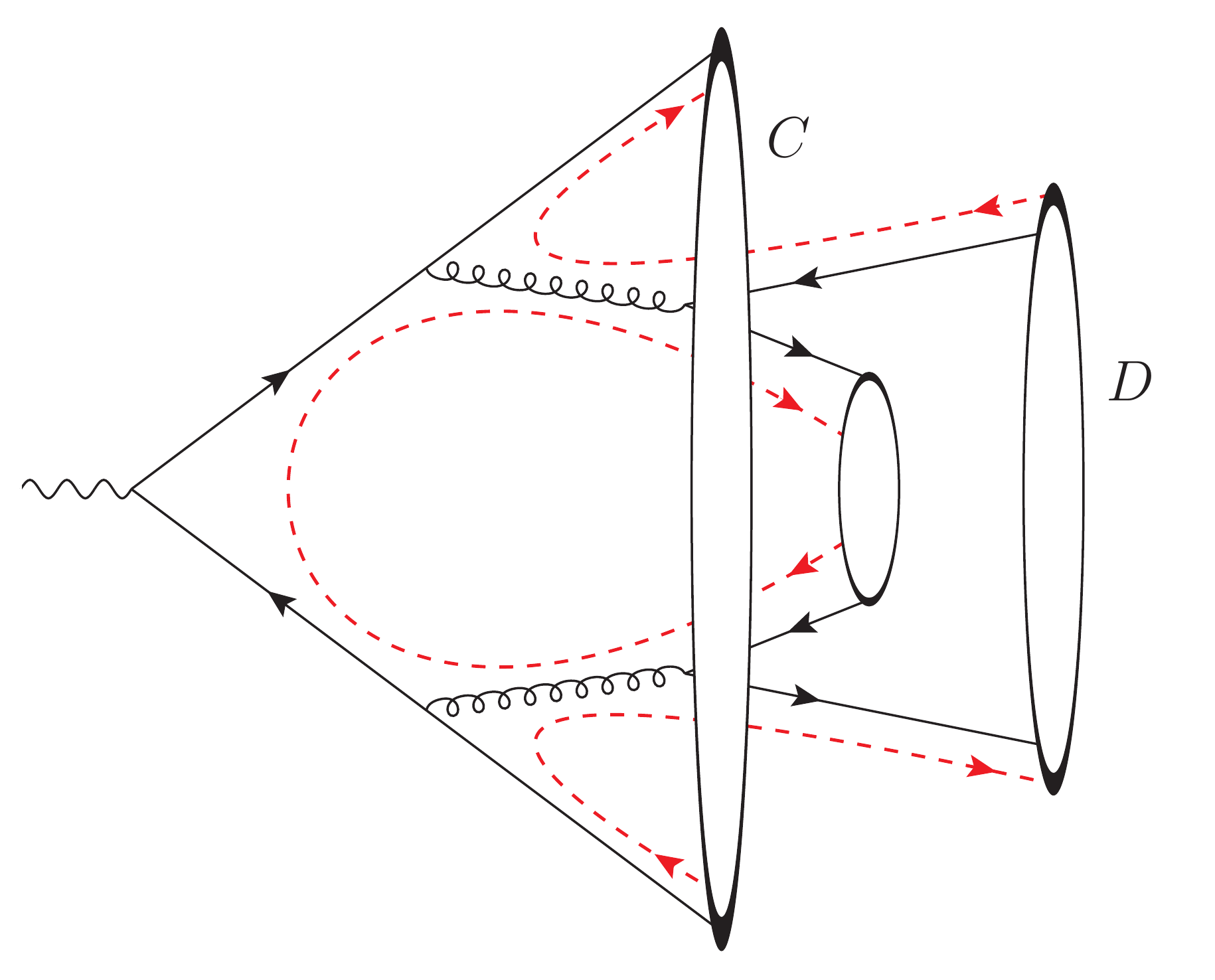}
  \end{minipage}
  }
  \caption{Formation of clusters, which we represent by ovals here. Colour lines
  are dashed. The \emph{left diagram} shows colour-singlet clusters formed
  according to the dominating colour structure in the $1/\NC$ expansion.
  The \emph{right diagram} shows a possible colour-reconnected state: the
  partons of the clusters $A$ and $B$ are arranged in new clusters, $C$ and
  $D$.}%
  \label{fig:crsketches}
\end{figure*}

As can be seen in the sketch in \figref{fig:clusterclasses} below, the parton
extraction model for the first and possible additional partonic subprocesses
introduces colour lines, which connect subprocesses to each other and to the
hadron remnants.  As a result, clusters emerge in hadronic collisions which link
different parts of the hadron collision. Clearly, these clusters cannot be
expected to feature the same invariant-mass distribution as the clusters in
$e^{+}e^{-}$ dijet events do.  Yet the cluster hadronization model for hadronic
collisions is adopted unchanged.  Colour reconnection intervenes at the stage
right \emph{before} hadrons are generated from the clusters. It provides the
possibility to create clusters in a way which does not strictly follow the
actual colour topology: The ends of the colour lines are reconnected, resulting
in a different \emph{cluster configuration}. This rearrangement of colour
charges is pictorially shown in \figref{fig:crsketches}. Based on the successful
role of preconfinement in $e^+e^-$ collisions, we designed two colour
reconnection models to work out colour singlets with invariant masses smaller
than a priori given. The colour reconnection models studied in this paper differ
in the underlying algorithm to find alternative cluster configurations.

\subsection{Plain colour reconnection}

A first model for colour reconnection has been implemented in \herwig{} as of
version~2.5 \cite{Gieseke:2011na}. We refer to it as the \emph{plain colour
reconnection} model (\pcr) in this paper. The following steps describe the full
procedure:
\begin{enumerate}
  \item Create a list of all quarks in the event, in \emph{random} order.
    Perform the subsequent steps exactly once for every quark in this list.
  \item The current quark is part of a cluster. Label this cluster $A$.
  \item Consider a colour reconnection with all other clusters that exist at
    that time. Label the potential reconnection partner $B$. For the possible
    new clusters $C$ and $D$, which would emerge when $A$ and $B$ are
    reconnected (cf.~\figref{fig:crsketches}), the following conditions must be
    satisfied: 
    \begin{itemize}
      \item The new clusters are lighter,
        \begin{equation}
          m_C+m_D < m_A+m_B\ ,
          \label{equ:pcr:condition}
        \end{equation}
	where $m_i$ denotes the invariant mass of cluster $i$.
      \item $C$ and $D$ are no colour octets.
    \end{itemize}
  \item If at least one reconnection possibility could be found in step 3,
    select the one which results in the \emph{smallest} sum of cluster masses,
    $m_C + m_D$.  Accept this colour reconnection with an adjustable probability
    \preco. In this case replace the clusters $A$ and $B$ by the newly formed
    clusters $C$ and $D$.
  \item Continue with the next quark in step 2.
\end{enumerate}
The parameter \preco steers the amount of colour reconnection in the \pcr model.
Because of the selection rule in step 4, the \pcr model tends to replace the
heaviest clusters by lighter ones. A priori the model is not guaranteed to be
generally valid because of the following reasons: The random ordering in the
first step makes this algorithm non-deterministic since a different order of the
initial clusters, generally speaking, leads to different reconnection
possibilities being tested. Moreover, apparently quarks and antiquarks are
treated differently in the algorithm described above.

\subsection{Statistical colour reconnection}

The other colour reconnection implementation studied in this paper overcomes the
conceptual drawbacks of the \pcr model. We refer to this model as
\emph{statistical colour reconnection} (\scr) throughout this work. In the first
place, the algorithm aims at finding a cluster configuration with a preferably
small colour length, defined as
\begin{equation}
  \lambda \equiv \sum_{i=1}^{\ncluster} m_{i}^2\ ,
\end{equation}
where $\ncluster$ is the number of clusters in the event and $m_i$ is the
invariant mass of cluster $i$. In the definition of the colour length we opt for
\emph{squared} masses to give cluster configurations with similarly heavy
clusters precedence over configurations with less equally distributed cluster
masses.

Clearly, it is impossible to locate the global minimum of $\lambda$, in general,
since an event with $100$ parton pairs, for instance, implies about $100!
\approx 10^{158}$ possible cluster configurations to be tested.  The Simulated
Annealing algorithm from Ref.~\cite{Kirkpatrick}, however, has proven useful in
solving optimisation problems like this approximately. The \scr model is an
application of this algorithm with $\lambda$ as the objective function to be
minimised.

The \scr algorithm selects random pairs of clusters and suggests them for colour
reconnection. Just like in the \pcr model, clusters consisting of splitting
products of a colour-octet state are vetoed. A reconnection step which reduces
$\lambda$ is always accepted. If the reconnection raises the colour length, it
is accepted with probability
\begin{equation}
  \label{eq:boltzmannfactor}
  p = \exp{ \left( -\frac{\lambda_2-\lambda_1}{T} \right) }\ ,
\end{equation}
where $\lambda_1$ and $\lambda_2$ denote the colour lengths before and after the
reconnection, respectively. This gives the system the possibility to escape
local minima in the colour length. The ``temperature'' $T$ is a control
parameter, which is gradually reduced during the procedure.  At high
temperatures, $T \geq \mathcal{O}(\lambda_2 - \lambda_1)$, the algorithm is
likely to accept steps which raise $\lambda$. By contrast, lower temperatures
imply a small probability for colour-length-increasing reconnection steps.

The transition from high to low temperatures is determined by the annealing
schedule, which flexibly adapts to the number of clusters, \ncluster, and to the
colour length in the event. First, a starting temperature is determined from the
typical change in the colour length, $\Delta \lambda = \lambda_2 - \lambda_1$.
To this end, a few random dry-run colour reconnections $S$ are performed, all
starting with the default cluster configuration. The initial temperature is set
to
\begin{equation}
  T_{\mathrm{init}} \equiv \SCRc \cdot \underset{i \, \in \, S}{\mathrm{median}} \{
  |\Delta \lambda|_i\}\ ,
\end{equation}
where $\SCRc$ is a free parameter of the model. Using the median makes this
definition less prone to outliers compared to the mean. The algorithm proceeds
in steps with fixed temperature. At the end of each temperature step $T$
decreases by a factor $\SCRf$, which is another free model parameter, with
$\SCRf \in (0,1)$. Each value of $T$ is held constant for $\SCRalpha\ncluster$
reconnection attempts with another free parameter $\SCRalpha$. The algorithm
stops as soon as no successful colour reconnections happen in a temperature
step, but at most $\SCRnsteps$ temperature steps are tested. We use the
parameters $\SCRc$, $\SCRalpha$, $\SCRf$ and $\SCRnsteps$, which are all related
to the annealing schedule, to tune the \scr model to data.  

We would like to stress that the annealing model is used only as a
numerical tool to minimize the colour length introduced above and hence
give no physical interpretation to the model parameters themselves.  We
argue later, that merely the  idea of minimizing the colour length is
indeed meaningful and physical.

\section{Characteristics of colour reconnection}
\label{sec:characteristics}

In this section we want to study hadronization-related quantities which allow us
to understand colour reconnection from an event generator--internal point of
view. Here, a set of typical values for $\SCRc$, $\SCRalpha$, $\SCRf$ and
$\SCRnsteps$ in the \scr model, as well as for \preco in the \pcr model, was
used, which was obtained from tunes to experimental data, as described below in
Sec.~\ref{sec:tuning}.

\subsection{Colour length drop}
\label{sec:colourlengthdrop}

To quantify the effect of colour reconnection at generator level, we define the
colour length drop
\begin{equation}
  \deltaif \equiv 1 - \frac{\lambdafinal}{\lambdainit} \ ,
  \label{equ:deltaif}
\end{equation}
where \lambdainit and \lambdafinal denote the colour length in an event before
and after colour reconnection, respectively. $\deltaif$ approximately vanishes
in events with $\lambdainit \approx \lambdafinal$, i.e.\ with no or only minor
changes in the colour length $\lambda$ due to colour reconnection. The other
extreme, $\deltaif \approx 1$, indicates a notable drop in $\lambda$.

\begin{figure*}[htb]
  \parbox[t]{0.32\textwidth}{
  \centering
  \includegraphics[width=0.32\textwidth]{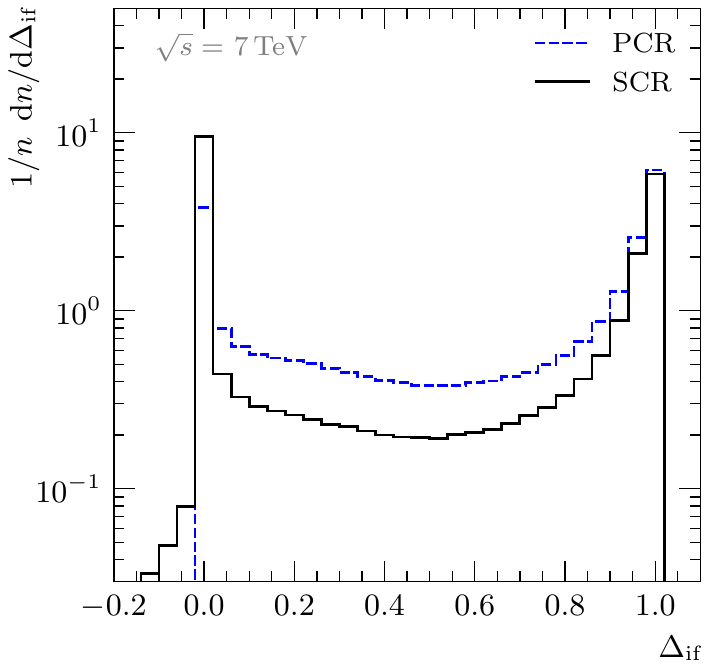}
  \\(a)}
  \hfill
  \parbox[t]{0.32\textwidth}{
  \centering
  \includegraphics[width=0.32\textwidth]{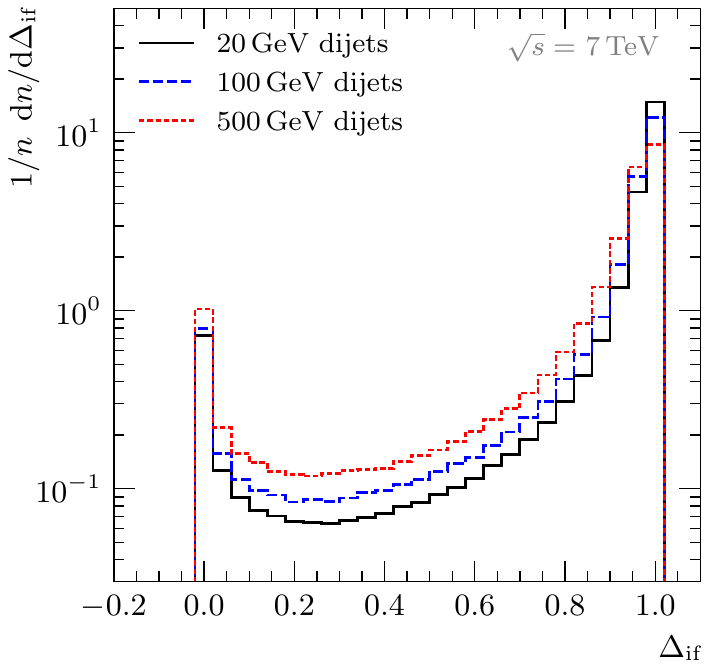}
  \\(b)}
  \hfill
  \parbox[t]{0.32\textwidth}{
  \centering
  \includegraphics[width=0.32\textwidth]{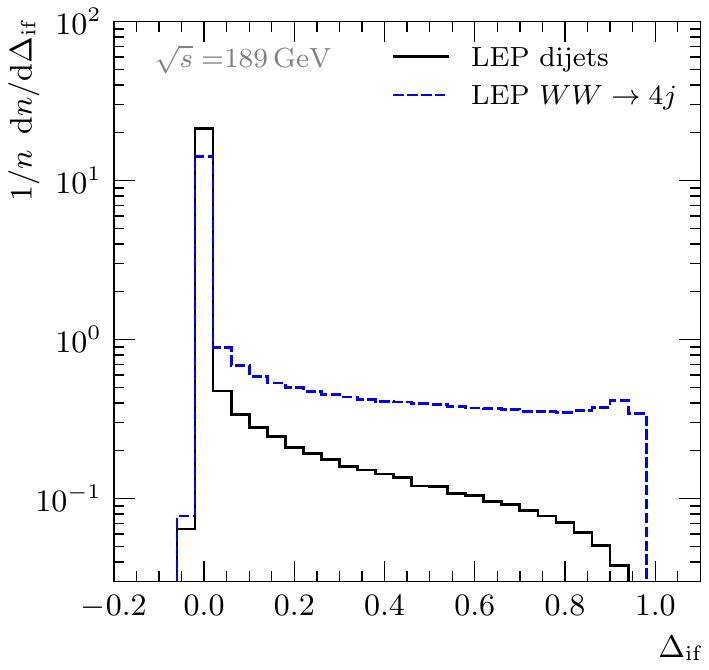}
  \\(c)}
  \caption{Colour length drop in $pp$ and $e^+e^-$ collisions. Figure (a) shows
  $\deltaif$ using the \pcr and the \scr models.  The events were generated with
  soft inclusive LHC generator settings at \unit{7}{\TeV}. In (b) we show the
  colour length drop within the \scr model in LHC dijet production with a number
  of $p_\perp$ cuts, where the c.m.\ energy is also \unit{7}{\TeV}. (c) shows
  the drop in the colour length (using the \scr model) with LEP generator setup
  running at \unit{189}{\GeV}. We compare dijet events to $W$ boson pair
  production with fully hadronic decays.}
  \label{fig:deltaif}
\end{figure*}

The distribution of \deltaif for soft inclusive LHC events at \unit{7}{\TeV} is
shown in \figref{fig:deltaif}(a). The plain and the statistical colour
reconnection models result in similar distributions with pronounced peaks at 0
and 1.  Note that Fig.~\ref{fig:deltaif} shows logarithmic plots, so the plateau
in between the peaks is really low.  There is also a small fraction of events
with negative \deltaif, though. The colour reconnection procedure actually
raises $\lambda$ in these events. In the \scr algorithm, this can happen since
$\lambda$-raising steps are explicitly allowed with a certain probability,
cf.~\equref{eq:boltzmannfactor}. However, also the \pcr algorithm might
potentially raise $\lambda$ since the reconnection condition,
\equref{equ:pcr:condition}, is formulated in terms of the \emph{first power} of
cluster masses, whereas $\lambda$ is defined as the sum of \emph{squared}
cluster masses. As these events are rare, we expect no impact on physical
observables. 

With soft inclusive hadron-hadron generator settings there are, generally
speaking, two important classes of events. One of the two are events where there
is no notable change in the sum of squared cluster masses, $\lambda$. In another
large fraction of events, however, colour reconnection causes an extreme drop in
$\lambda$. An obvious interpretation for this drop is that the colour
reconnection procedure replaces disproportionally heavy clusters by way lighter
ones.

\begin{figure*}[htb]
  \parbox[t]{\colwidth}{
  \centering
  \includegraphics{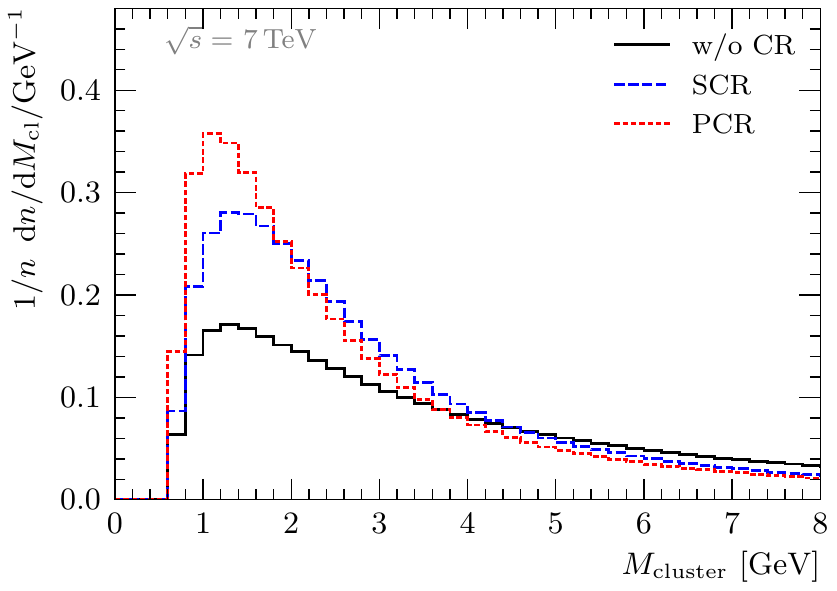}
  \\(a)}
  \hfill
  \parbox[t]{\colwidth}{
  \centering
  \includegraphics{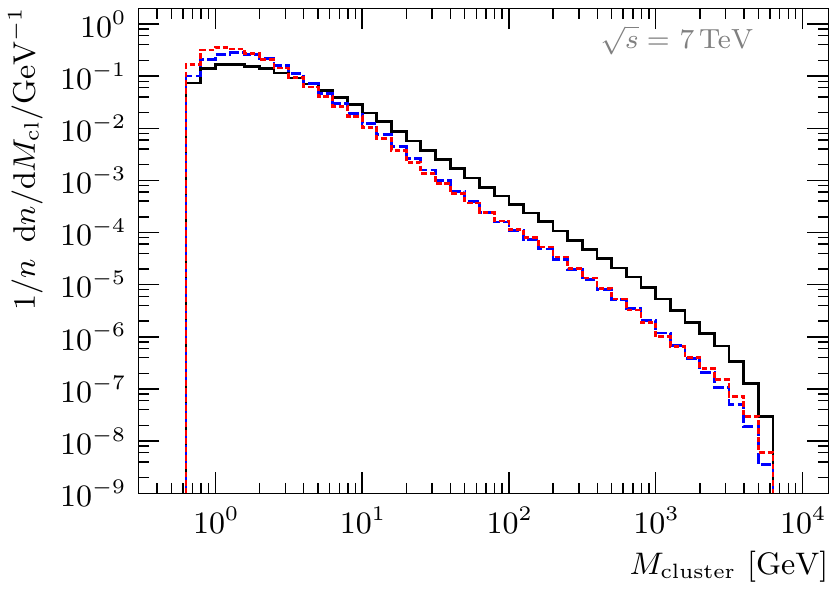}
  \\(b)}
  \caption{Invariant mass of primary clusters in soft inclusive LHC events at
  \unit{7}{\TeV}. The histograms are normalized to unity, where also invisible
  bins are taken into account. The histograms in (b) differ from the ones in (a)
  only in their binning.}
  \label{fig:clustermass}
\end{figure*}

This shift in the cluster mass spectrum, which both models aim at by
construction, can also be observed directly. Figure \ref{fig:clustermass} shows
the cluster mass distribution before and after colour reconnection. As expected
and also intended, both CR procedures cause the distribution to be enhanced in
the low-mass peak region and suppressed in its, potentially unphysical,
high-mass tail.

In Fig.~\ref{fig:deltaif}(b) we show the colour length drop in hard dijet events
in $pp$ collisions. We observe a notable decrease of large colour length drops,
$\deltaif = 1$, with increasing cut on the jet transverse momentum at parton
level.  The reason for this decrease is that higher momentum fractions are
required for the hard dijet subprocess, whereas in soft events the remaining
momentum fraction of the proton remnants is higher. Hence clusters containing a
proton remnant are less massive in hard events, which implies less need for
colour reconnection.

The distribution of the colour length drop in $e^+e^-$ annihilation events looks
completely different, as shown in Fig.~\ref{fig:deltaif}(c). We find that colour
reconnection has no impact on the colour length in the bulk of dijet events.  We
show only the \deltaif distribution from the \scr model here, but the \pcr model
yields similar results. These results confirm that due to colour preconfinement
partons nearby in momentum space in most cases are combined to colour singlets
already. In events with hadronic $W$ pair decays, however, hadrons emerge from
two separate colour singlets. If there is a phase space overlap of the two
parton jet pairs, the production of hadrons is expected to be sensitive to
colour reconnection. We address this question later on in Sec.~\ref{sec:lep}.
Here we want to remark that the fraction of $WW$ events with non-vanishing
colour length drop is slightly higher than for the dijet case.  Nevertheless,
the vast majority of $WW$ events is not affected by colour reconnection, too.

\subsection{Classification of clusters}
\label{sec:clusterclassification}

\begin{figure}[htb]
  \centering
  \includegraphics[width=\linewidth,bb=8 0 345 282]{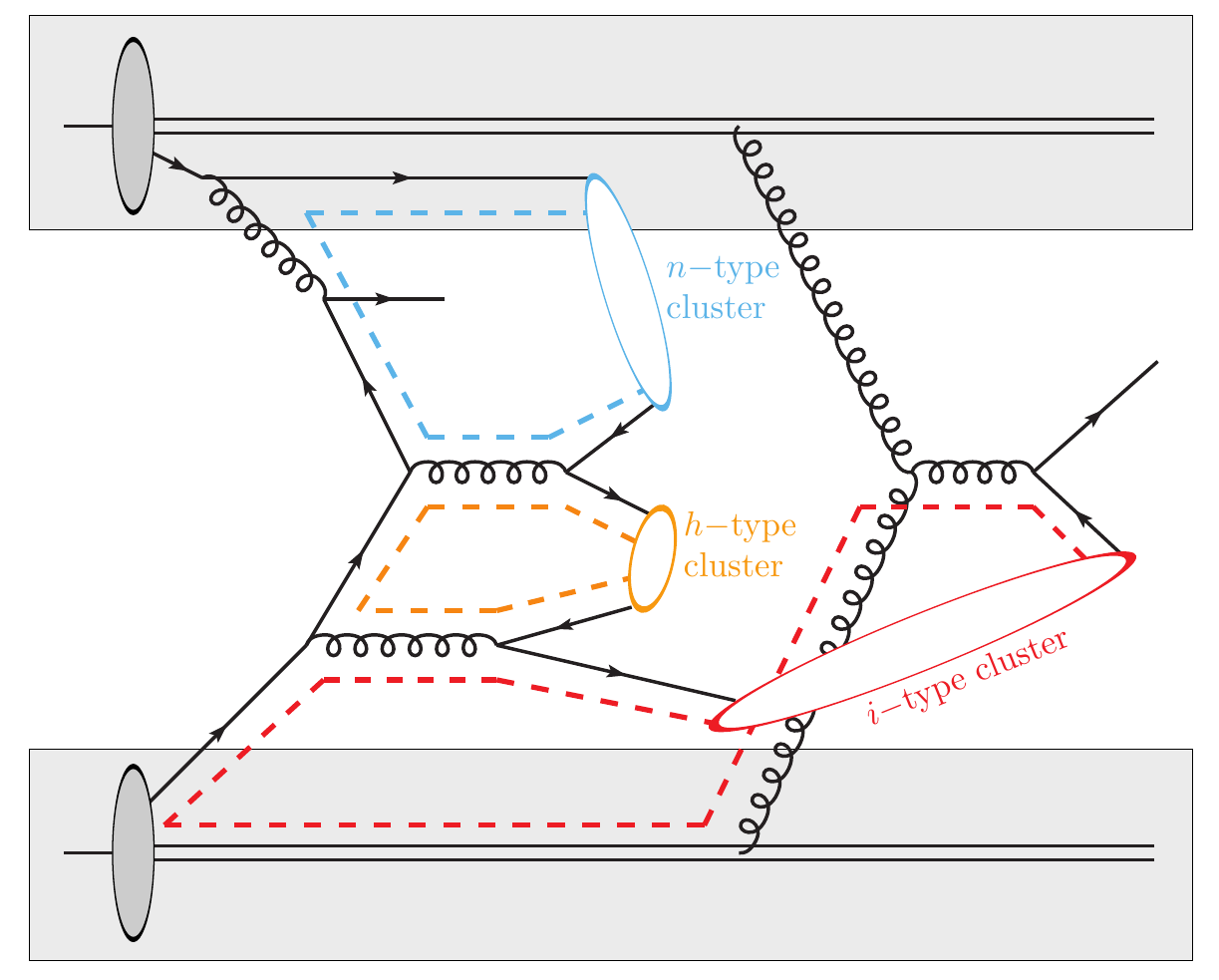}%
  \caption{Classification of colour clusters in a hadron collision event, which,
  in this example, consists of the primary subprocess (left) and one additional
  parton interaction. The grey-shaded area denotes non-perturbative parts of the
  simulation.  The three clusters represent the cluster classes defined in
  \secref{sec:clusterclassification}: $n$-type (blue), $i$-type (red) and
  $h$-type clusters (orange).}
  \label{fig:clusterclasses}
\end{figure}

These results generically raise the question which mechanism in the hadron event
generation is responsible for these overly heavy clusters.  To gain access to
this issue, we classify all clusters by their ancestors in the event history. A
sketch of the three types of clusters in shown in \figref{fig:clusterclasses}.
\begin{itemize}
  \item The first class are the clusters consisting of partons emitted
    perturbatively in the same partonic subprocess. We call them \emph{$h$-type}
    (\textbf{h}ard) clusters.
  \item The second class of clusters are the subprocesses-interconnecting
    clusters, which combine partons generated perturbatively in different
    partonic subprocesses. They are labelled as \emph{$i$-type}
    (\textbf{i}nterconnecting) clusters.
  \item The remaining clusters, which can occur in hadron collision events, are
    composed of at least one parton created non-perturbatively, i.e.\ during the
    extraction of partons from the hadrons or in soft scatters. In what follows,
    these clusters are called \emph{$n$-type} (\textbf{n}on-perturbative)
    clusters.
\end{itemize}

\begin{figure}[t]
  \centering
  \includegraphics[width=\linewidth, bb=12 0 320 204]{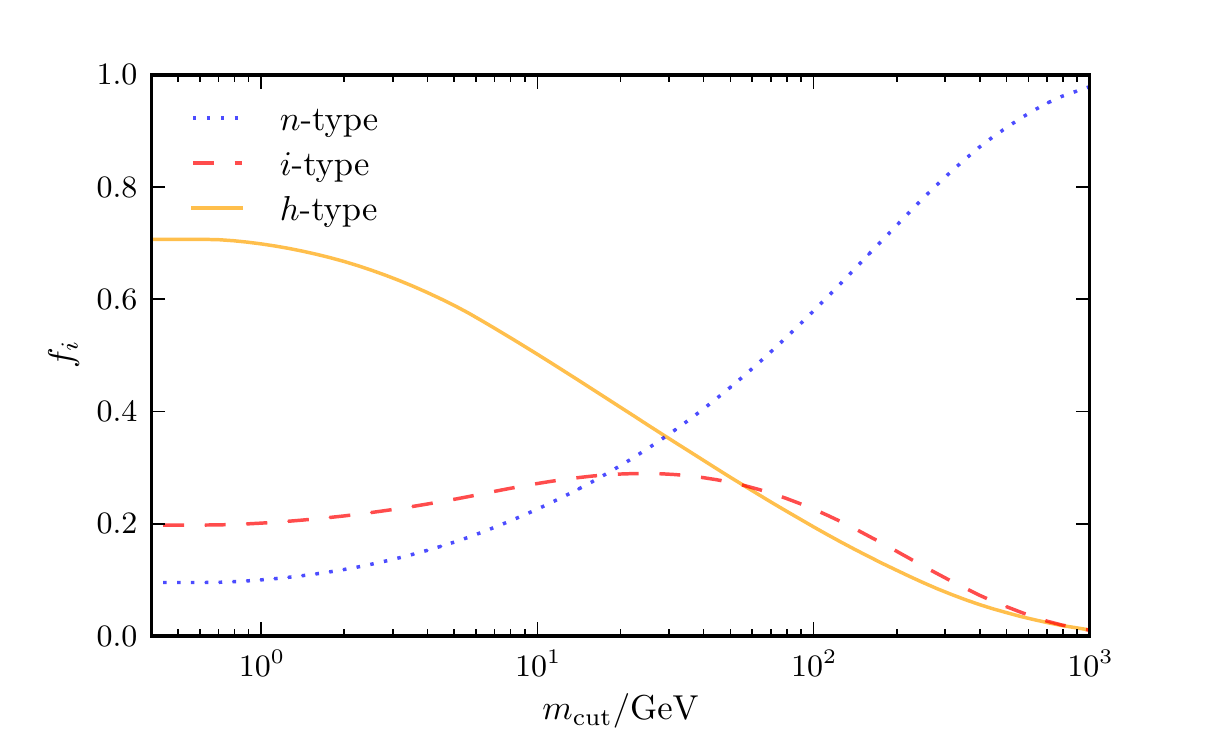}
  \caption{Cluster fraction functions, defined in \equref{eq:fractionfunction},
    for LHC dijet events at \unit{7}{\TeV}.}
  \label{fig:clusterratio}
\end{figure}
First we use this classification to analyse hadron collision events as they are
immediately before colour rearrangement. For that purpose, we define the cluster
fraction functions
\begin{equation}
  f_a(\mcut) \equiv N_a(\mcut) \Big/ \sum_{b=h,i,n} N_b(\mcut) =
  \frac{N_a(\mcut)}{\ncluster}\, ,
  \label{eq:fractionfunction}
\end{equation}
where $N_a(\mcut)$ is the number of $a$-type clusters ($a = h, i, n$) with $m
\geq \mcut$, counted in a sufficiently large number of events\footnotemark. For
instance, $f_{i} (\unit{100}{\GeV}) = 0.15$ says 15~\% of all clusters with a
mass larger than \unit{100}{\GeV} are subprocess-interconnecting clusters. By
construction, $f_a(\mcut)$ is a number between 0 and 1 for every class $a$.
Moreover, the cluster fraction functions satisfy
\begin{equation*}
  \sum_{a=h,i,n} f_a(\mcut) = 1 .
\end{equation*}
Figure \ref{fig:clusterratio} shows the cluster fraction functions for LHC dijet
events at $\sqrt{s} = \unit{7}{\TeV}$. The fraction of non-perturbative clusters
increases with $\mcut$ and exceeds 0.5 at $\mcut \approx \unit{70}{\GeV}$. So
for an increasing threshold $\mcut$ up to values well beyond physically
reasonable cluster masses of a few GeV, the contribution of $n$-type clusters
becomes more and more dominant.
\footnotetext{Apparently, $f_a(\mcut)$ is only well-defined for $\mcut$ less
than the maximum cluster mass. On this interval, the series $\left( f_{a,n}
\right)$, with $n$ the number of events taken into account, converges pointwise
to the function $f_a$. This is a more formal definition of the cluster fraction
functions.}

\begin{figure*}[htb]
  \parbox[t]{\colwidth}{
  \centering
  \includegraphics{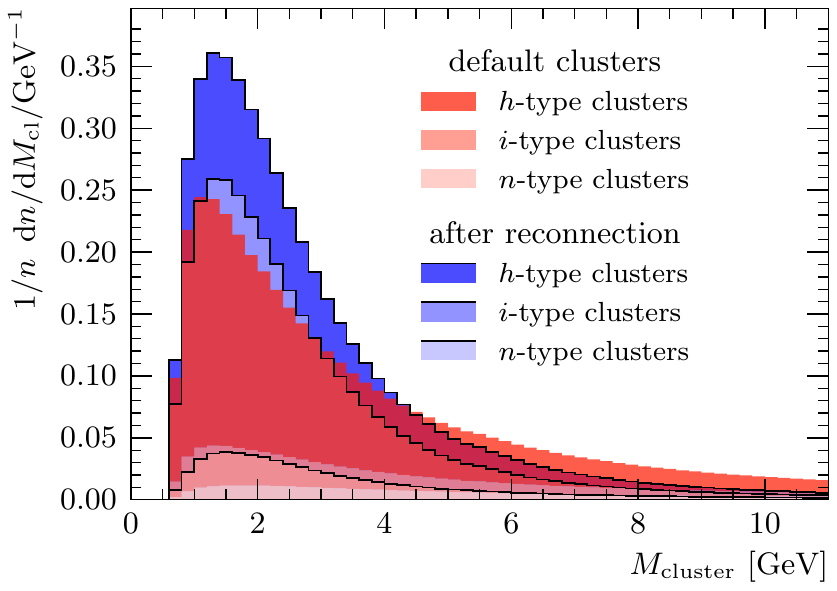}
  \\(a)}
  \hfill
  \parbox[t]{\colwidth}{
  \centering
  \includegraphics{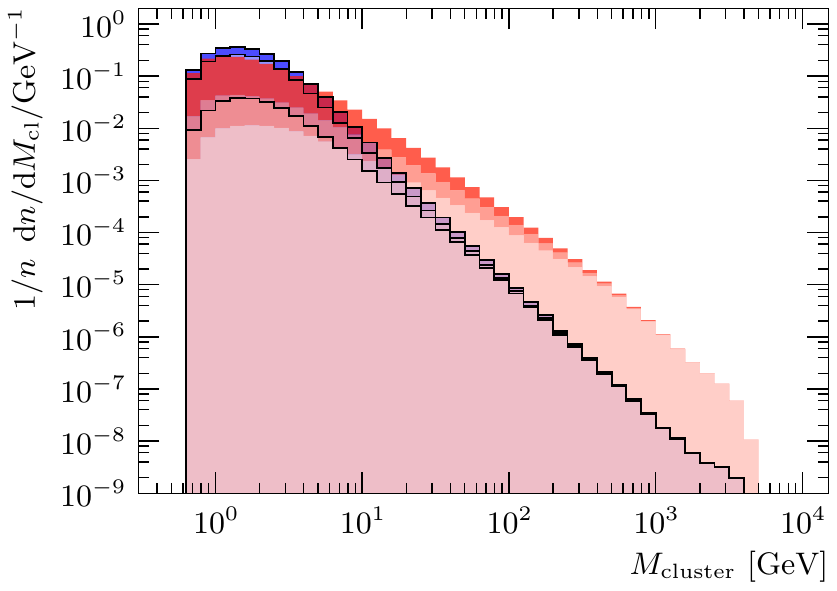}
  \\(b)}
  \caption{Primary cluster mass spectrum in LHC dijet events at \unit{7}{\TeV}.
  Figure (a) compares the mass distribution in the pre-colour-reconnection stage
  to the distribution after colour reconnection. The contributions of the three
  cluster classes are stacked. The histograms in (b) merely differ from the ones
  in (a) in their binning.}
  \label{fig:clustermass:stacked}
\end{figure*}

A bin-by-bin breakdown to the contributions of the various cluster types to the
total cluster mass distribution is shown in \figref{fig:clustermass:stacked}.
There are several things to learn from those plots.  First, non-perturbative
$n$-type clusters do not contribute as much to the peak region, say below
\unit{6}{\GeV}, as perturbative $h$-type and $i$-type clusters do.  In the
high-mass tail, however, $n$-type clusters clearly dominate, as already
indicated by the cluster fraction functions discussed above.  Both their minor
contribution at low masses and their large contribution at high masses do not
change after colour reconnection.  In total, however, the mass distribution is
more peaked after colour reconnection and the high-mass tail is suppressed by a
factor larger than 10.

\subsection{Resulting physics implications}
\label{sec:physimp}

The characteristics of clusters that have been studied in this section
clearly confirm the physical picture we have started out with.  The
colour reconnection model in fact reduces the invariant masses of
clusters that are mostly of non-perturbative origin.  These arise as an
artefact of the way we colour-connect additional hard scatters in the
MPI model with the rest of the event.  

At this non-perturbative level we have no handle on the colour
information from theory, hence we have modelled it.  First in a very
na\"{i}ve way when we extract the `first' parton from the proton, but
only to account for a more physical picture later, where we use colour
preconfinement as a guiding principle.  We therefore conclude that our
ansatz to model colour reconnections in the way we have done it
reproduces a meaningful physical picture.  

\section{Tuning and comparison of the model results with data}
\label{sec:tuning}

In this section we address the question of whether the MPI model in \herwig{},
equipped with the new CR model, can improve the description of the \mbox{ATLAS}
MB and UE data, see Fig.~\ref{fig:ATLAS900_Nch_1}. To that end we need to find
values of free parameters (tune parameters) of the MPI model with CR that allow
to get the best possible description of the experimental data.  Since both CR
models can be regarded as an extension of the cluster model
\cite{Webber:1983if}, which is used for hadronization in \herwig{}, the tune of
\herwig{} with CR models may require a simultaneous re-tuning of the
hadronization model parameters to a wide range of experimental data, primarily
from LEP (see Appendix D from Ref.~\cite{Bahr:2008pv}). Therefore, we start this
section by examining whether the description of LEP data is sensitive to CR
parameters.

\subsection{Validation against $e^{+}e^{-}$ LEP data}
\label{sec:lep}
  
\begin{figure*}[htb]
  \includegraphics[width=\colwidth]{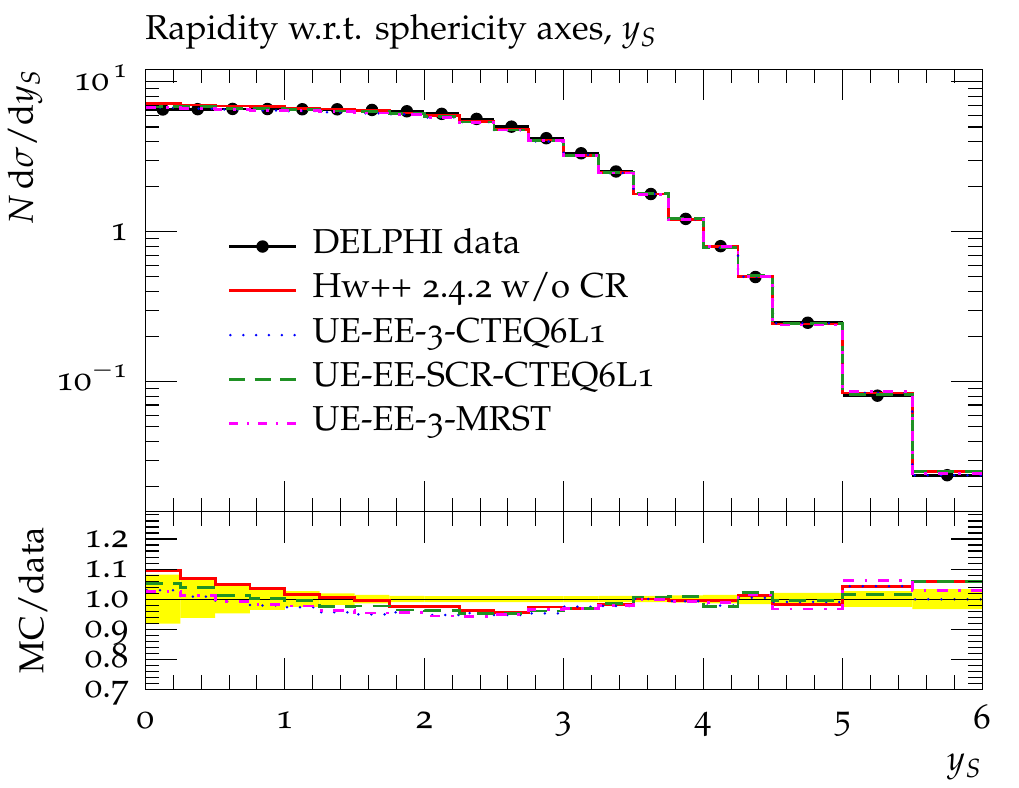}
  \hfill
  \includegraphics[width=\colwidth]{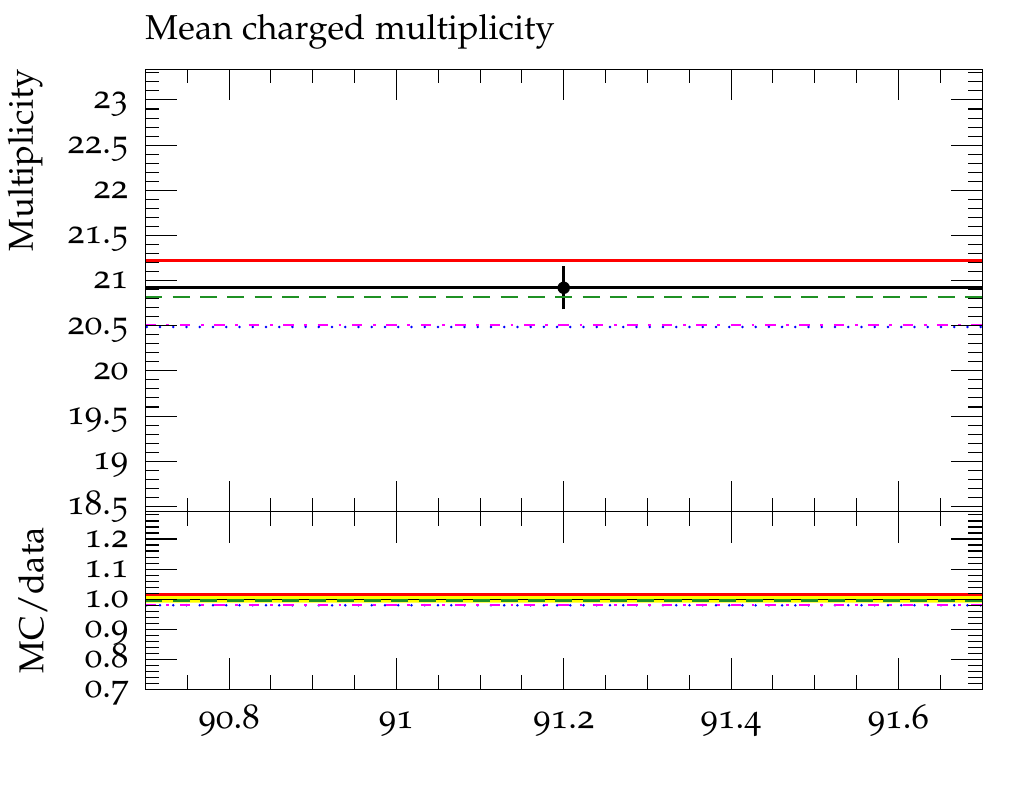}
  \caption{Comparison of \herwig{} without CR (red line) and with CR (using the
  main tunes for both CR methods presented in this paper) to exemplary
  measurements from the DELPHI detector at LEP.} 
  \label{fig:LEP}
\end{figure*}

Already in Section~\ref{sec:characteristics} we have seen that the colour
structure of LEP final states is well-defined by the perturbative parton shower
evolution.  Moreover, the CR model does not change this structure significantly.
Therefore, although CR is an extension of hadronization, we can expect that the
default hadronization parameters are still valid in combination with CR. This
was confirmed by comparing \herwig{} results with and without CR against a wide
range of experimental data from LEP~\cite{Abbiendi:2004qz, Abbiendi:2001qn,
Ackerstaff:1998hz, Pfeifenschneider:1999rz, Abreu:1995qx, Abreu:1996na,
Barate:1996fi, Decamp:1991uz, Heister:2003aj}.  As an example we show a
comparison of \herwig{} without and with CR (using the main tunes for both CR
methods presented in this paper) to two LEP observables in  Fig.~\ref{fig:LEP}.
The full set of plots, showing that the LEP data description in \herwig{} with
and without CR is of the same quality, can be found on the \herwig{} and MCplots
web pages \cite{tune_wiki, mcplots}.  These results allow us to factorize the
tuning procedure: The well-tested default \herwig{} tune for parton shower and
hadronization parameters is retained, and only the parameters from the CR and
MPI models are tuned to hadron collider data. However, we have checked each tune
presented in this paper against LEP results.

\begin{figure}[htb]
  \begin{center}
    \includegraphics[width=\linewidth]{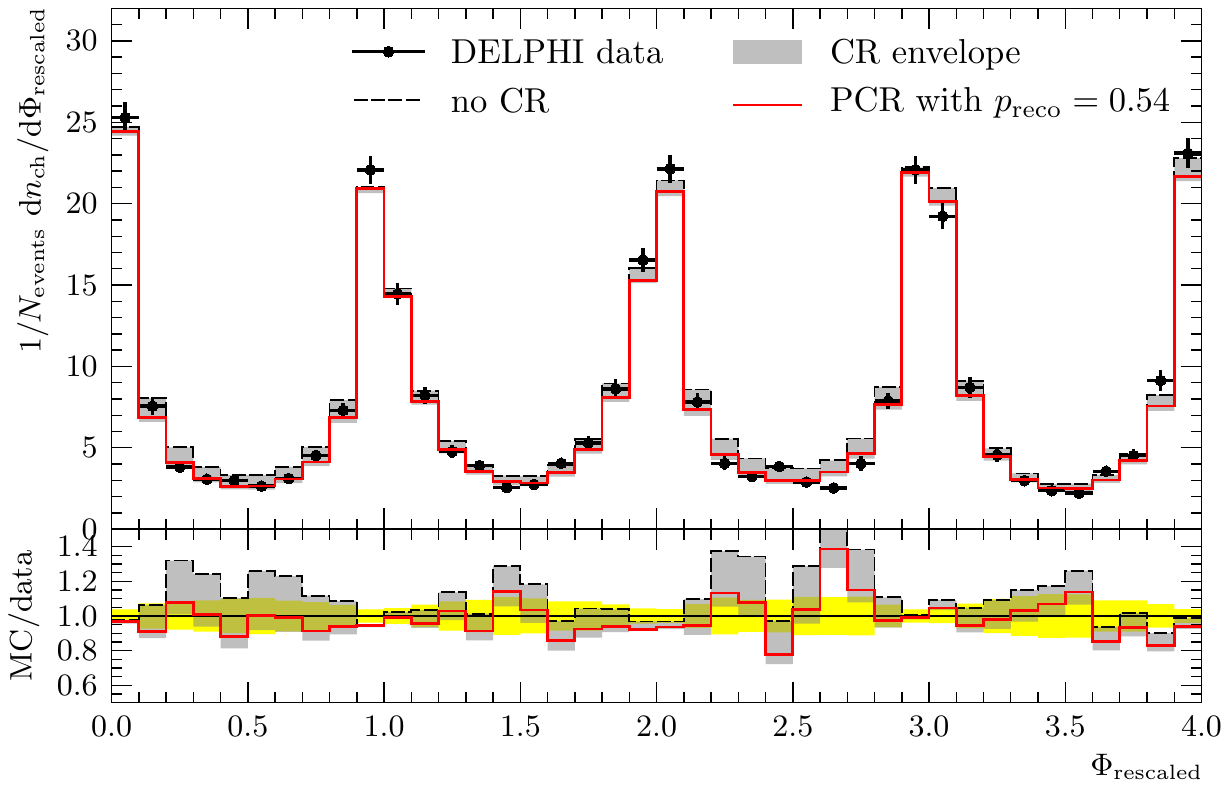}
    \caption{Charged-particle flow in hadronic $WW$ events at LEP with
    $\sqrt{s}= \unit{189}{\GeV}$. The grey band indicates the range which is
    covered by varying the colour reconnection strength \preco in the \pcr
    model.  The definition of the rescaled angle, $\Phi_{\rm rescaled}$, along
    with a detailed description of the analysis can be found in
    Ref.~\cite{Abdallah:2006uq}.} 
    \label{fig:LEP-WW}
  \end{center}
\end{figure}

In addition to the analyses used for the hadronization tuning, there are LEP
analyses dedicated to colour reconnection in $W^+W^- \to (q\bar{q})(q\bar{q})$
events \cite{Abdallah:2006uq, Achard:2003pe, Ziegler:2001jj, Abbiendi:2005es},
originally proposed in Ref.~\cite{Duchesneau:2000yn}.  In those analyses the $W$
bosons are reconstructed via kinematic cuts on all possible jet pairs in
four-jet events. The particle flow between jets originating from different
bosons was expected to be enhanced in Monte Carlo models including colour
reconnection. However, only moderate sensitivity to the tested CR models could
be found at the time. We have confirmed this with our colour reconnection
implementations. In Fig.~\ref{fig:LEP-WW} we show the sensitivity of the
particle flow between the identified jets to the reconnection strength in the
\pcr model, compared to DELPHI data from Ref.~\cite{Abdallah:2006uq}. We observe
a slight improvement in the description of the data. A number of apparent
outliers in the experimental data, however, indicate possibly too optimistic
systematic errors in the experimental analysis. For that reason, no clear
constraints on the model can be deduced from the data.

As the $W$ bosons are produced on shell and significantly boosted at
$\sqrt{s}=\unit{189}{\GeV}$, the finite $W$ width can cause the two $W$ bosons
to travel long distances before decaying. In the limit of a very small $W$
width, large reconnection effects between the two $W$ systems should thus be
suppressed in the model.  The moderate sensitivity of the particle flow to
colour reconnections implies, however, that colour reconnection effects are
small in $WW$ events.  Note that also the largely vanishing colour length drop
in $WW$ events, cf.\ Fig.~\ref{fig:deltaif}(c) and the discussion in
Sec.~\ref{sec:colourlengthdrop}, supports this conclusion.  Hence we retain the
described generic reconnection models also for $WW$ events and do not introduce
an extra suppression mechanism.

\subsection{Tuning to data from hadron colliders}

Now that we have validated the CR models by comparison against LEP data, we are
ready to tune their parameters to data provided by hadron colliders.  Before LHC
data was available, the MPI model in \herwig{} \cite{Bahr:2009ek} was tuned by
subdividing the two-dimensional parameter space, spanned by the model's main
parameters, the inverse proton radius squared $\mu^2$ and the minimum transverse
momentum $\ptmin$, into a grid. For each of the parameter points on this grid,
the total $\chi^2$ against the Tevatron underlying-event
data~\cite{Affolder:2001xt,Acosta:2004wqa} was calculated. A region in the
parameter plane was found, where similarly good values for the overall $\chi^2$
could be obtained.

While tuning the MPI models including colour reconnection we are dealing with a
larger number $N$ of tunable parameters $p_i$, where $N=4$ in case of the \pcr
($\pdisrupt$, $\preco$, $\ptmin$ and $\mu^2$) and $N=7$ in case of the \scr
model ($\pdisrupt$, $\ptmin$, $\mu^2$, $\SCRalpha$, $\SCRc$, $\SCRf$ and
$\SCRnsteps$).  Hence the simple tuning strategy from above is ineffective. A
comprehensive scan of 7 parameters, with 10 divisions in each parameter would
require too much CPU time.

Instead, we use a parametrization-based tune method which is much more efficient
for our case.  The starting point for this tuning procedure is the selection of
a range $[p_i^{\rm min},\, p_i^{\rm max}]$ for each of the $N$ tuning parameters
$p_i$. Event samples are generated for random points of this $N$-dimensional
hypercube in the parameter space.  The number of different points depends on the
number of input parameters to ensure a well converging behaviour of the final
tune. Each generated event is directly handed over to the Rivet
package~\cite{Buckley:2010ar} to analyse the generated events.  This allows the
computation of observables for each parameter point, which construct the input
for the tuning process. The obtained distributions of observables for each
parameter variation are the starting point for the main part of the tune, which
is achieved using the Professor framework~\cite{Buckley:2009bj}.  Professor
parametrizes the generator response to the probed parameter points. In that way
it finds the set of parameters, which fits the selected observables best. The
user is able to affect the tuning by applying a weight for each observable,
which specifies the impact of the variable for the tuning process.

\subsubsection{Tuning to minimum-bias data}

\begin{figure*}[htb]
  \includegraphics[width=\colwidth]{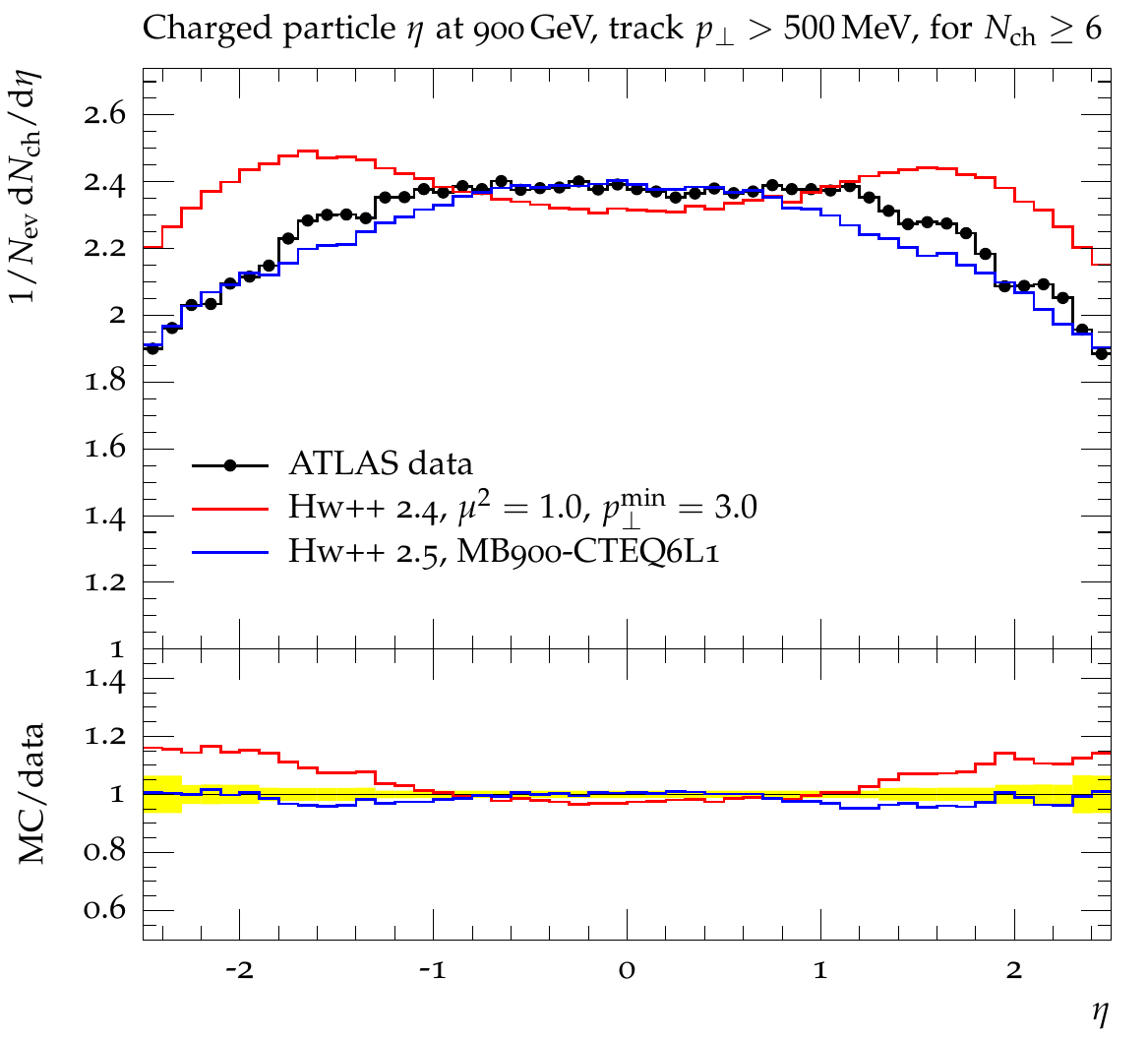}\hfill
  \includegraphics[width=\colwidth]{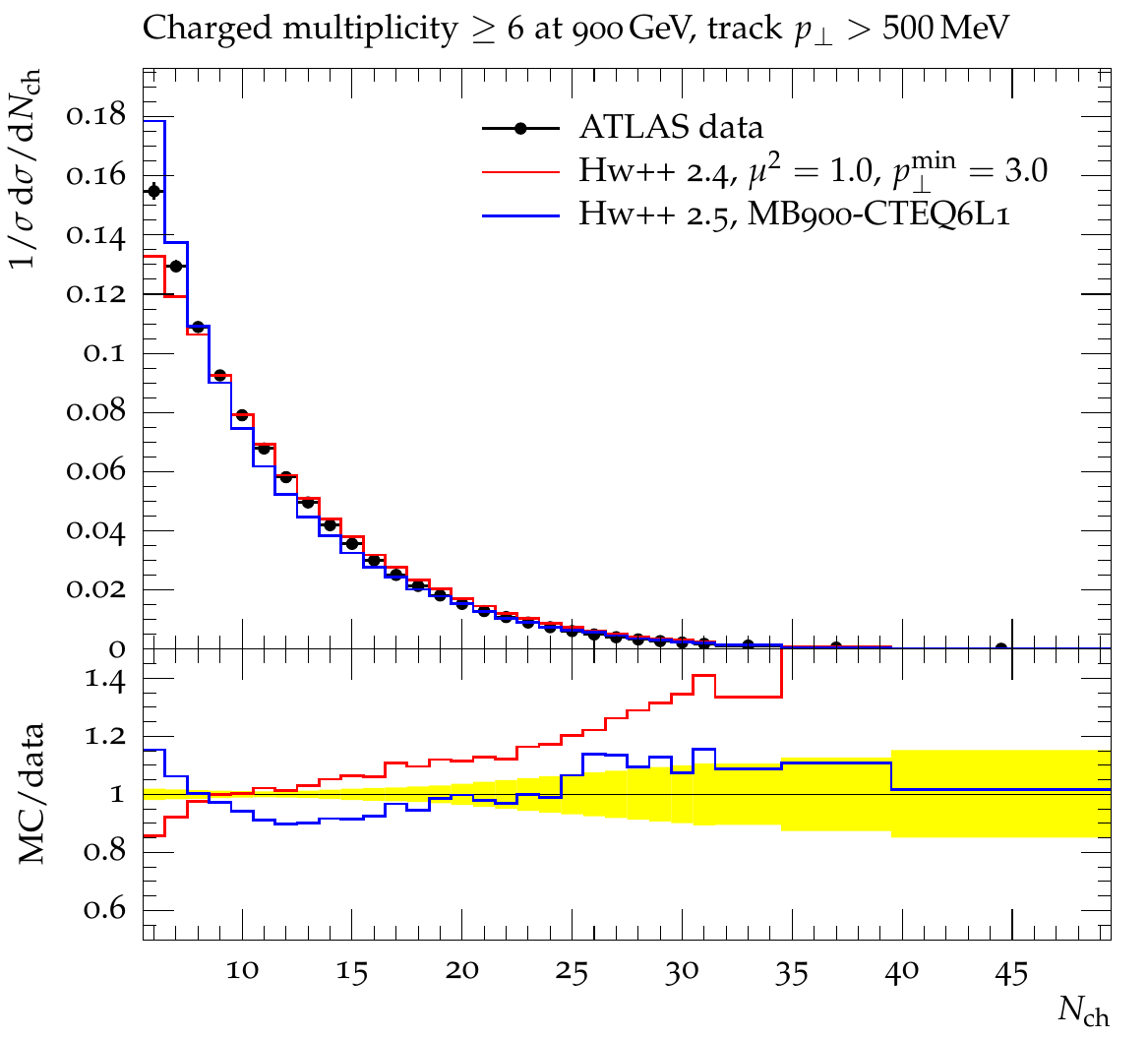}\\
  \includegraphics[width=\colwidth]{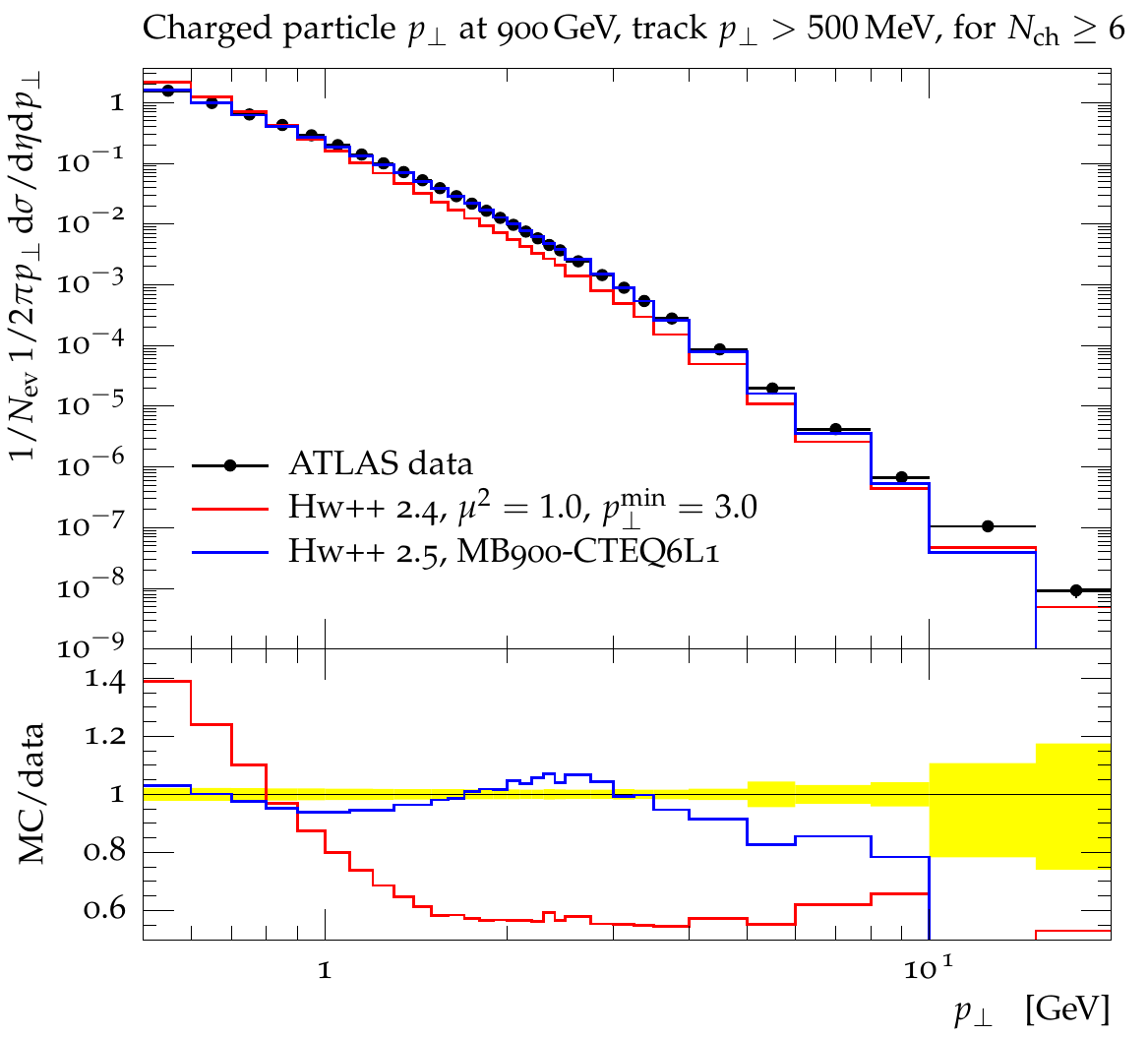}\hfill
  \includegraphics[width=\colwidth]{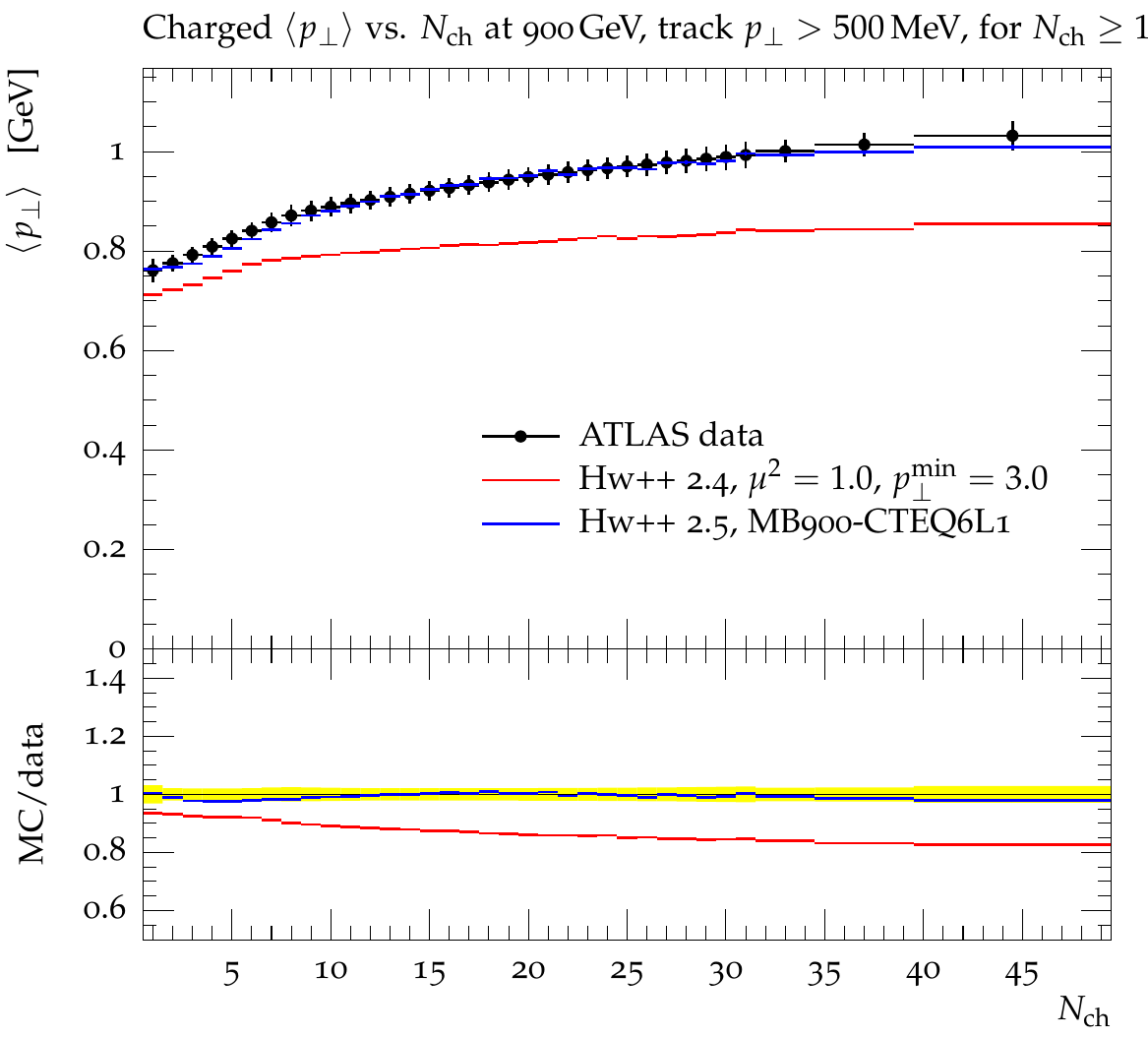}
  \caption{Comparison of \herwig{} 2.4.2 without CR and \herwig{} 2.5 with \pcr
  to ATLAS minimum-bias distributions at $\sqrt{s}=\unit{0.9}{\TeV}$ with
  $N_{\mathrm{ch}} \ge 6$, $p_{\perp} > \unit{500}{\MeV}$ and $|\eta| < 2.5$.
  The ATLAS data was published in Ref.~\cite{:2010ir}.}
  \label{fig:ATLAS_900_Nch6}
\end{figure*}

As we initially were primarily aiming at an improved description of MB data, we
started by tuning the \pcr model to ATLAS MB data. Since currently there is  no
model for soft diffractive physics in \herwig{}, we use the diffraction-reduced
\mbox{ATLAS} MB measurement with an additional cut on the number of charged
particles, $N_{\rm ch} \ge 6$. The observables we used for the tune are the
pseudorapidity distribution of the charged particles, the charged multiplicity,
the charged-particle transverse momentum spectrum and the average transverse
momentum measured as a function of the number of charged particles.  All four
available MB observables entered the tune with equal weights.  The results of
this tune are shown by the blue lines in Fig.~\ref{fig:ATLAS_900_Nch6}.  The
bottom right figure shows that colour reconnection helps to achieve a better
description of $\langle p_T\rangle (N_{\rm ch})$.  Also the other three
distributions are now well described.  We conclude that the CR model was the
missing piece of the MPI model in \herwig{}++.  We clearly improve the
description of the pseudorapidity distribution.  

\subsubsection{Tuning to underlying-event data}

The next important question was whether the new model is able to describe the UE
data collected by ATLAS at \unit{7}{\TeV} \cite{Aad:2010fh}.  The measurements
are made relative to a leading object (the hardest charged track in this case).
Then, the transverse plane is subdivided in azimuthal angle $\phi$ relative to
this leading object at $\phi=0$.  The region around the leading object,
$|\phi|<\pi/3$, is called the ``towards'' region.  The opposite region, where we
usually find a recoiling hard jet, $|\phi| > 2\pi/3$, is called ``away'' region,
while the remaining region, transverse to the leading object and its recoil,
where the underlying event is expected to be least `contaminated' by activity
from the hard subprocess, is called ``transverse'' region.  Again, we only focus
on the tuning of the \pcr model here. For the underlying-event tune two
observables were used: The mean number of stable charged particles per unit of
$\eta$-$\phi$, $\dNchgdetadphi$, and the mean scalar $p_{\perp}$ sum of stable
particles per unit of $\eta$-$\phi$, $\dpTsumdetadphi$, both as a function of
$\ptlead$, with charged particles in the kinematic range $p_{\perp} >
\unit{500}{\MeV}$ and $|\eta| < 2.5$.

\begin{figure*}[f]
  \parbox[t]{\subfigwidth}{
  \centering
  \includegraphics[width=\subfigwidth]{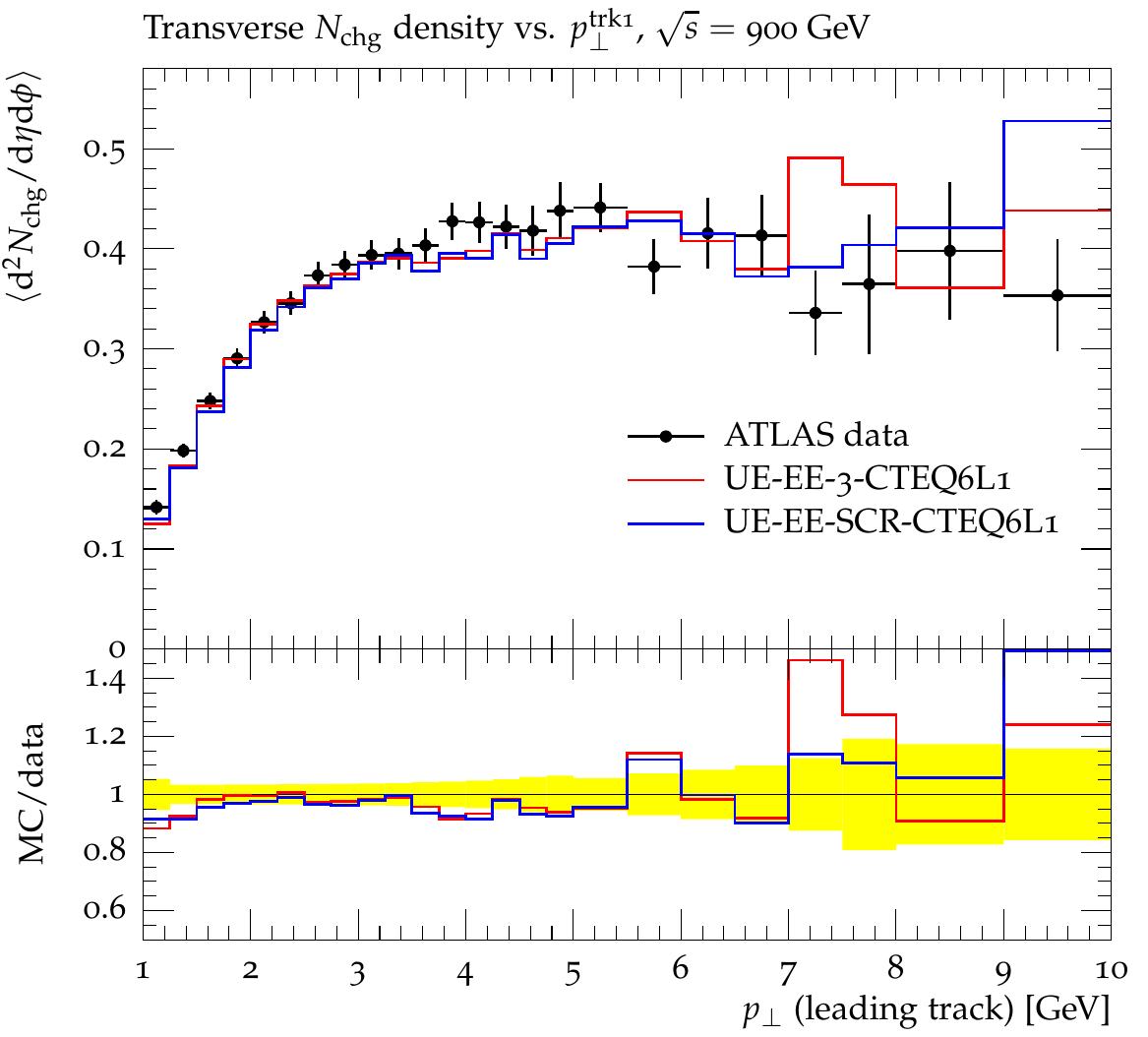}\\
  \includegraphics[width=\subfigwidth]{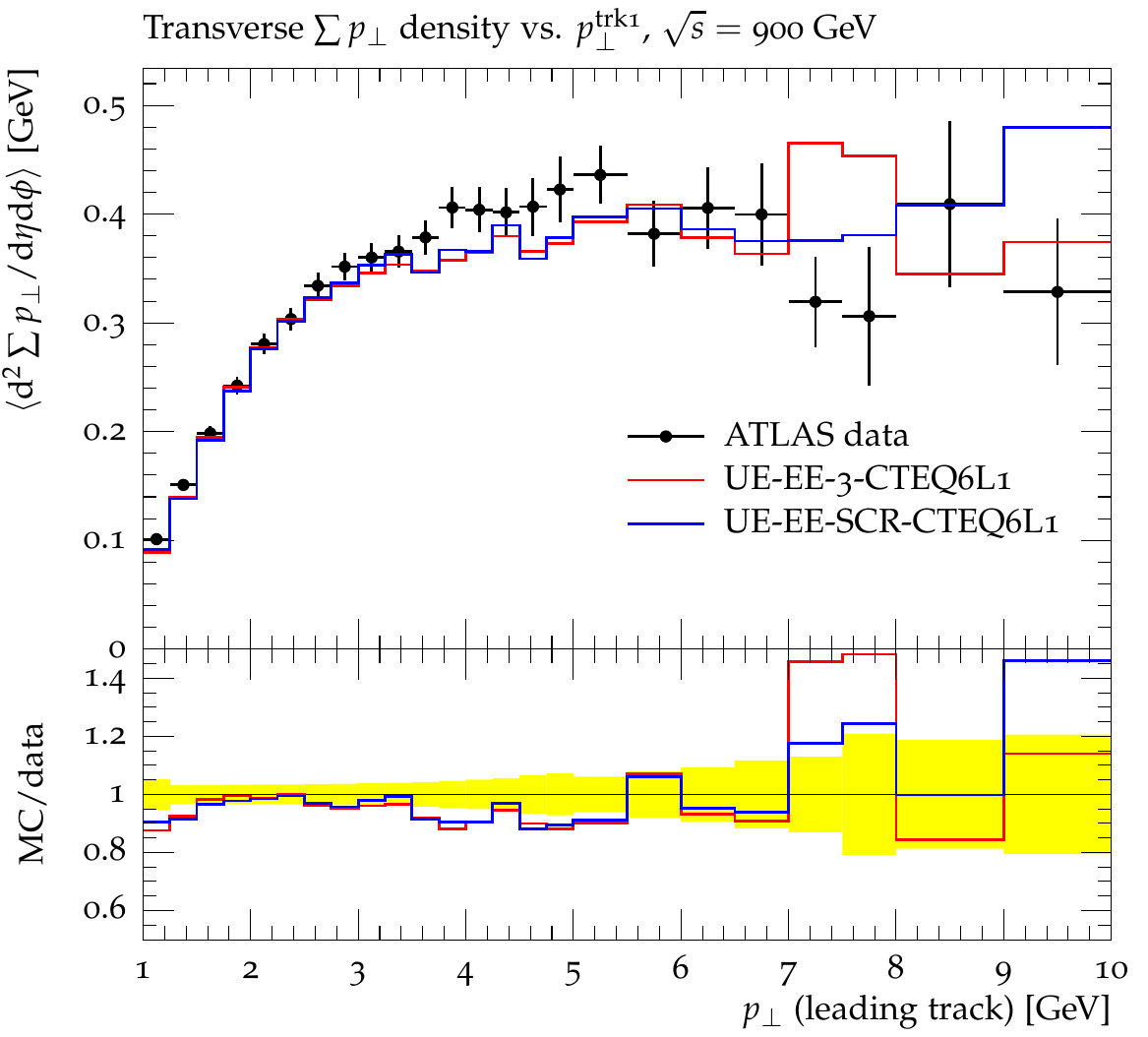}
  \\(a)}
  \hfill
  \parbox[t]{\subfigwidth}{
  \centering
  \includegraphics[width=\subfigwidth]{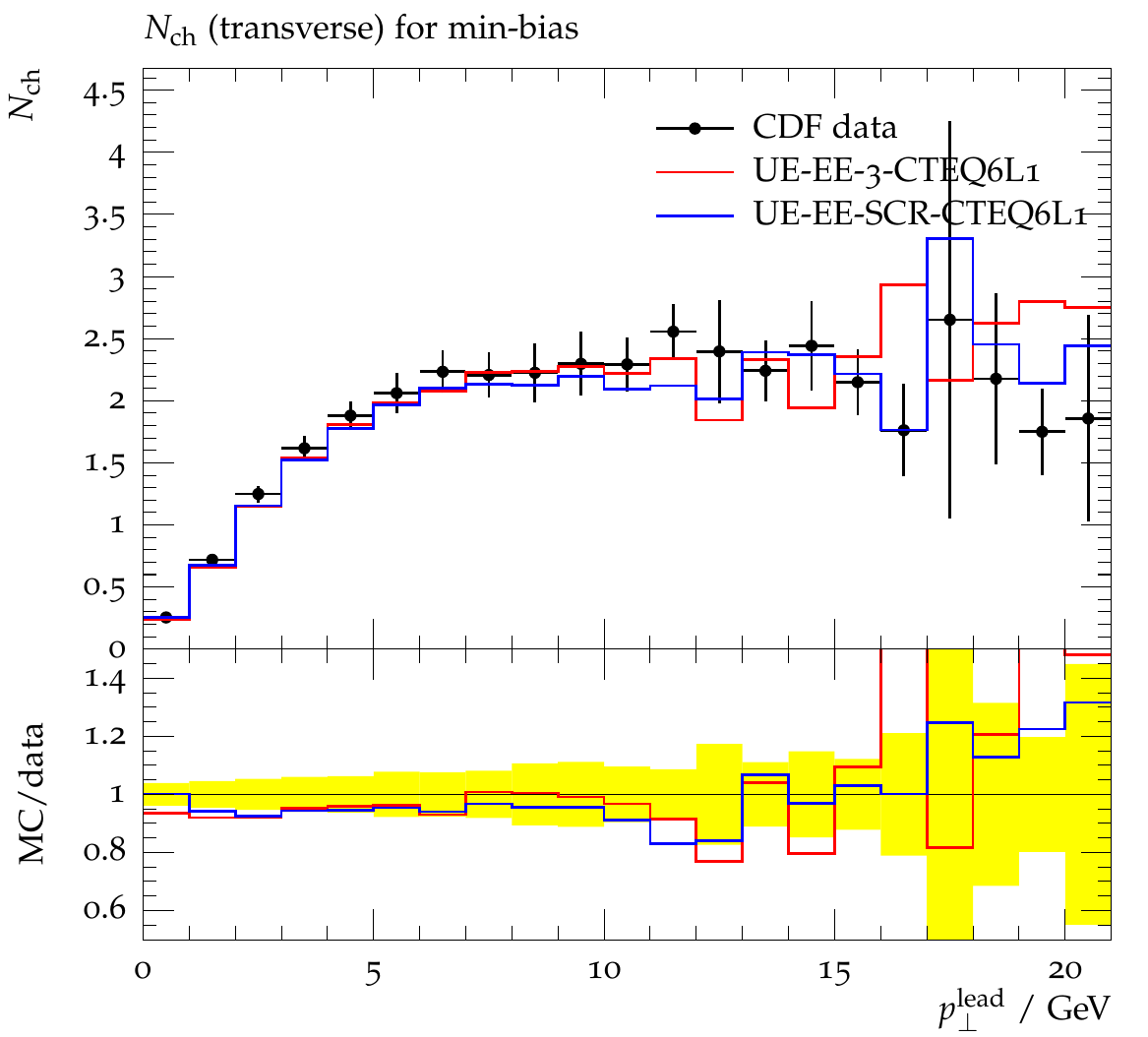}\\
  \includegraphics[width=\subfigwidth]{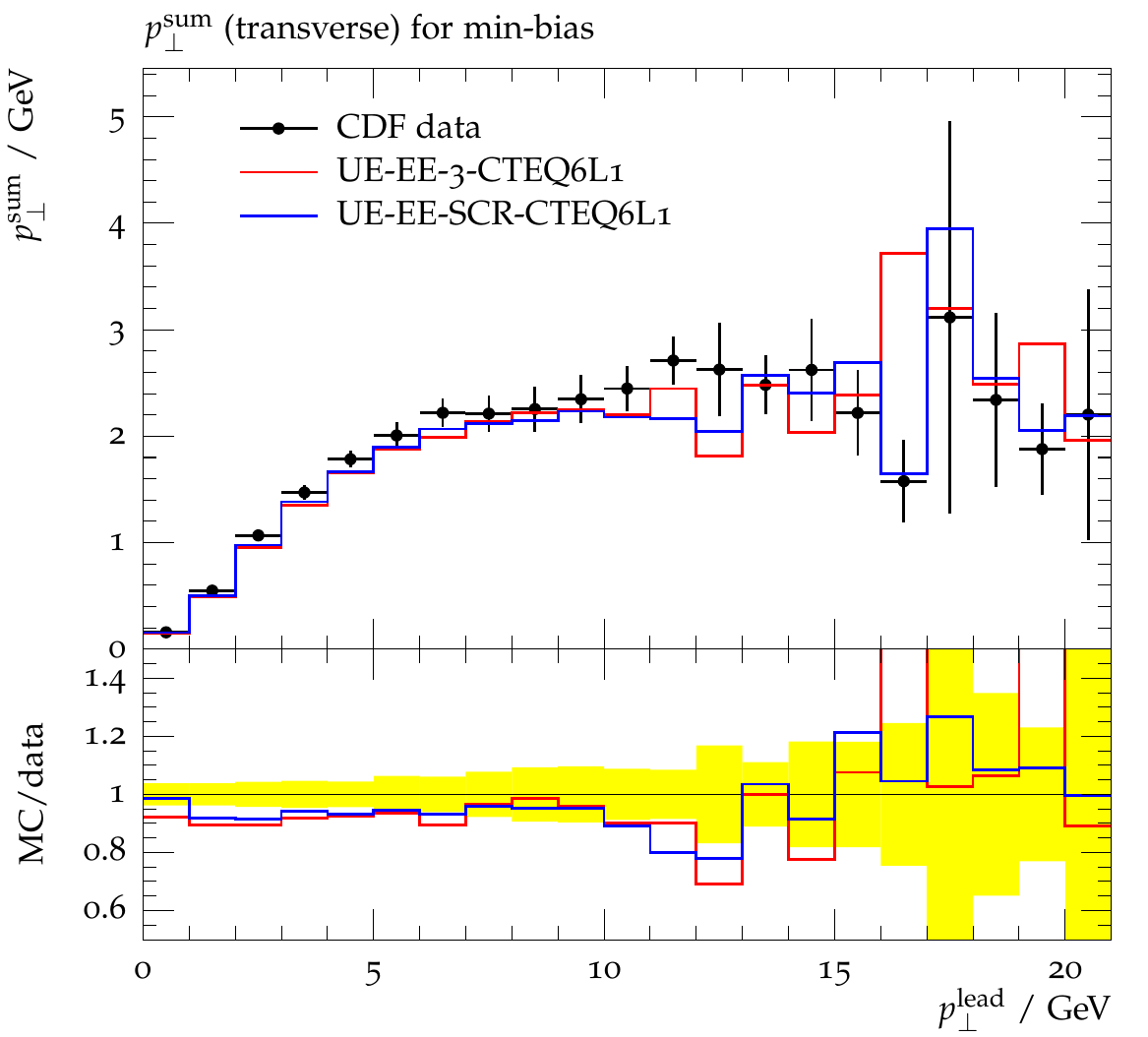}
  \\(b)}
  \hfill
  \parbox[t]{\subfigwidth}{
  \centering
  \includegraphics[width=\subfigwidth]{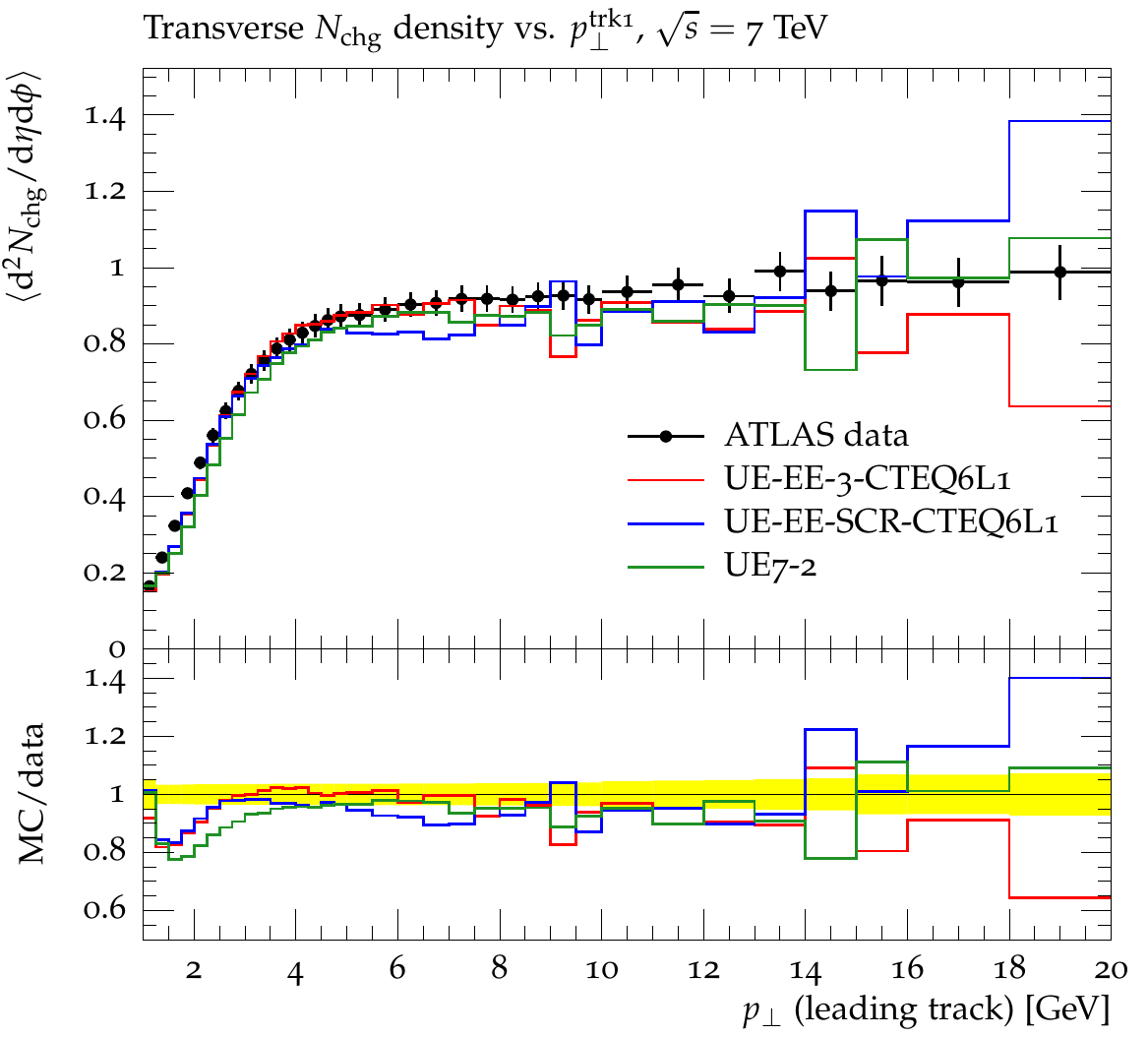}\\
  \includegraphics[width=\subfigwidth]{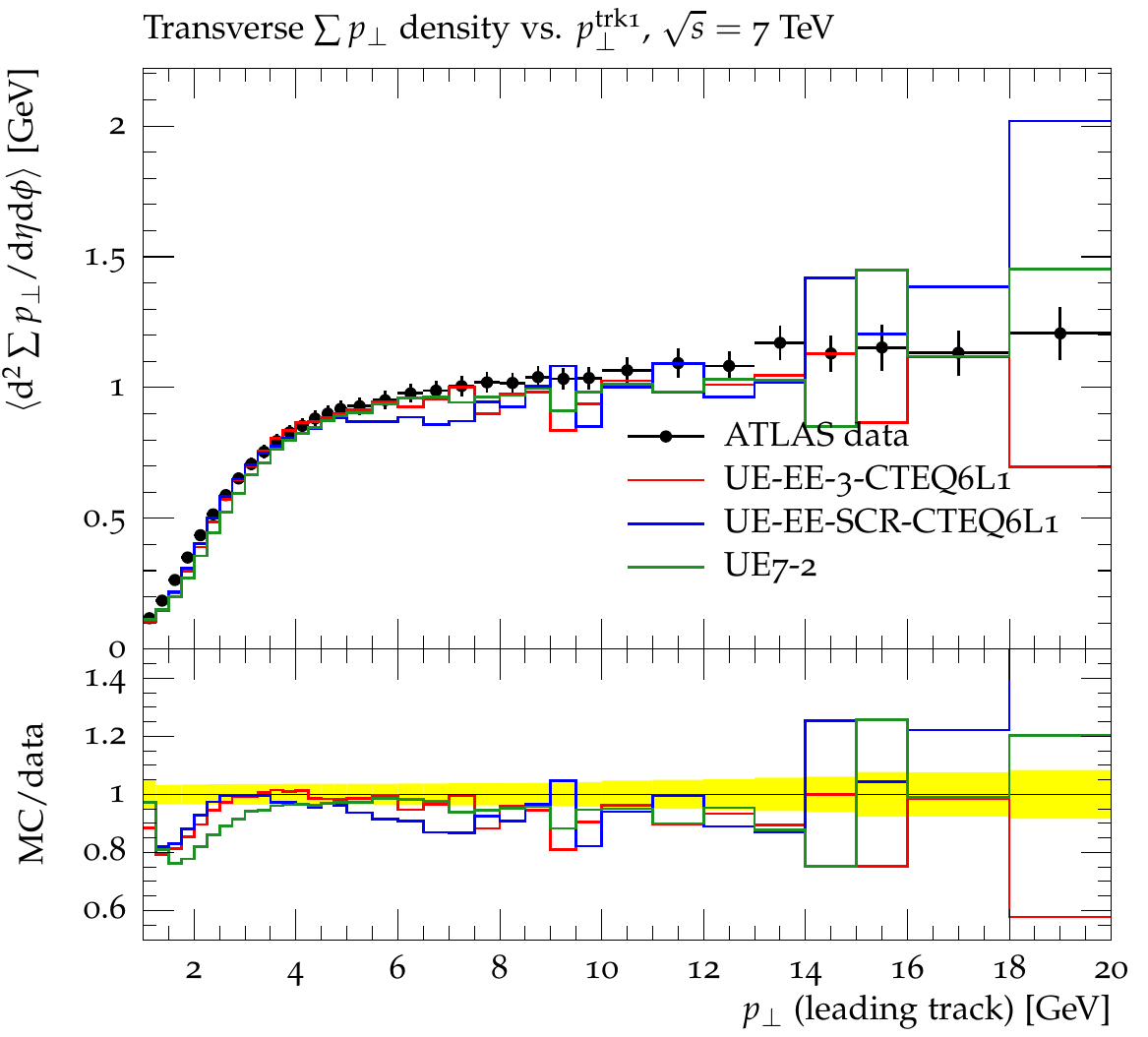}
  \\(c)}
  \caption{ATLAS data at \unit{900}{\GeV} (1st column), CDF data at
  \unit{1800}{\GeV} (2nd column) and ATLAS data at \unit{7}{\TeV} (3rd column),
  showing the multiplicity density and $\ptsum$ of the charged particles in the
  ``transverse'' area as a function of $\ptlead$. The data is compared to the
  \UEvii, \EEiiiCTEQ and \EESCRCTEQ tunes.}
  \label{fig:UE-comparison-transverse}
\end{figure*}

\begin{figure*}[f]
  \parbox[t]{\subfigwidth}{
  \centering
  \includegraphics[width=\subfigwidth]{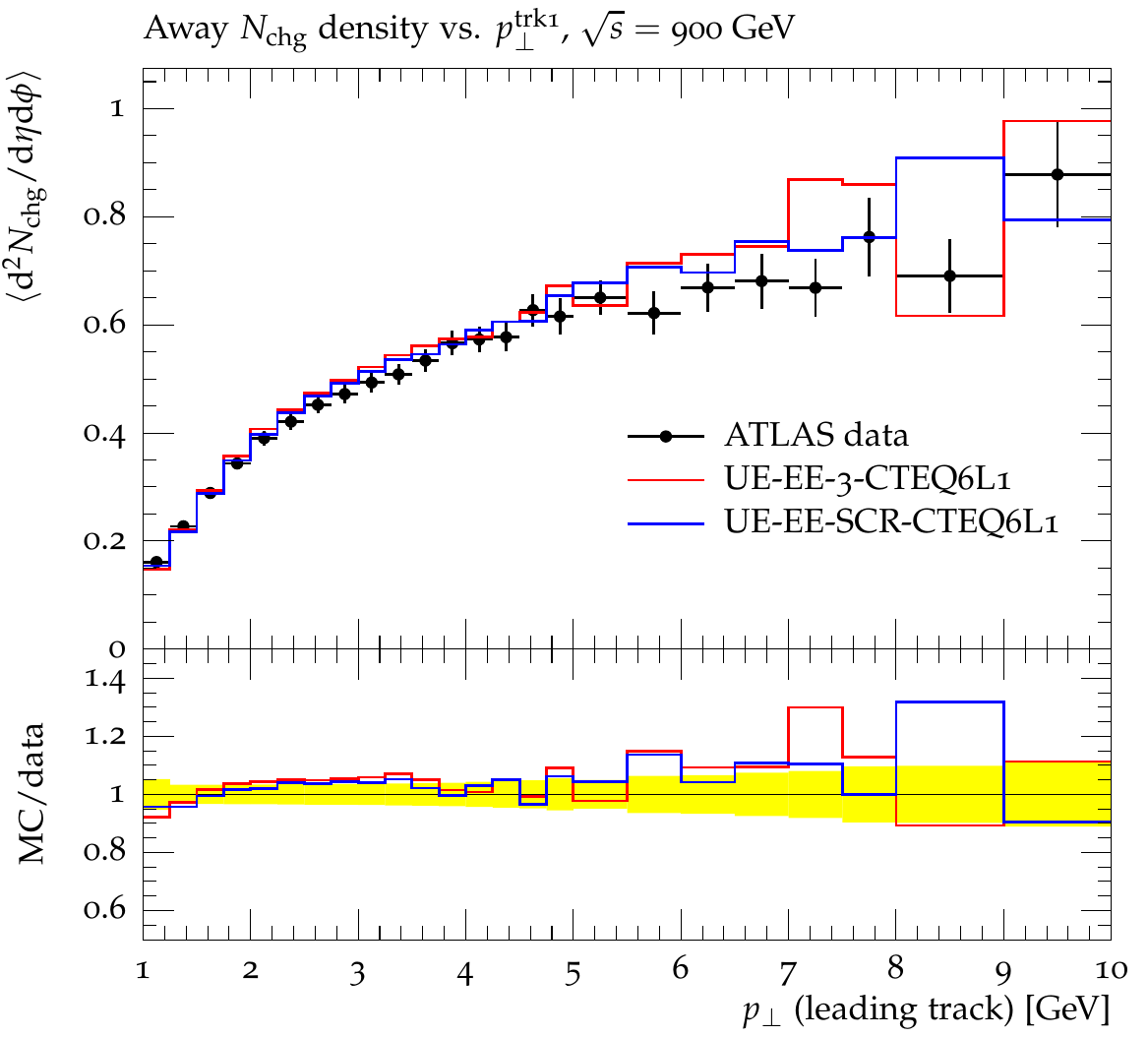}\\
  \includegraphics[width=\subfigwidth]{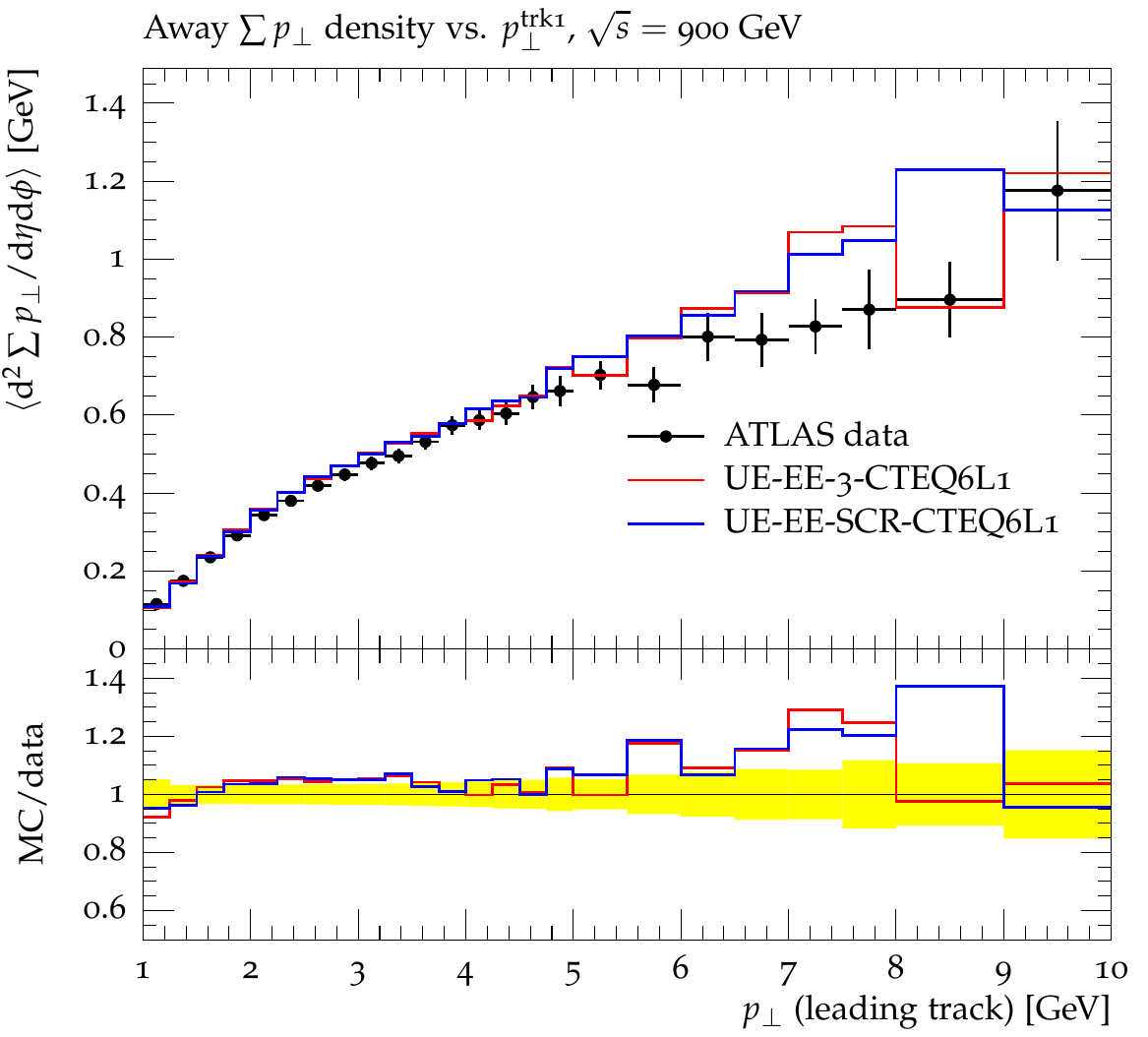}
  \\(a)}
  \hfill
  \parbox[t]{\subfigwidth}{
  \centering
  \includegraphics[width=\subfigwidth]{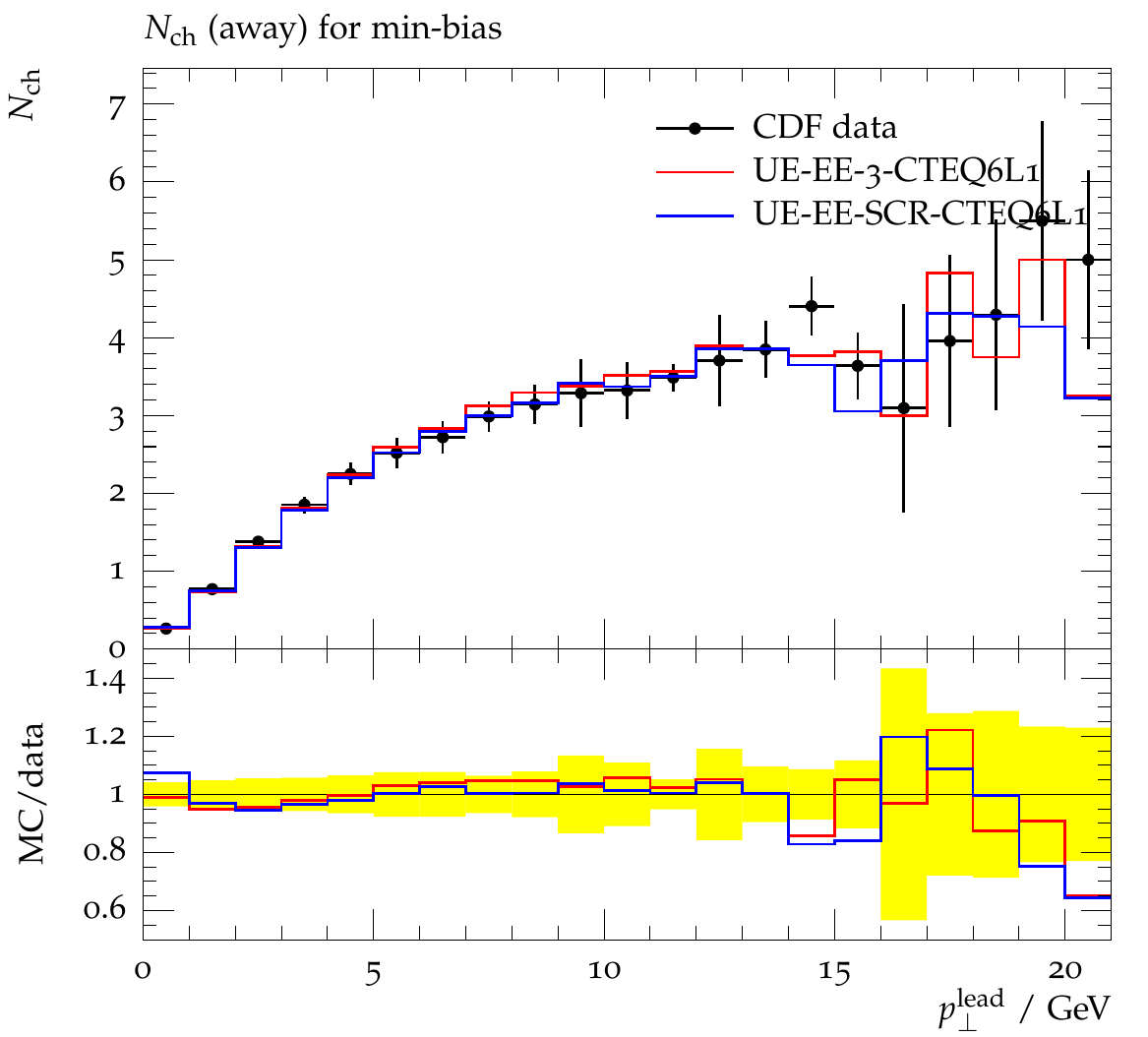}\\
  \includegraphics[width=\subfigwidth]{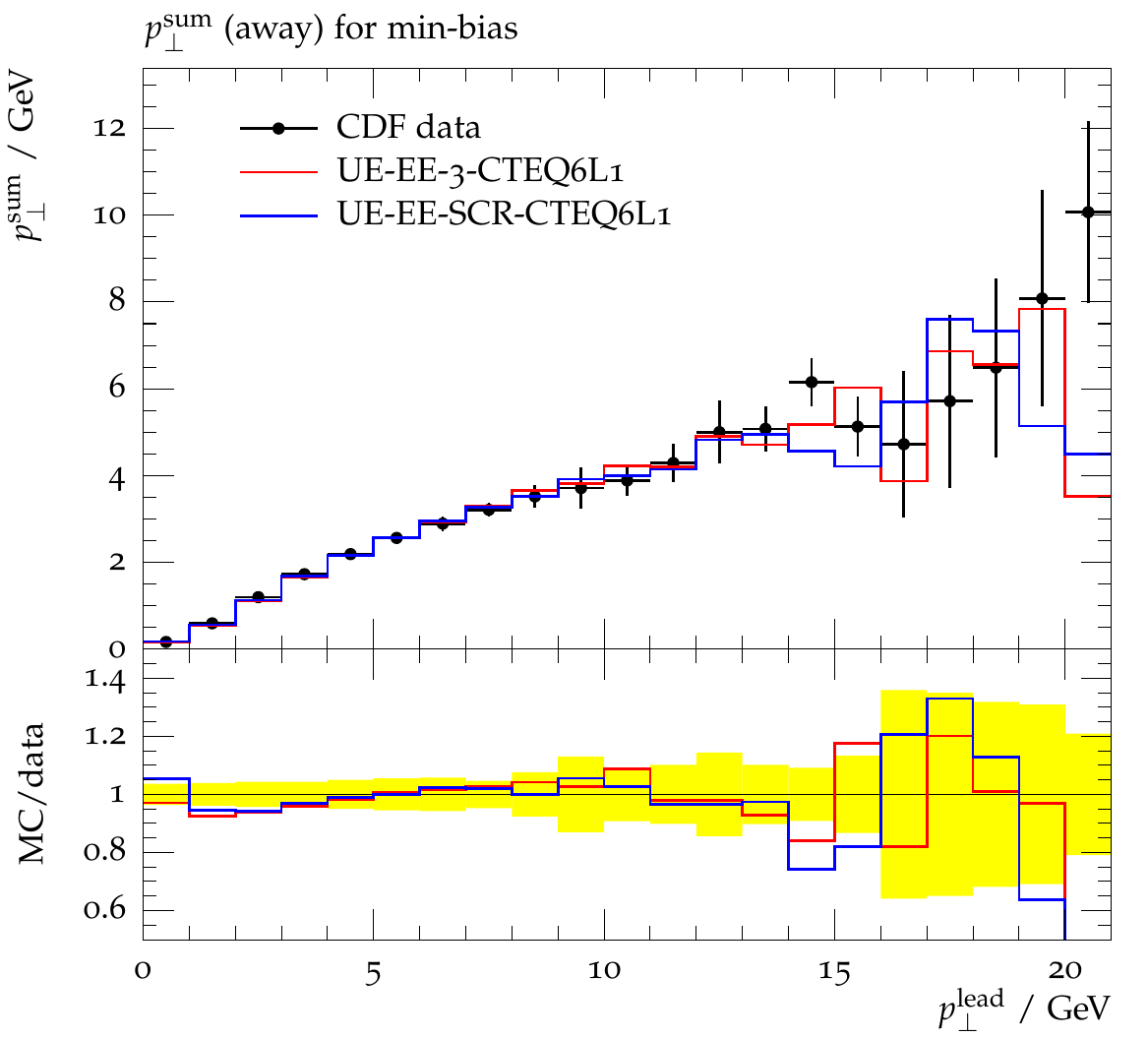}
  \\(b)}
  \hfill
  \parbox[t]{\subfigwidth}{
  \centering
  \includegraphics[width=\subfigwidth]{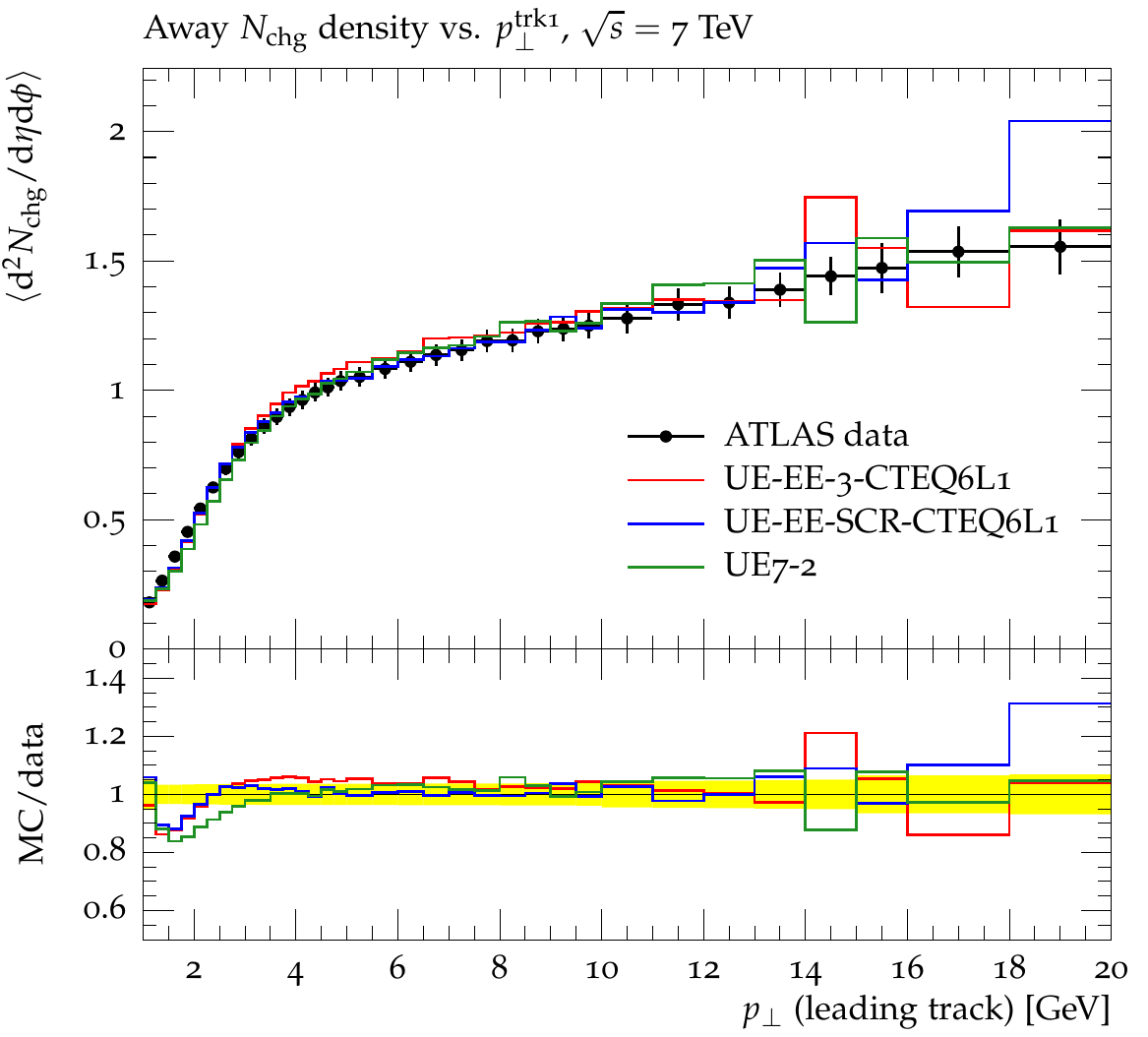}\\
  \includegraphics[width=\subfigwidth]{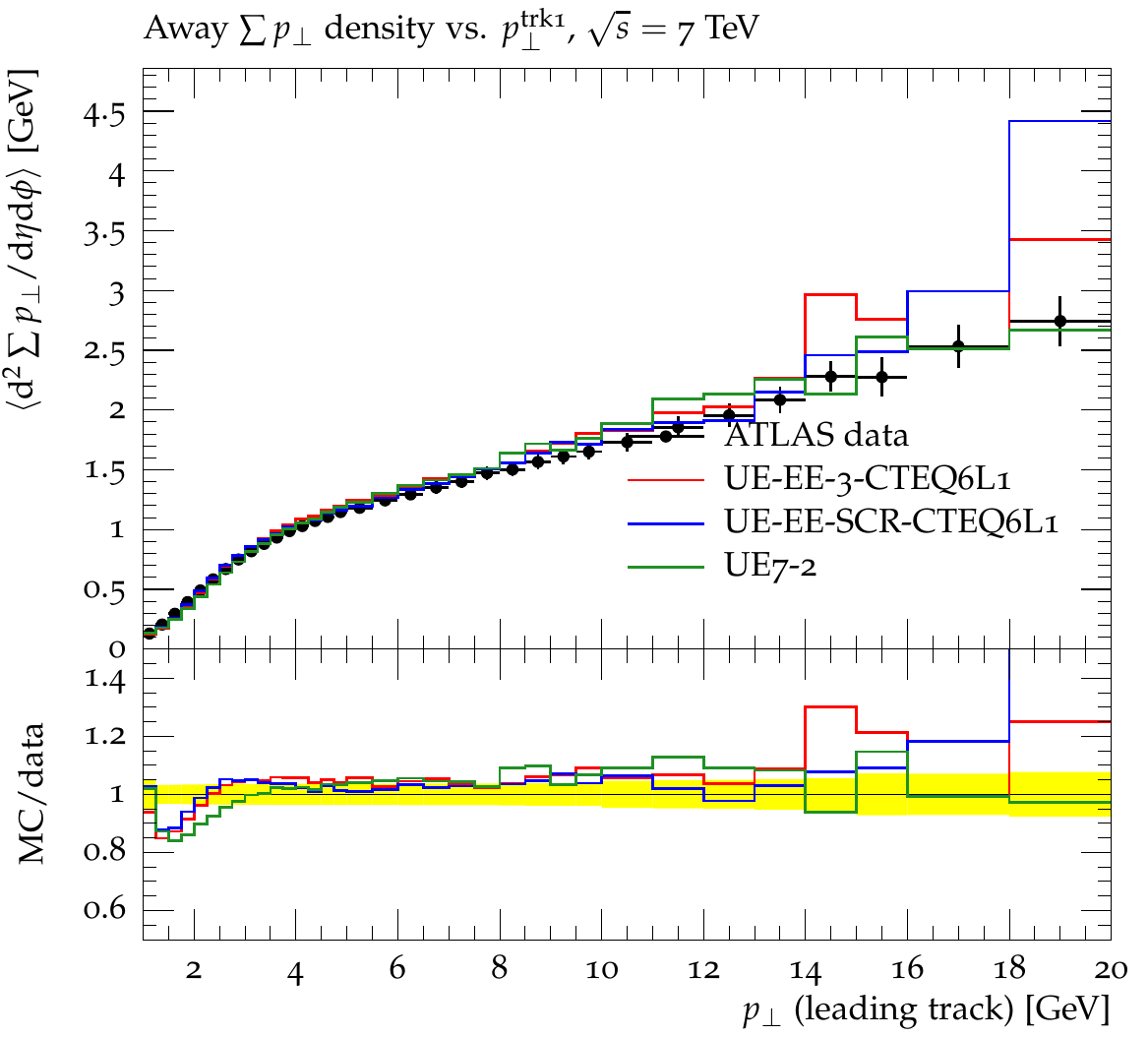}
  \\(c)}
  \caption{Same as Fig.~\ref{fig:UE-comparison-transverse}, but with the
  observables measured in the ``away'' region.}
  \label{fig:UE-comparison-away}
\end{figure*}

\begin{figure*}[f]
  \parbox[t]{\subfigwidth}{
  \centering
  \includegraphics[width=\subfigwidth]{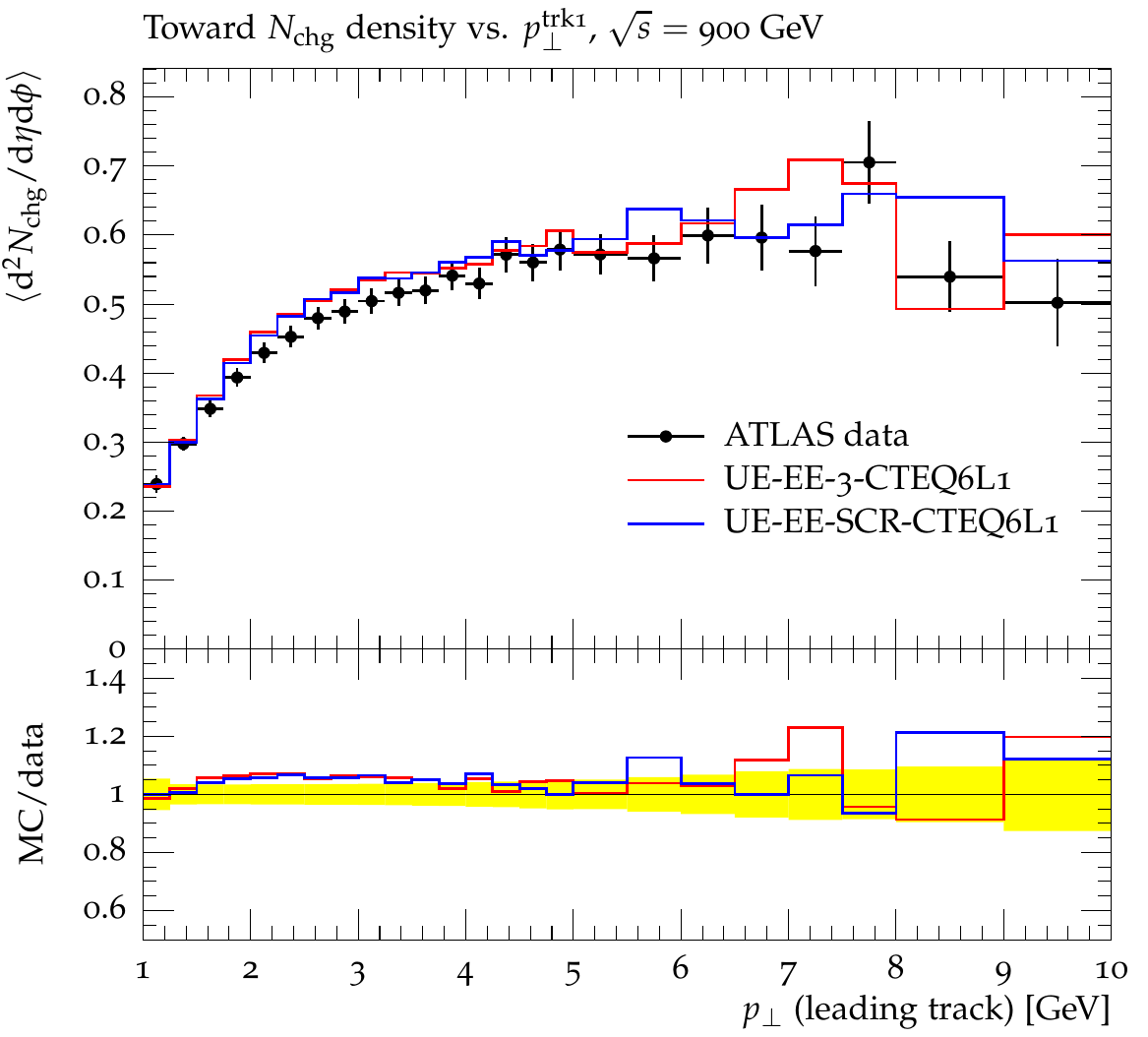}\\
  \includegraphics[width=\subfigwidth]{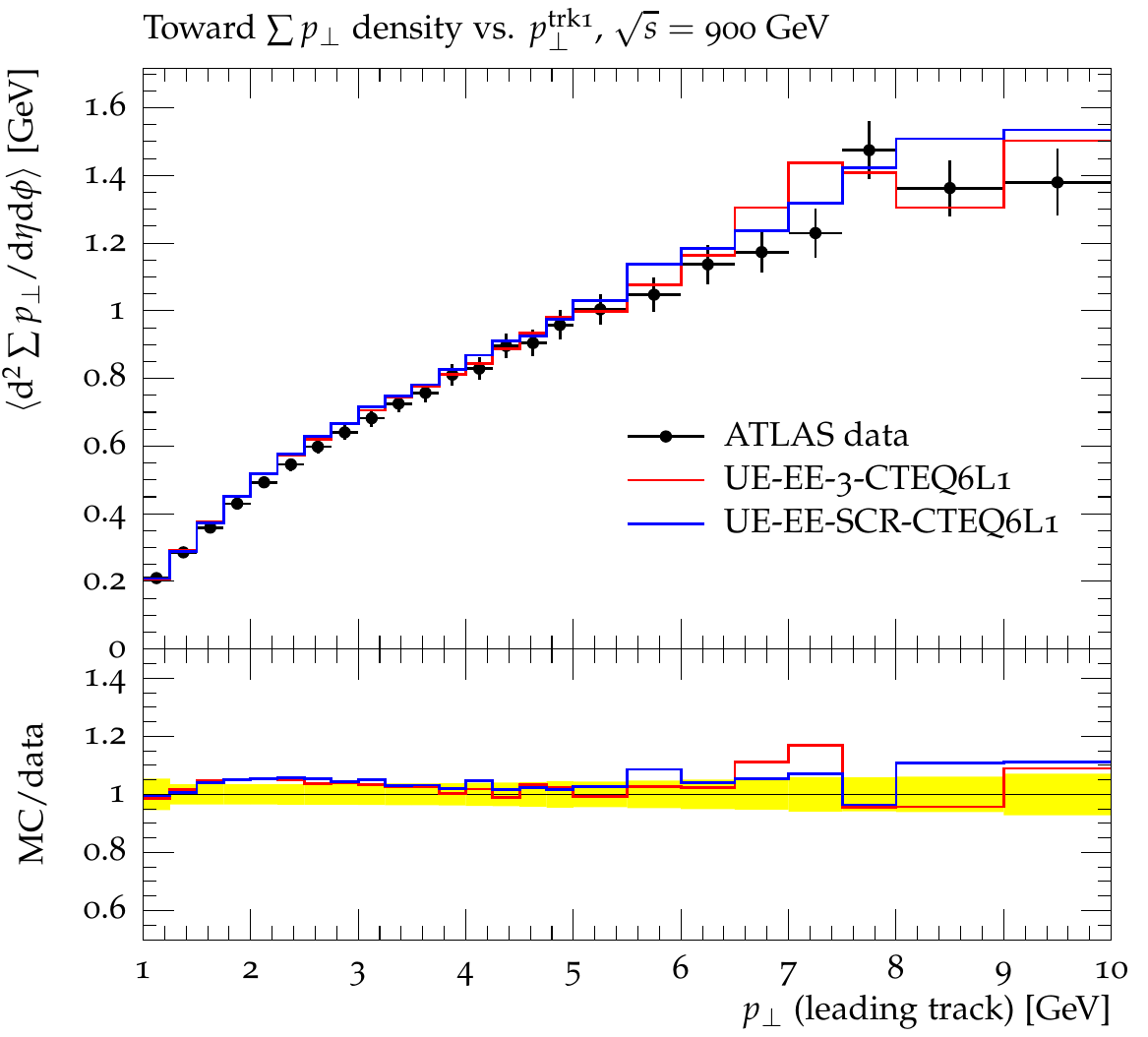}
  \\(a)}
  \hfill
  \parbox[t]{\subfigwidth}{
  \centering
  \includegraphics[width=\subfigwidth]{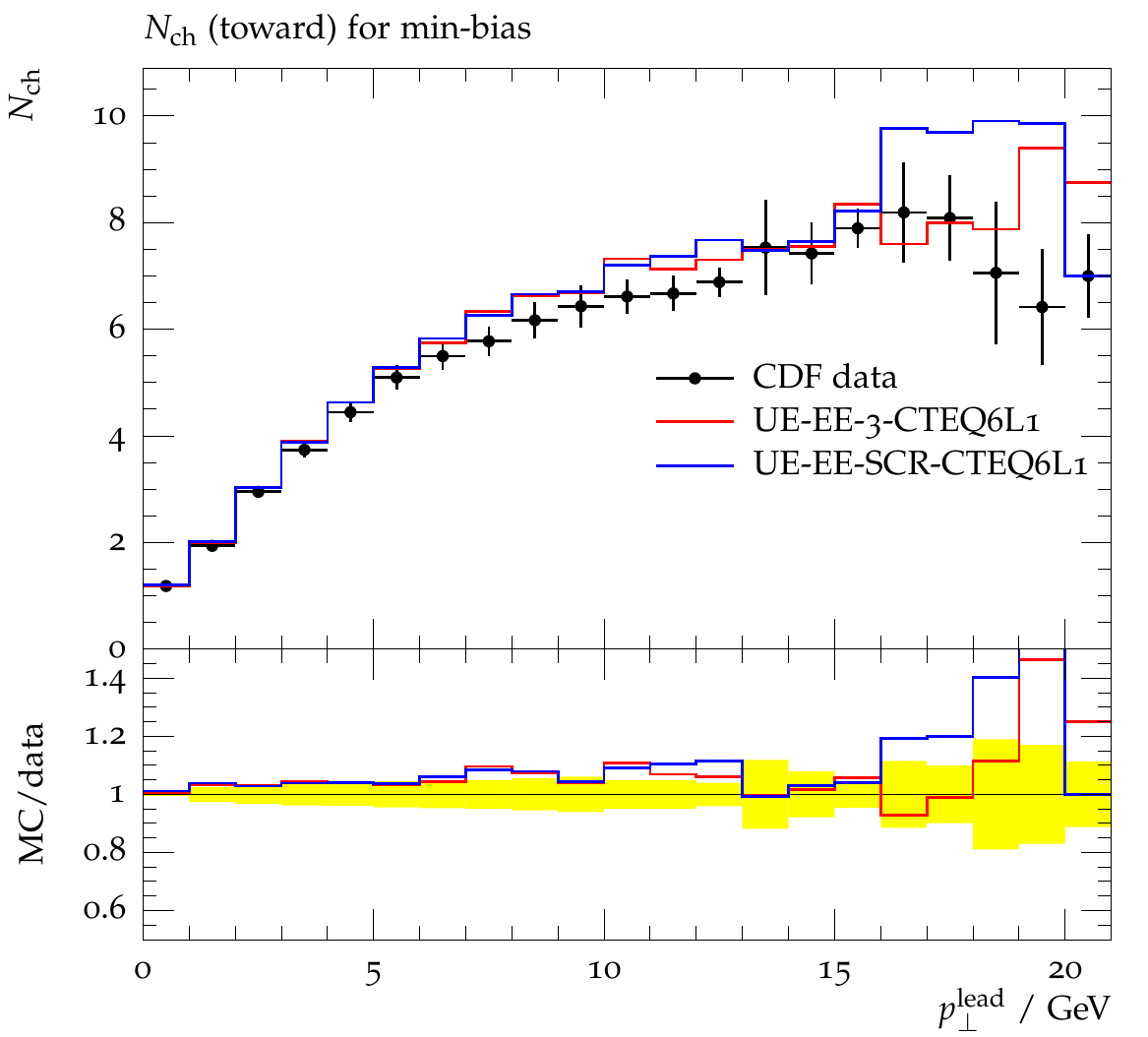}\\
  \includegraphics[width=\subfigwidth]{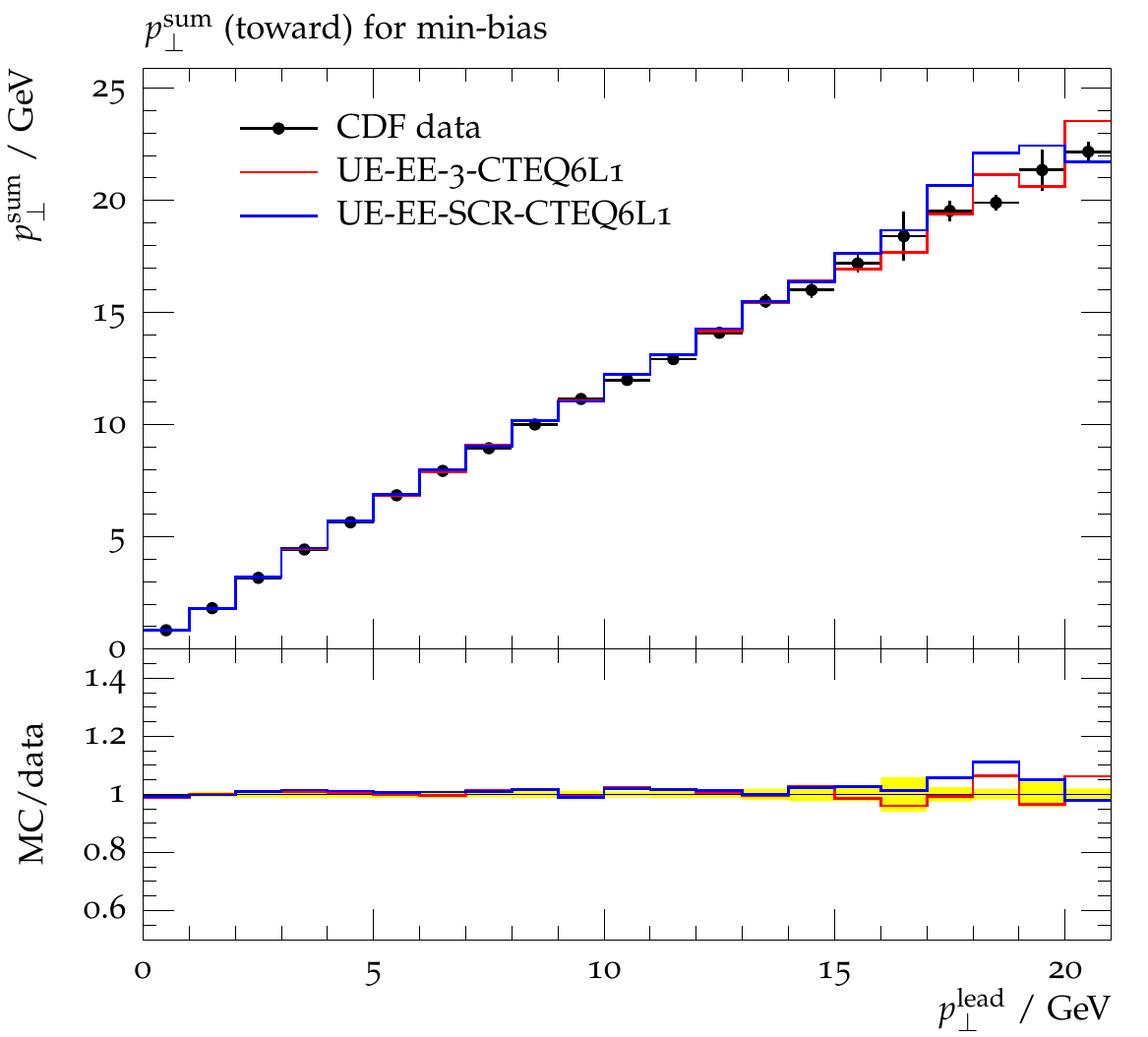}
  \\(b)}
  \hfill
  \parbox[t]{\subfigwidth}{
  \centering
  \includegraphics[width=\subfigwidth]{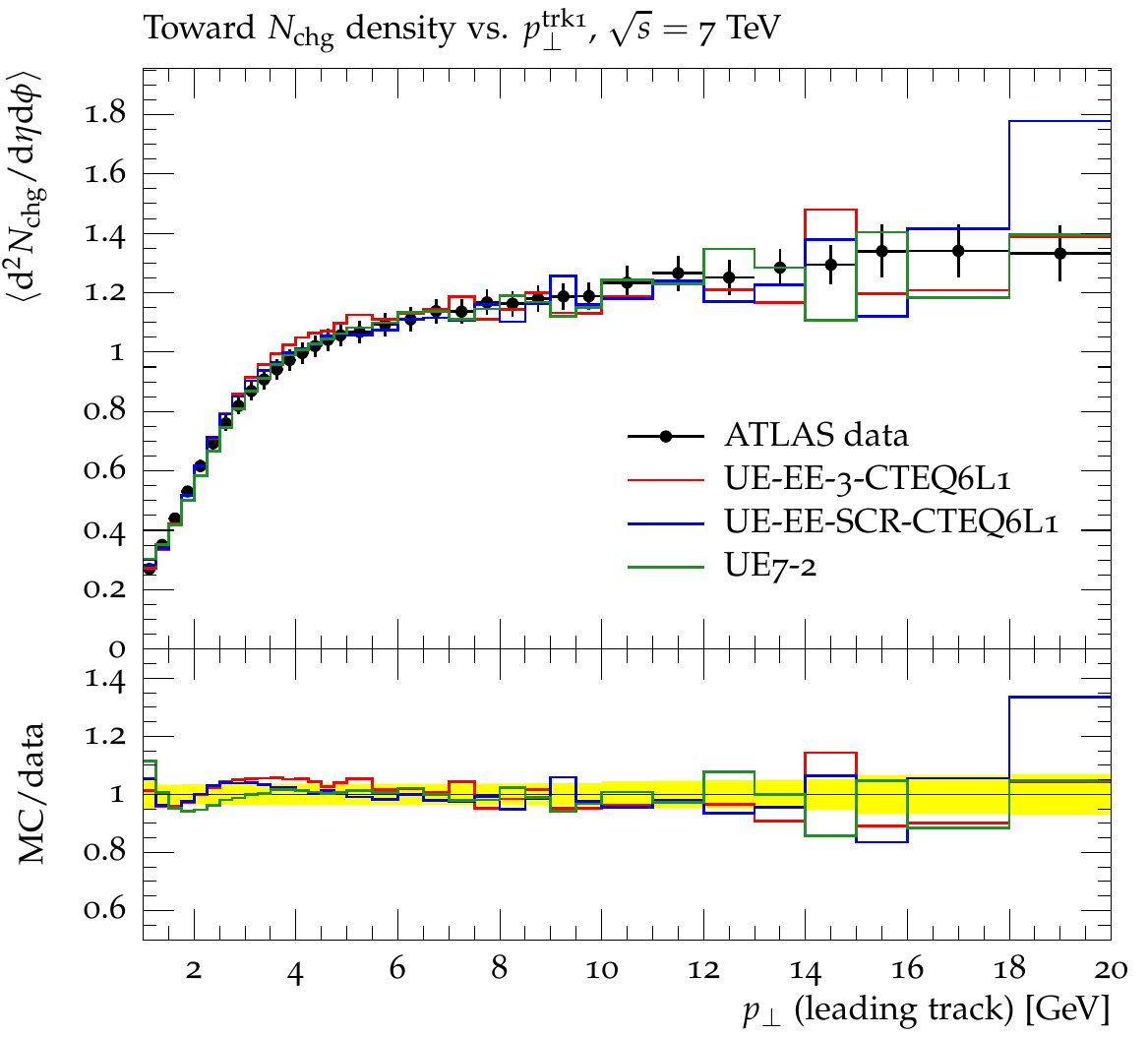}\\
  \includegraphics[width=\subfigwidth]{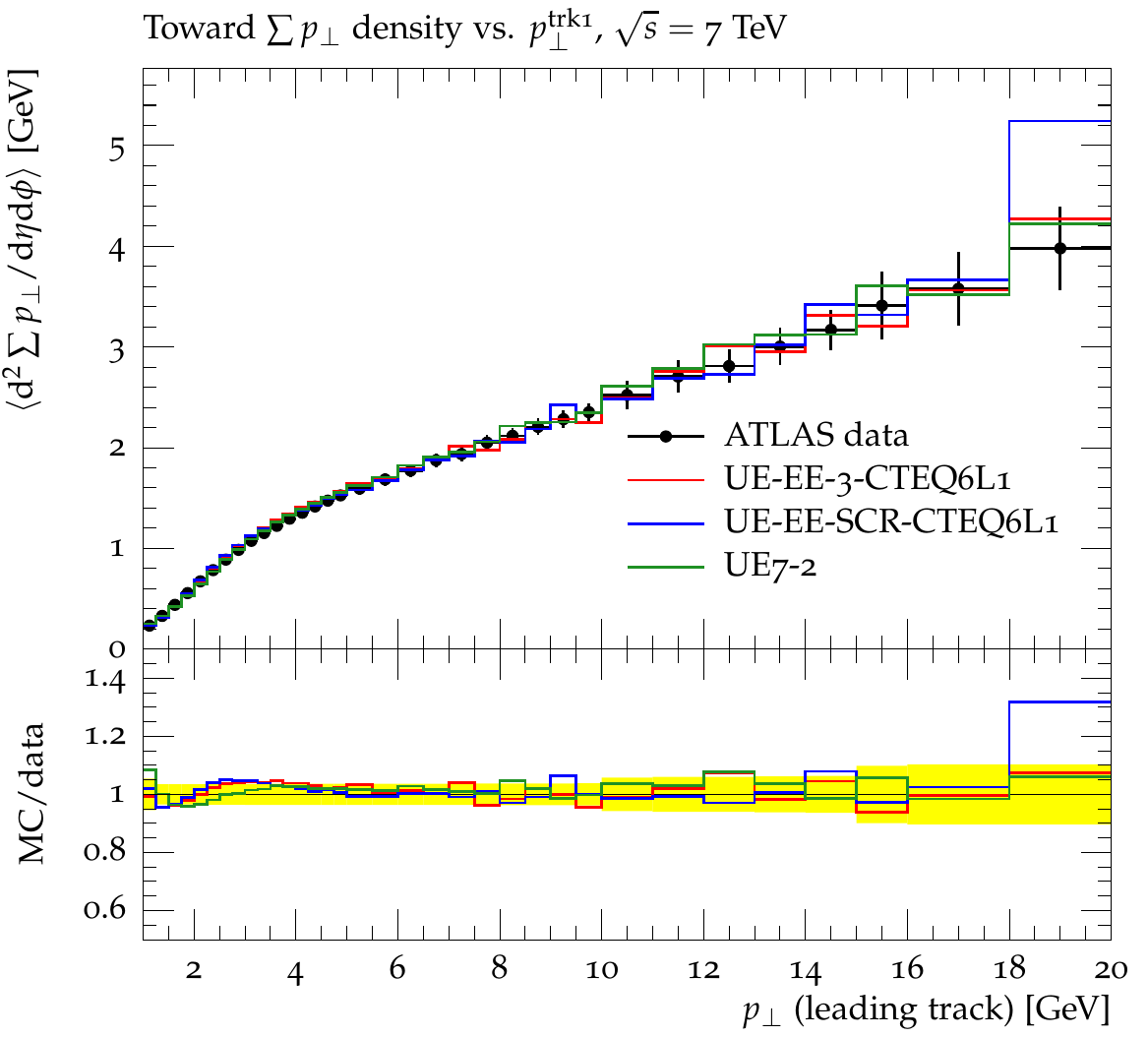}
  \\(c)}
  \caption{Same as Fig.~\ref{fig:UE-comparison-transverse}, but with the
  observables measured in the ``toward'' region.}
  \label{fig:UE-comparison-toward}
\end{figure*}

The resulting tune, named \UEvii, gives very satisfactory results not only for
the tuned observables but also for all other observables provided by ATLAS in
Ref.~\cite{Aad:2010fh}.  In Figs.~\ref{fig:UE-comparison-transverse}(c),
\ref{fig:UE-comparison-away}(c) and \ref{fig:UE-comparison-toward}(c),  we
show $\dNchgdetadphi$ and $\dpTsumdetadphi$ as a function of $\ptlead$ for
$p_{\perp} > \unit{500}{\MeV}$ in the ``transverse'', ``away'' and ``toward''
regions, compared to the \herwig{}++ \UEvii results (green line).

\begin{figure*}[f]
  \parbox[t]{\subfigwidth}{
  \centering
  \includegraphics[width=\subfigwidth]{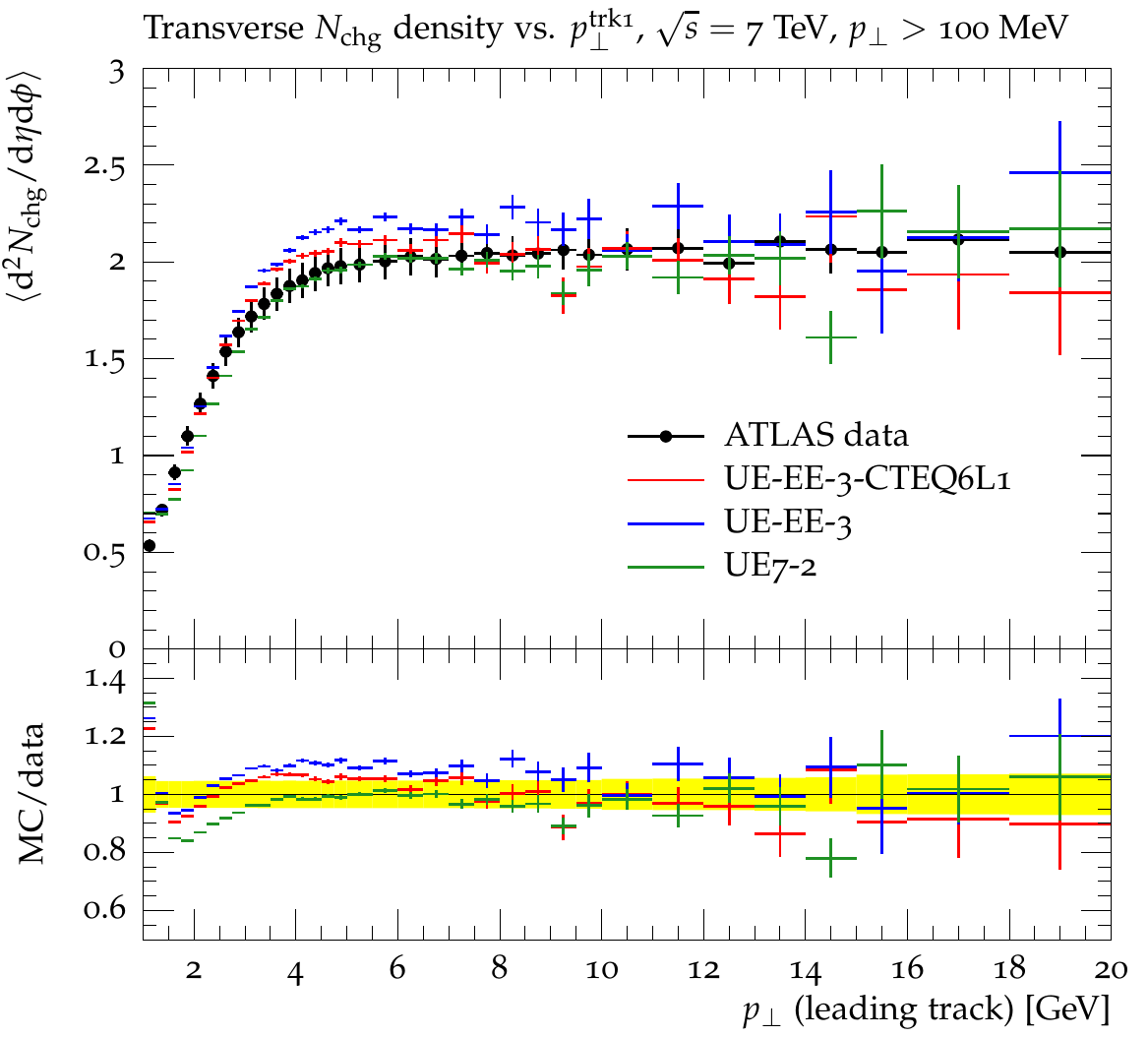}\\
  \includegraphics[width=\subfigwidth]{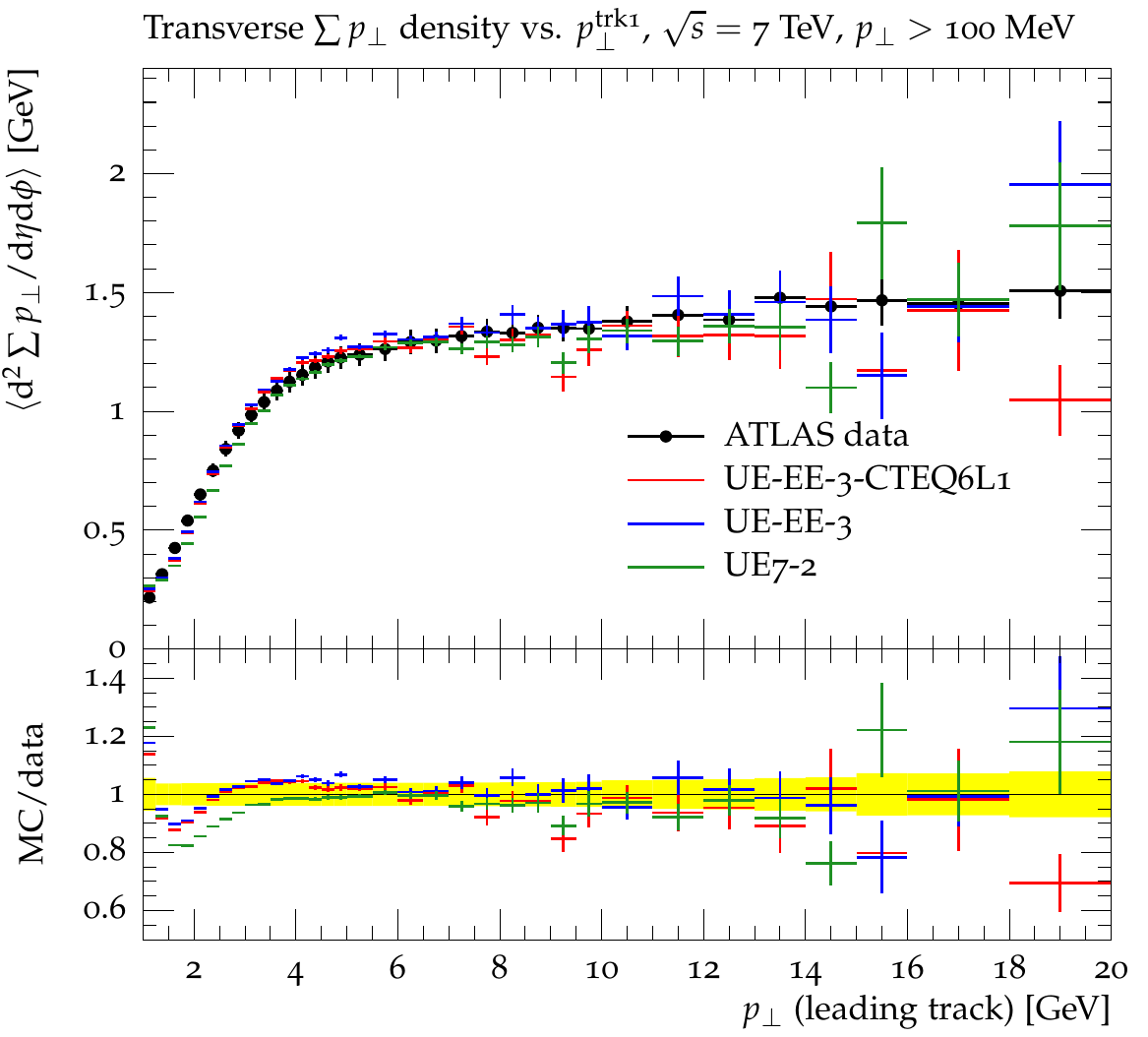}
  \\(a)}
  \hfill
  \parbox[t]{\subfigwidth}{
  \centering
  \includegraphics[width=\subfigwidth]{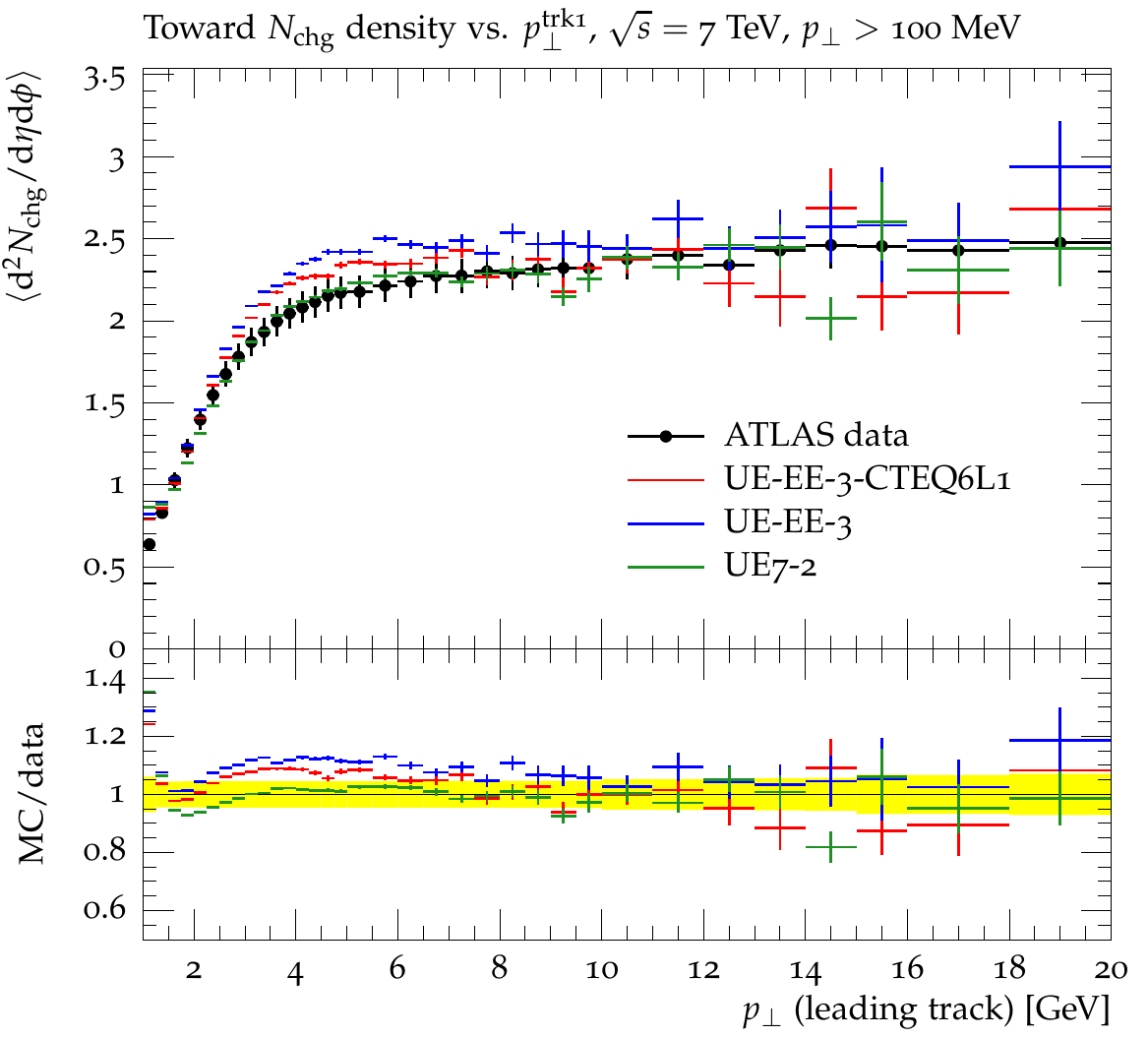}\\
  \includegraphics[width=\subfigwidth]{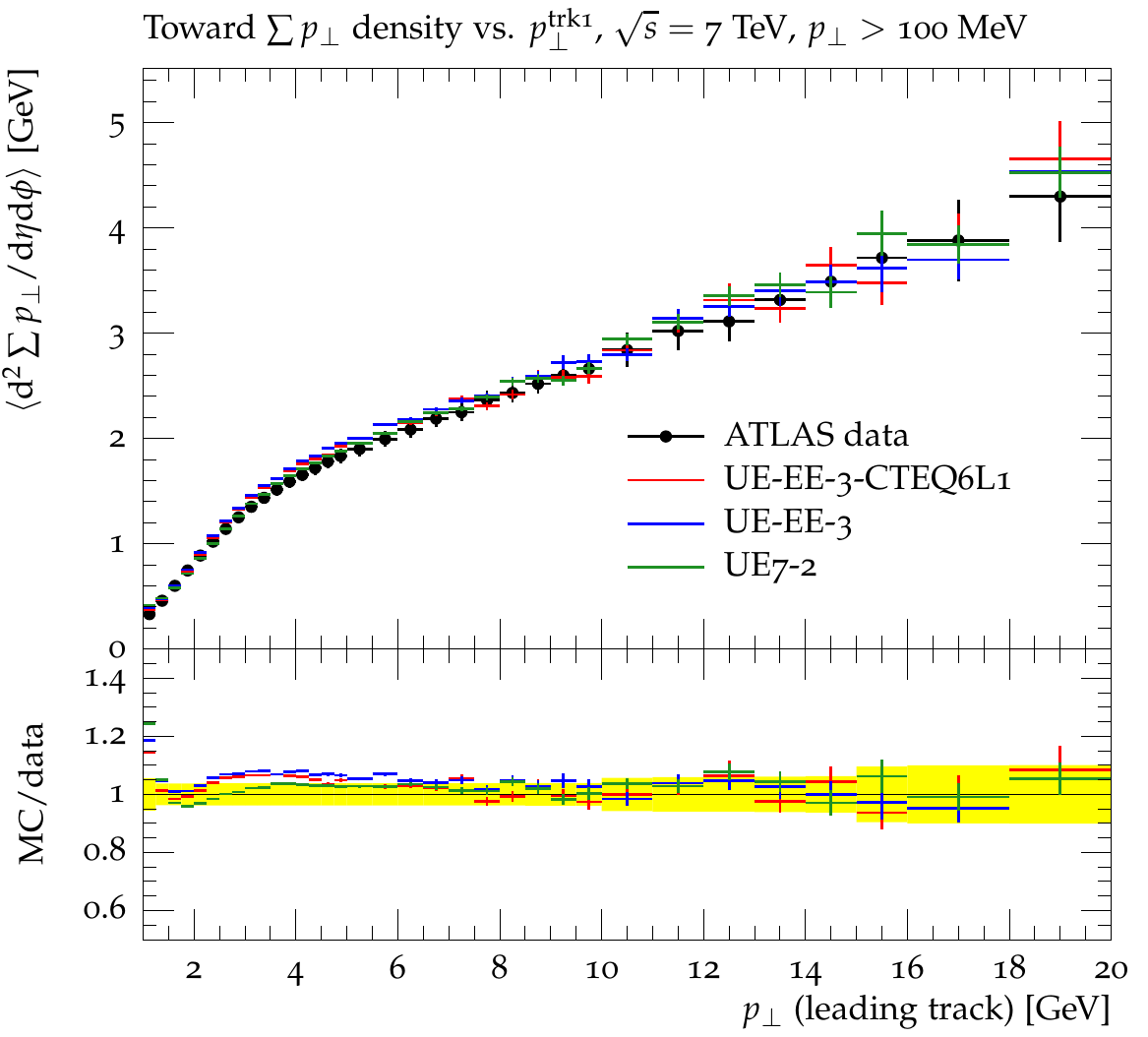}
  \\(b)}
  \hfill
  \parbox[t]{\subfigwidth}{
  \centering
  \includegraphics[width=\subfigwidth]{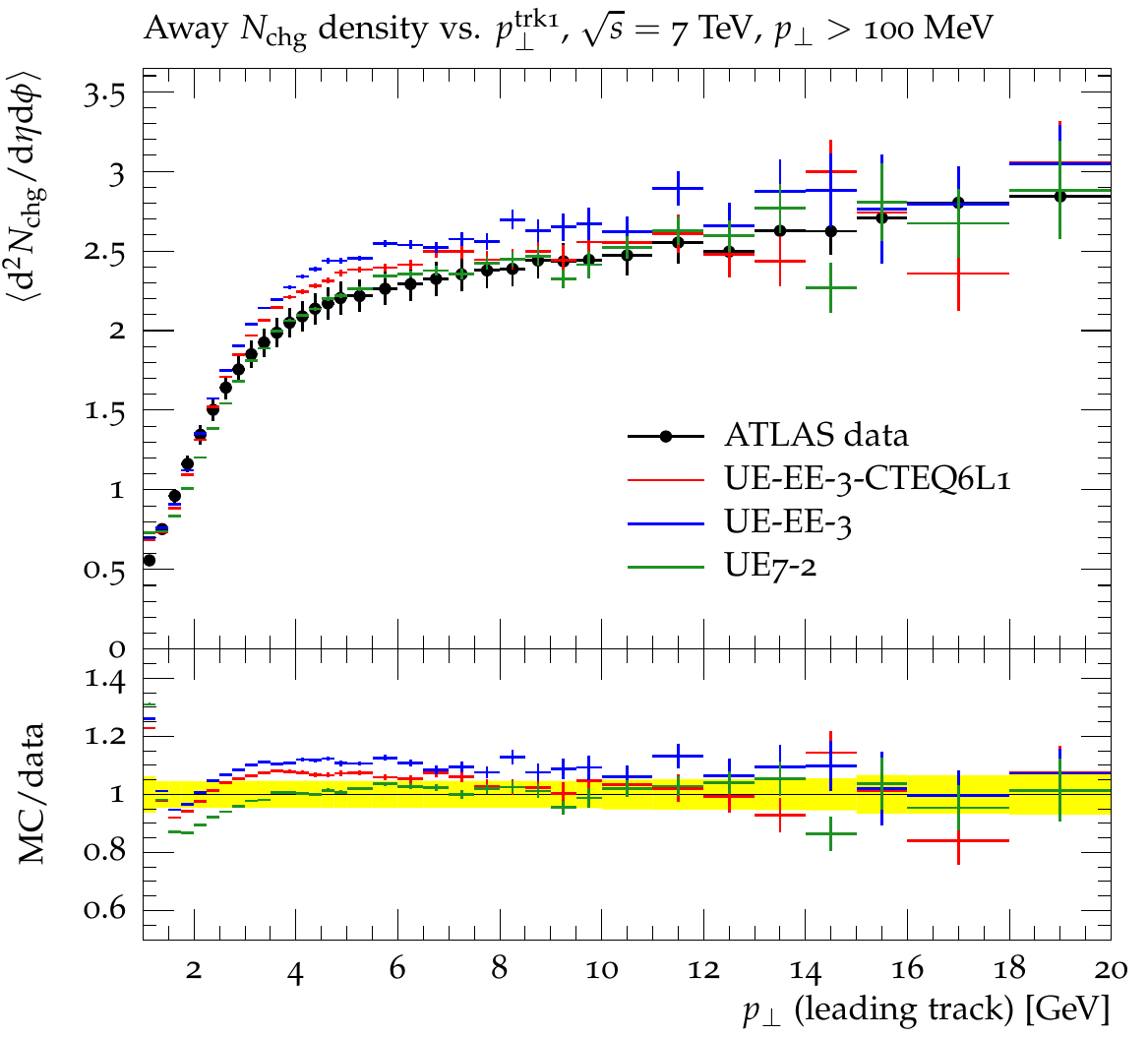}\\
  \includegraphics[width=\subfigwidth]{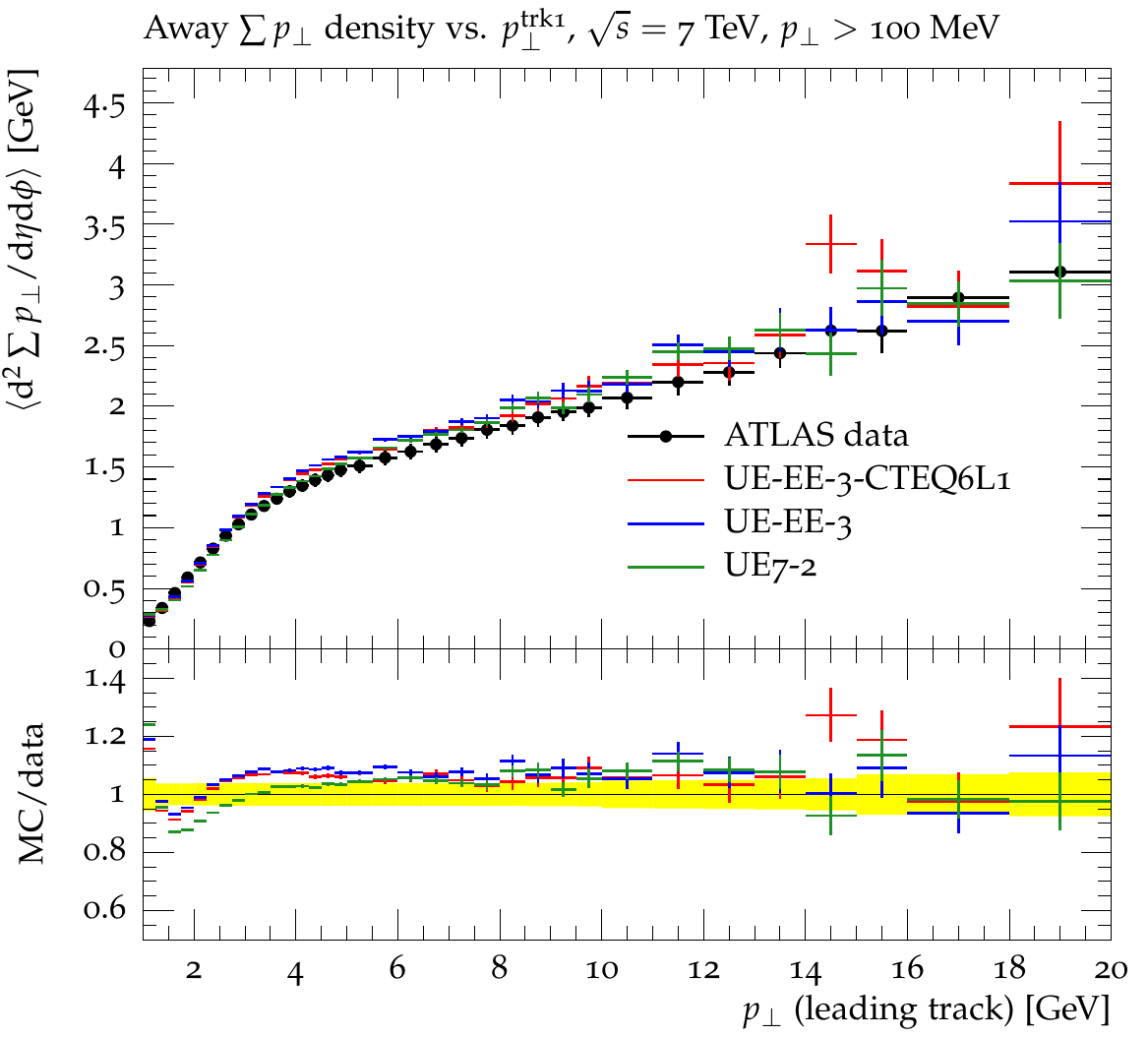}
  \\(c)}
  \caption{ ATLAS UE data at \unit{7}{\TeV} for the lower $p_{\perp}$ cut
  ($p_{\perp} > \unit{100}{\MeV}$) for the transverse (1st column), towards (2nd
  column) and away (3rd column) areas, showing the multiplicity density and
  $\ptsum$ of the charged particles as a function of $\ptlead$. The data is
  compared to the \UEvii, \EEiii and \EEiiiCTEQ tunes.}
  \label{fig:UE-comparison_100MeV}
\end{figure*}

We repeated the tuning process for the UE data collected by ATLAS at
\unit{900}{\GeV} and CDF at \unit{1800}{\GeV}, and obtained as good results as
for \unit{7}{\TeV} (not shown in
Figs.~\ref{fig:UE-comparison-transverse}--\ref{fig:UE-comparison-toward} for the
sake of simplicity).  It is worth mentioning that the ATLAS UE observables with
the lower $p_{\perp}$ cut on the charged particles, $p_{\perp} >
\unit{100}{\MeV}$, were not available during the preparation of the \UEvii tune
but are also well described by the tune, see
Fig.~\ref{fig:UE-comparison_100MeV}(c). These results can therefore be
considered as a prediction of the model.

\begin{figure*}[f]
  \includegraphics[width=\colwidth]{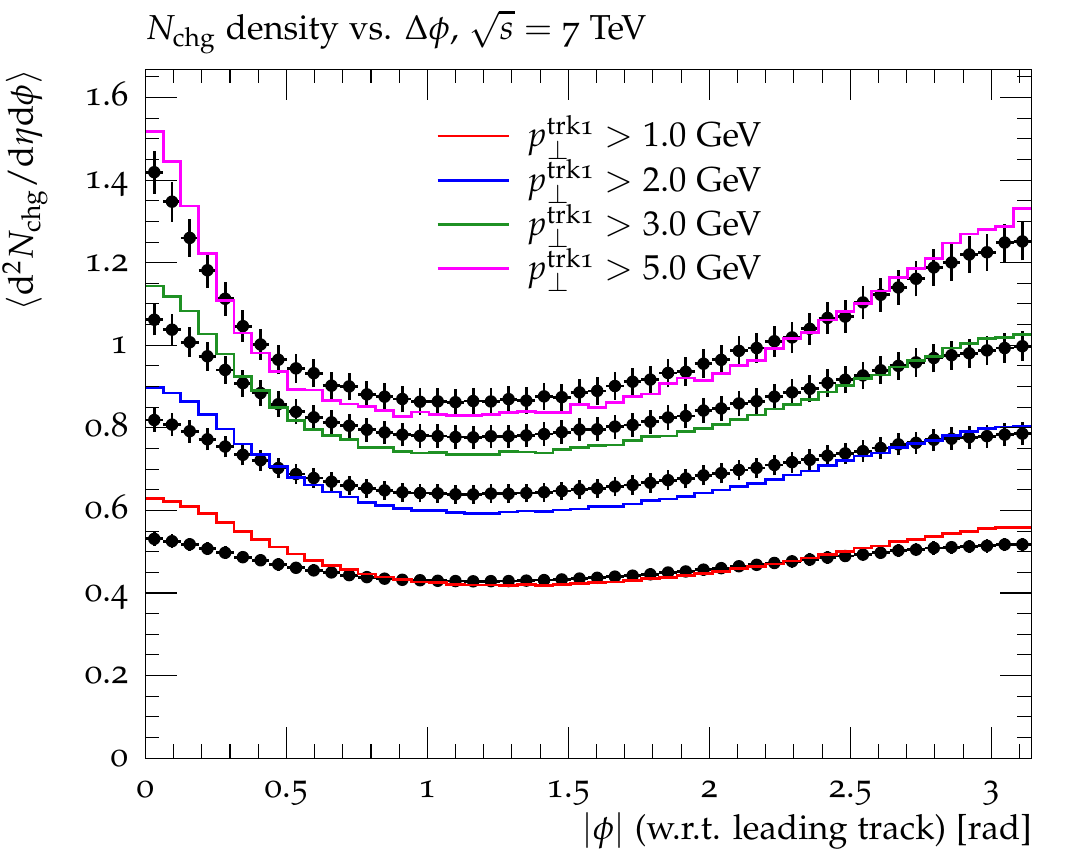}\hfill
  \includegraphics[width=\colwidth]{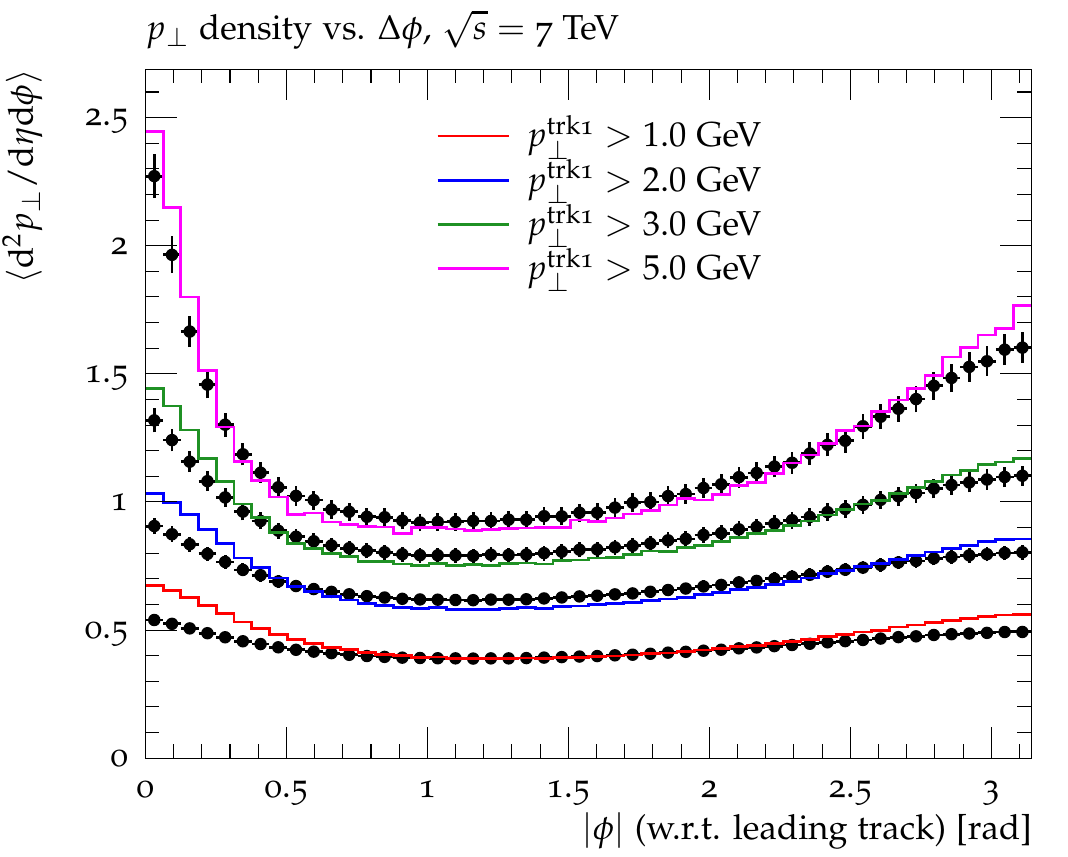}%
  \caption{Azimuthal distribution of the charged particle multiplicity
  (\emph{left panel}) and \ptsum densities (\emph{right panel}), with respect to
  the direction of the leading charged particle (at $\phi = 0$), for $|\eta| <
  2.5$.  The densities are shown for $\ptlead>\unit{1}{\GeV}$,
  $\ptlead>\unit{2}{\GeV}$, $\ptlead>\unit{3}{\GeV}$ and
  $\ptlead>\unit{5}{\GeV}$. The data is compared to the \UEvii tune.}%
  \label{fig:UE7000phi}
\end{figure*}

Figure~\ref{fig:UE7000phi} shows the angular distributions of the
charged-particle multiplicity and $\sum p_{\perp}$, with respect to the leading
charged particle (at $\phi=0$).  The data sets are shown for four different cut
values in the transverse momentum of the leading charged particle, $\ptlead$.
With increasing cut on $\ptlead$, the development of a jet-like structure can be
observed.  The overall description of the data is satisfactory but we can also
see that the description improves as the lower cut value in \ptlead increases as
then the description is more driven by perturbation theory.  The full comparison
with all ATLAS UE and MB data sets is available on the \herwig{} tune page
\cite{tune_wiki}.  At this stage different UE tunes were mandatory for different
hadronic centre-of-mass energies $\sqrt{s}$.  In the next section we address the
question of whether an energy-independent UE tune can be obtained using the
present model.

\begin{figure*}[f]
  \begin{minipage}[]{0.245\textwidth}
    \includegraphics[width=\textwidth]{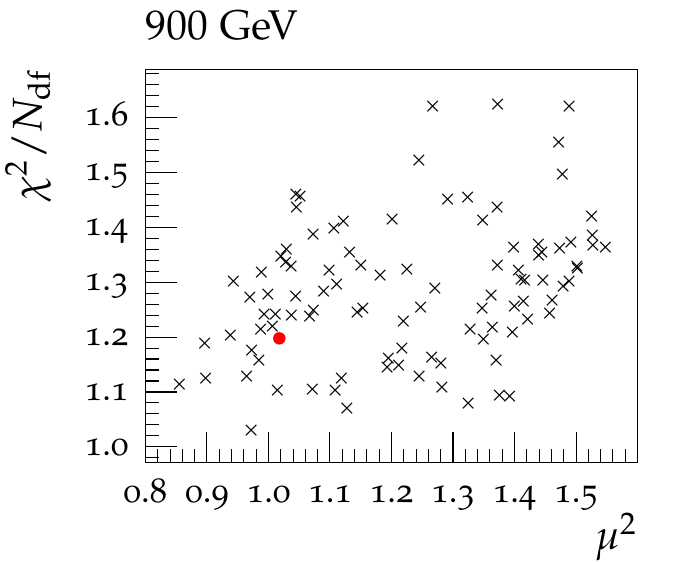}\\
    \includegraphics[width=\textwidth]{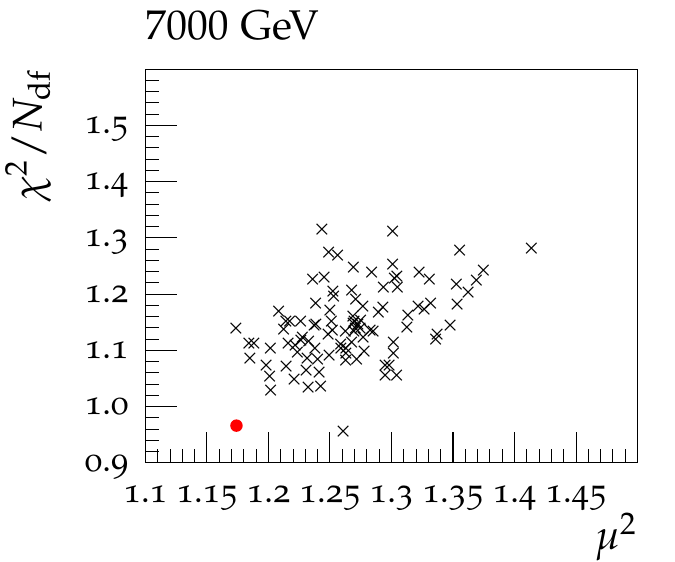}
  \end{minipage}
  \begin{minipage}[]{0.245\textwidth}
    \includegraphics[width=\textwidth]{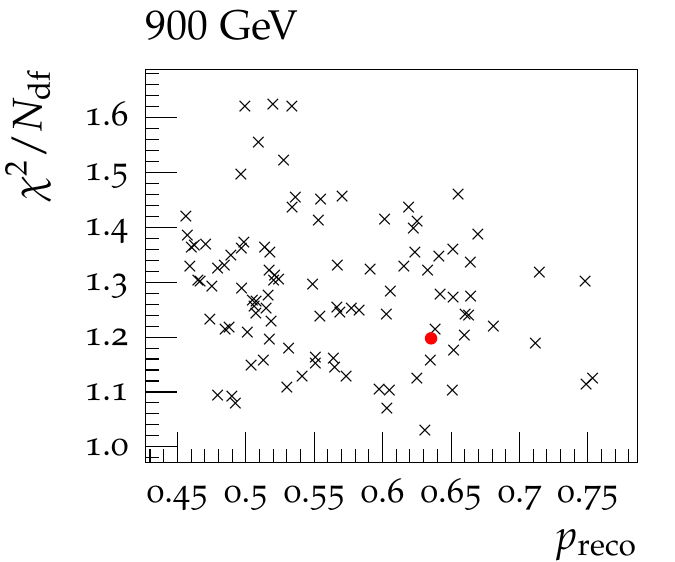}\\
    \includegraphics[width=\textwidth]{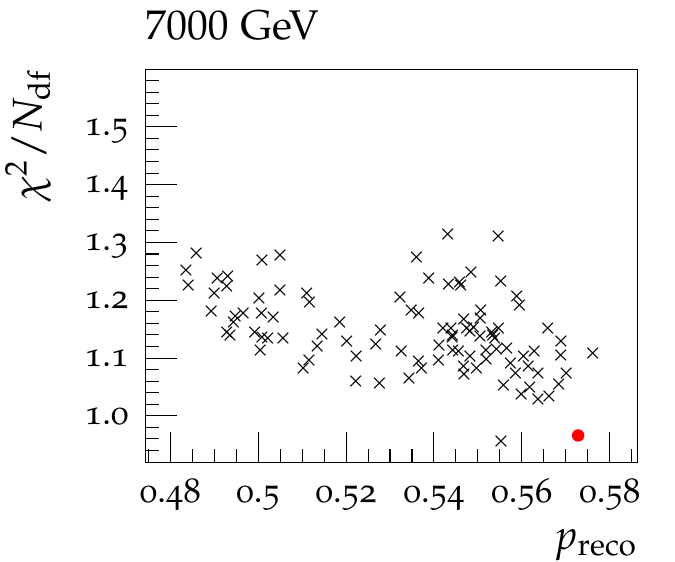}
  \end{minipage}
  \begin{minipage}[]{0.245\textwidth}
    \includegraphics[width=\textwidth]{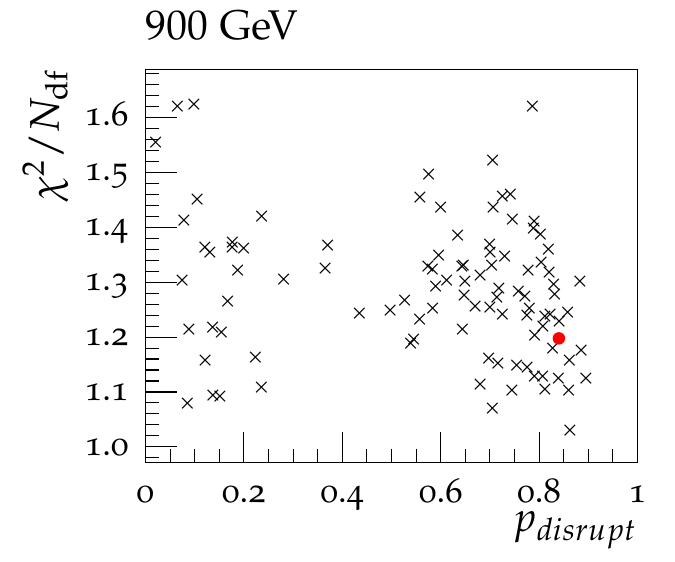}\\
    \includegraphics[width=\textwidth]{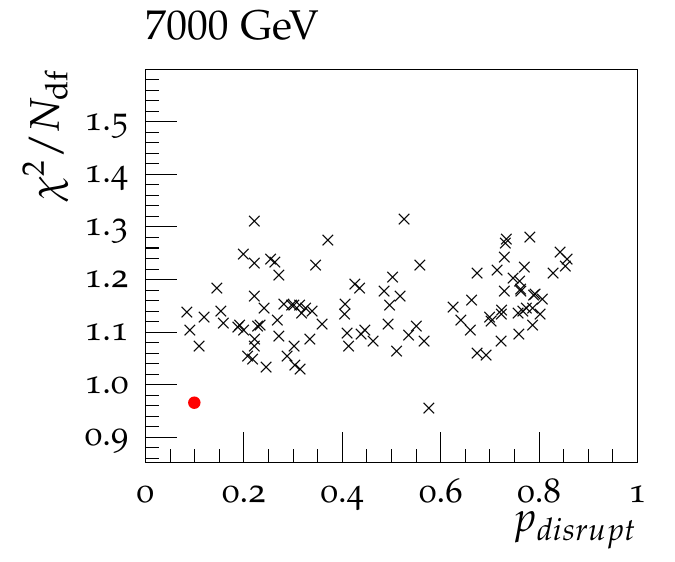}
  \end{minipage}
  \begin{minipage}[]{0.245\textwidth}
    \includegraphics[width=\textwidth]{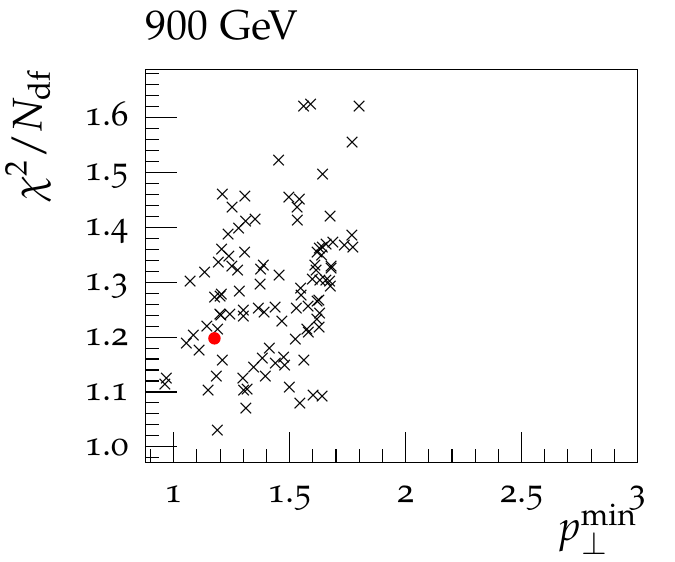}\\
    \includegraphics[width=\textwidth]{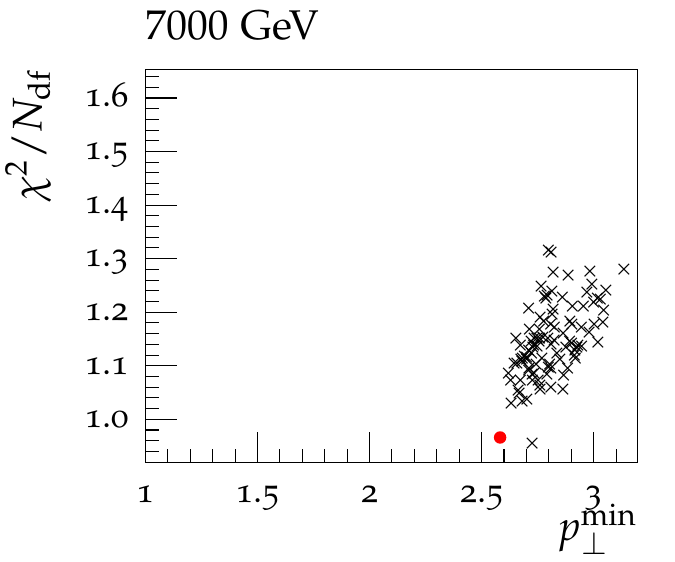}
  \end{minipage}
  \caption{The spread of \EEiiiCTEQ tuning results for the parameters $\mu^2$,
  $\preco$, $\pdisrupt $ and $\ptmin $, using cubic generator response
  parametrizations with all generator runs (red circles) and with subsets of
  generator runs (black crosses). The first row shows results for tunes to data
  at \unit{900}{\GeV} and the second at \unit{7}{\TeV}. }
  \label{fig:UE-EEscatter}
\end{figure*}

\subsubsection{Centre-of-mass energy dependence of UE tunes}

To study the energy dependence of the parameters properly, we examine a set of
observables at different collider energies, whose description is sensitive to
the MPI model parameters. The experimental data should be measured at all
energies in similar phase-space regions and under not too different trigger
conditions.  These conditions were met by two UE observables: $\dNchgdetadphi$
and $\dpTsumdetadphi$, both measured as a function of $\ptlead$ (with $\ptlead <
\unit{20}{\GeV}$) by ATLAS at \unit{900}{\GeV}  and \unit{7000}{\GeV} (with
$p_{\perp} > \unit{500}{\MeV}$) and by CDF at \unit{1800}{\GeV}. Let us first
focus on the \pcr model. In this case we have four free model parameters,
$\pdisrupt$, $\preco$, $\ptmin$ and $\mu^2$.  For each hadronic centre-of-mass
energy we performed independent four-dimensional tunings.  Note that $\ptlead$
denotes the transverse momentum of the hardest track in the case of ATLAS,
whereas the CDF underlying-event analysis uses the $p_\perp$ of the leading jet,
which we call $\ptlead$ here, as well.

Figure~\ref{fig:UE-EEscatter} shows the spread of the tuning results for each
parameter against Professor's heuristic $\chi^2$.  In the first row we present
results for \unit{900}{\GeV} and in the second row for \unit{7}{\TeV}.  Each
point is from a separate tune, made using various combinations of generator runs
at different points in the parameter space.  We see that the parameters are not
well constrained and are sensitive to the input Monte Carlo (MC) runs.  This is
due to what we have already seen during the tuning of the MPI model without
CR~\cite{Bahr:2008dy, Bahr:2008wk, Bahr:2009ek} to Tevatron data, namely the
strong and constant correlation between $\ptmin$ and $\mu^2$.  This correlation
reflects the fact that a smaller hadron radius always balances against a larger
$p_{\perp}$ cutoff, as far as the underlying-event activity is concerned.  With
one of these two parameters fixed, the remaining parameters are much less
sensitive to the input MC runs.

The most important information we can see on these figures is that the
experimental data for the two different c.m.\ energies (\unit{900}{\GeV} and
\unit{7}{\TeV}) cannot be described by the same set of model parameters.  More
precisely, the experimental data prefers different $\ptmin$ values for different
hadronic centre-of-mass energies, while the rest of the parameters may perhaps
remain independent of the energy.  This observation led us to the creation of
energy-extrapolated UE tunes, named \EEiii, in which all parameters are fixed
except for $\ptmin$, which varies with energy.  We summarize the tune values for
$\ptmin$ at different energies in Tab.~\ref{tab:ptmin}.  The other model
parameters, which do not depend on the c.m.\ energy, are given in Table
\ref{tab:fixedparams}.

Since by construction the MPI model depends on the PDF set, we created two
separate energy-extrapolated tunes for the CTEQ6L1 and MRST LO** PDFs. In
general, both tunes yield similar and satisfactory descriptions of experimental
data\footnote{The only difference is that the CTEQ6L1 gives more flexibility in
the choice of the model parameters.}. As an example see
Fig.~\ref{fig:UE-comparison_100MeV}, in which we compare the \EEiii and
\EEiiiCTEQ tunes to ATLAS UE observables, measured in all three regions (toward,
transverse and away). 

\begin{table}[tb]
  \caption{Tune values for $\ptmin$. All other model parameters, which do not
  depend on the c.m.\ energy, are summarized in Tab.~\ref{tab:fixedparams}.}
  \label{tab:ptmin}
  \centering
  \begin{tabular}{lrrr}
    \toprule
    \multicolumn{4}{c}{$\ptmin\enspace [\GeV]$} \\
    \midrule
    $\sqrt{s}\enspace [\GeV]$ & 900  & 1800 & 7000 \\
    \midrule
    \EEiii                    & 1.55 & 2.26 & 2.75 \\
    \EEiiiCTEQ                & 1.86 & 2.55 & 3.06 \\
    \EESCRCTEQ                & 1.58 & 2.14 & 2.60 \\
    \bottomrule
  \end{tabular}
\end{table}

\begin{table*}[htb]
  \sidecaption
  \begin{tabular}{lccc} \toprule & \EEiii & \EEiiiCTEQ & \EESCRCTEQ \\
    \midrule
    $\mu^2\enspace [\GeV^2]$      &  1.11  & 1.35       & 1.5        \\
    $\pdisrupt$                   &  0.80  & 0.75       & 0.8        \\
    $\preco$                      &  0.54  & 0.61       & ---        \\
    $\SCRc$                       &  ---   & ---        & 0.01       \\
    $\SCRf$                       &  ---   & ---        & 0.21       \\
    $\SCRnsteps$                  &  ---   & ---        &  10        \\
    $\SCRalpha$                   &  ---   & ---        & 0.66       \\
    \midrule
    $\ptminnought\enspace [\GeV]$ &  3.11  & 2.81       & 2.64       \\
    $b$                           &  0.21  & 0.24       & 0.21       \\
    \bottomrule
  \end{tabular}
  \caption{Parameters of the energy-extrapolating underlying-event tunes. The
  last two parameters describe the running of $\ptmin$ according to
  Eq.~(\ref{eq:power}).}
  \label{tab:fixedparams}
\end{table*}

We repeated this procedure also for the \scr model. However, since in this case
the tuning procedure was more complicated, as explained below, we concentrated
on one PDF set only, namely CTEQ6L1.  The first obvious complication was the
larger number of parameters to tune. The second complication was associated with
the fact that one of the tuning parameters, $\SCRnsteps$, is an integer number.
The current version of Professor, however, does not provide such an option,
instead it treats all parameters as real numbers.  Therefore, we decided to
carry out fifty separate tunes for different fixed values of $\SCRnsteps$,
starting from 1 to 50. The last problem that we encountered, which is probably
associated with the two previously mentioned problems, was that for some
parameter values the predictions from Professor were significantly different
from the results we received directly from \herwig{}++ runs.  Initially, we
increased the order of the interpolating polynomials from second to fourth,
which should improve Professor's predictions, but this did not improve the
situation.  Therefore, we first identified regions of the parameter space where
this problem appeared most frequently and then excluded these from the tuning
procedure. As a result, we obtained an energy-extrapolated underlying-event tune
for the \scr model, which we call \EESCRCTEQ.

\afterpage{\clearpage}

In Figures~\ref{fig:UE-comparison-transverse}, \ref{fig:UE-comparison-away} and
\ref{fig:UE-comparison-toward} we show a comparison of the \pcr{} and \scr{}
energy-extrapolated (CTEQ6L1) tunes and the \UEvii tune against $\dNchgdetadphi$
and $\dpTsumdetadphi$ as a function of $\ptlead$ for $p_{\perp} >
\unit{500}{\MeV}$ in all three regions (toward, transverse and away) and at
three different collider energies.  We can see that the quality of the data
description is high and at the same level for all tunes.  Nevertheless, we
favour the \scr{} model as here we have a clearer physics picture and a more
flexible model.

\begin{figure}[htb]
  \begin{center}
    \includegraphics[width=\linewidth]{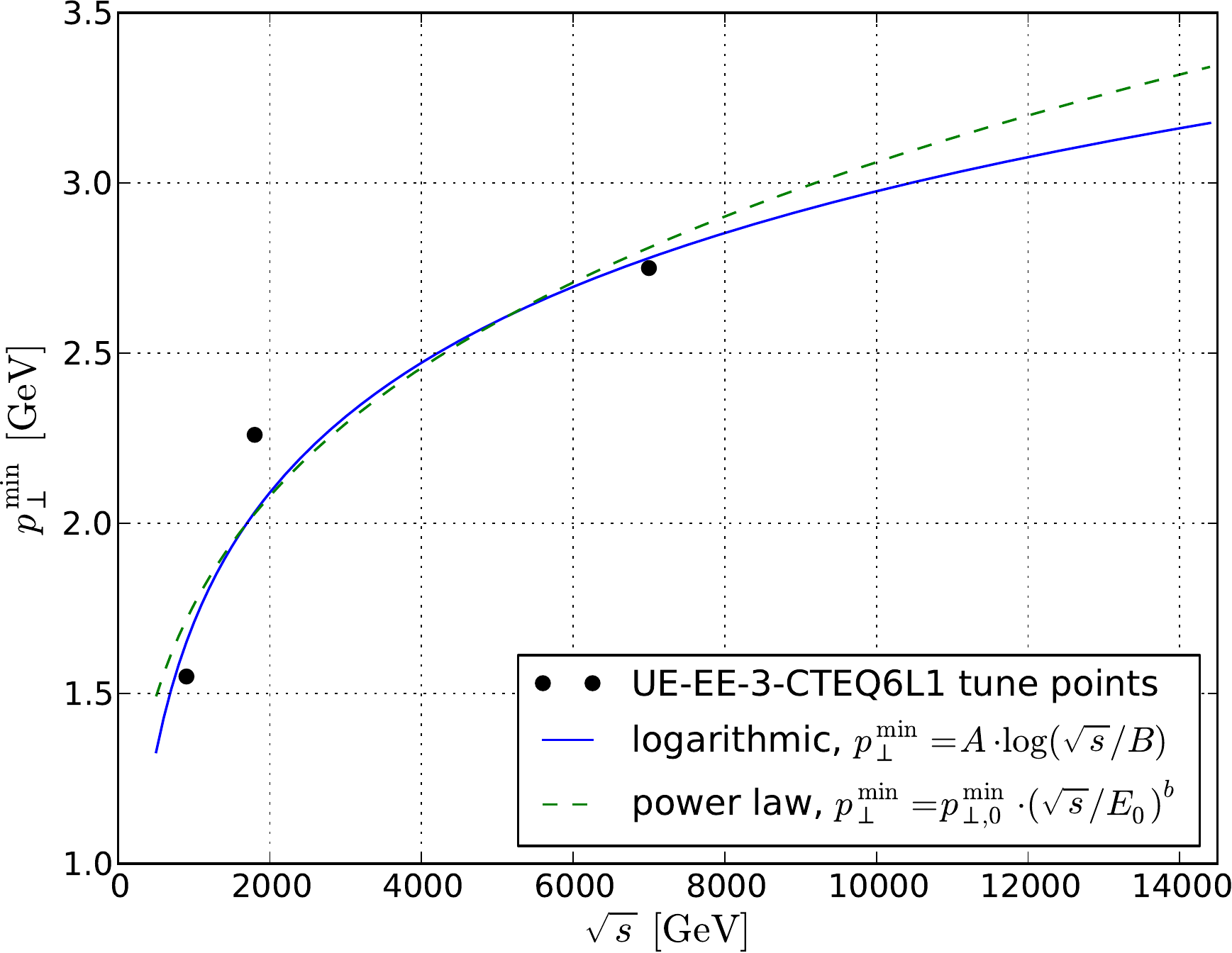}
    \caption{ Energy extrapolation of $\ptmin$ in the \EEiiiCTEQ tune.}
    \label{fig:fit}
  \end{center}
\end{figure}

In the last step, we parametrized the $\ptmin$ dependence.  In a first attempt
we have chosen a logarithmic function to extrapolate $\ptmin$ to energies
different from the tune energies. Therefore we fitted a function of the form
$\ptmin(s) = A\,\log(\sqrt{s}/B)$, where $A$ and $B$ are free fit parameters, to
the three $\ptmin$ values obtained in the \EEiii tune. The fit is shown in
Fig.~\ref{fig:fit}.  Based on this, we provide UE tunes for c.m.\ energies the
LHC was or will be operating at. Since the logarithmic form is not very stable
for lower energies, we have replaced this ansatz with a power law, see
also e.g.\ \cite{Ryskin:2011qe}, 
\begin{equation}
  \label{eq:power}
  \ptmin(s) = \ptminnought \left(\frac{\sqrt{s}}{E_0}\right)^b\ .
\end{equation}
This is the default parametrization of the energy dependence from \herwig{}++
release 2.6 \cite{release26}.  The default value of $E_0$ is \unit{7}{\TeV}.
For the collider energies at consideration in our tunes there are no significant
differences in all observables due to this change.  The values for $b$ and
$\ptminnought$, which we find by fitting Eq.~(\ref{eq:power}) to the $\ptmin$
values from Tab.~\ref{tab:ptmin}, are summarized in the last two rows of
Table~\ref{tab:fixedparams}.

\begin{figure*}[f]
  \begin{minipage}[t]{0.44\textwidth}
    \includegraphics[width=\textwidth,keepaspectratio]{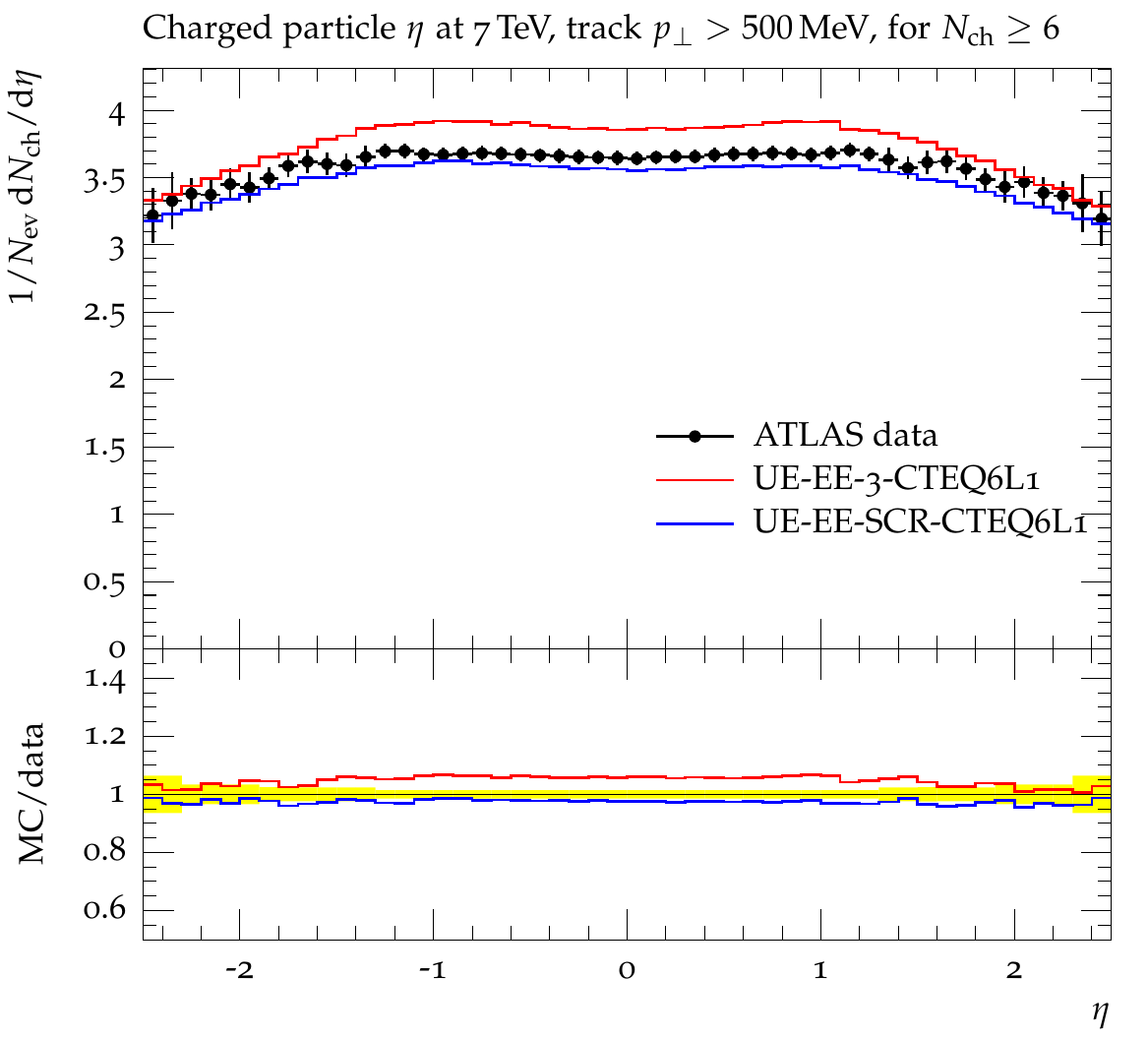}\\
    \includegraphics[width=\textwidth,keepaspectratio]{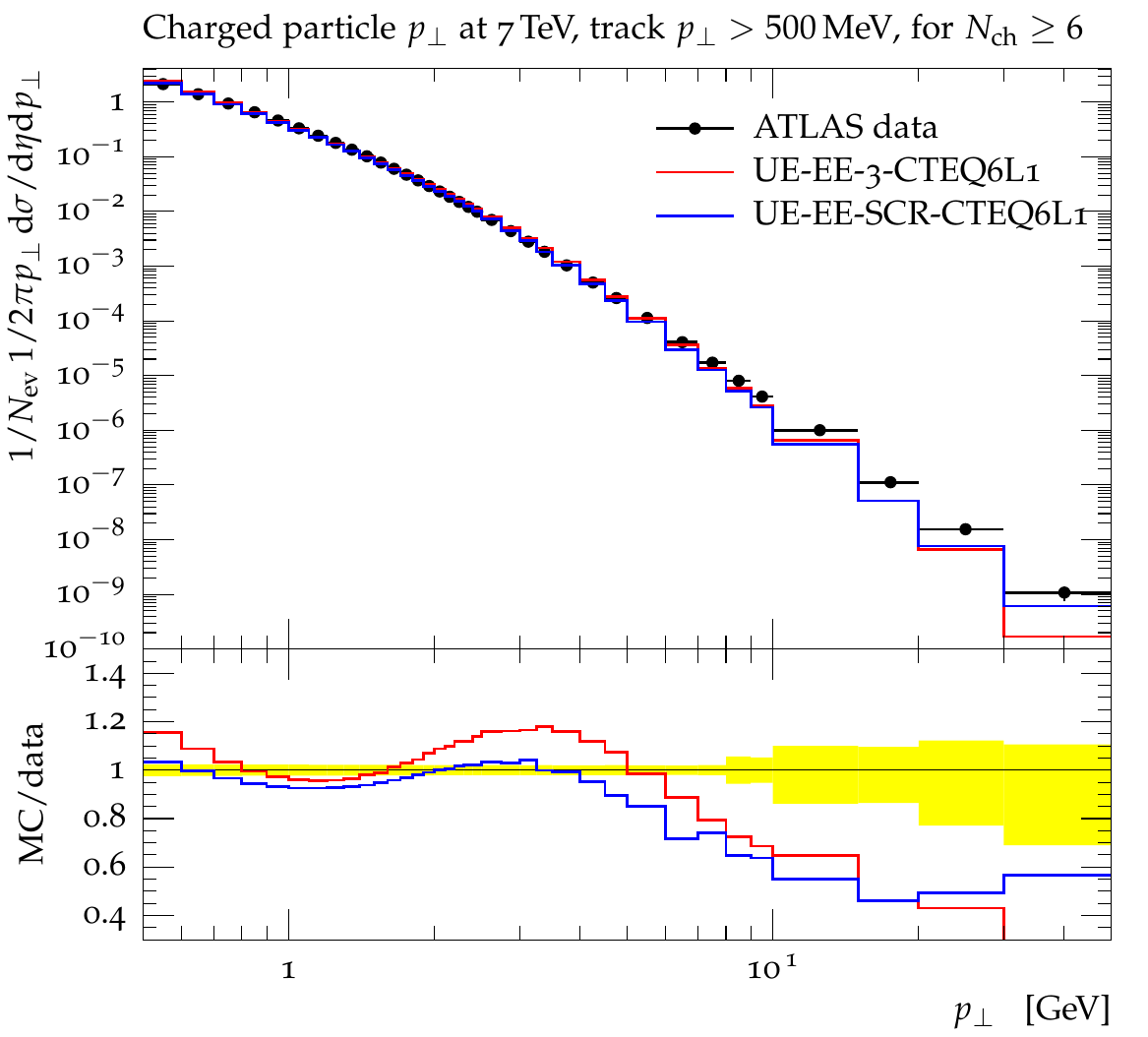}
  \end{minipage}
  \hfill%
  \begin{minipage}[t]{0.44\textwidth}
    \includegraphics[width=\textwidth,keepaspectratio]{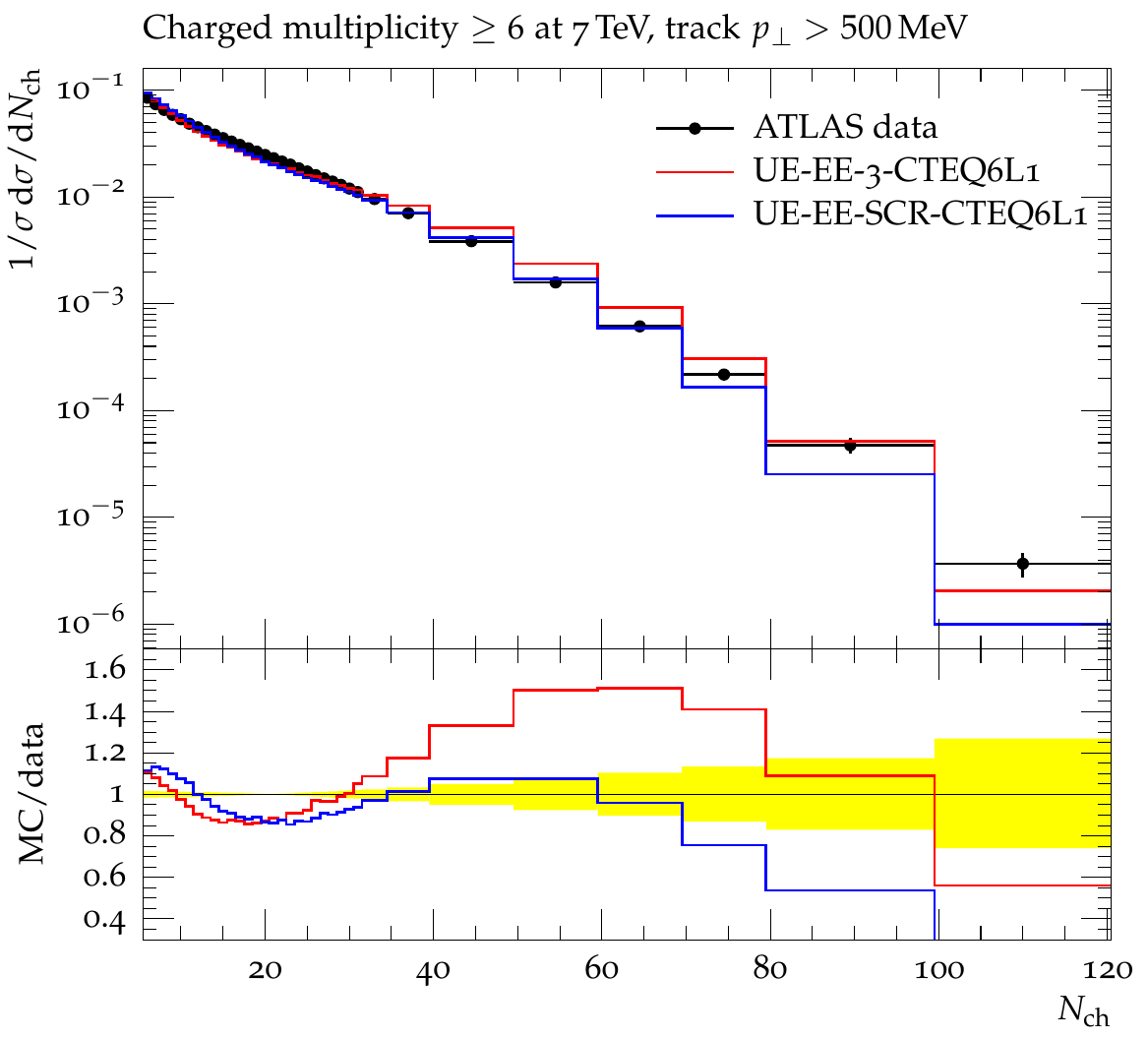}\\
    \includegraphics[width=\textwidth,keepaspectratio]{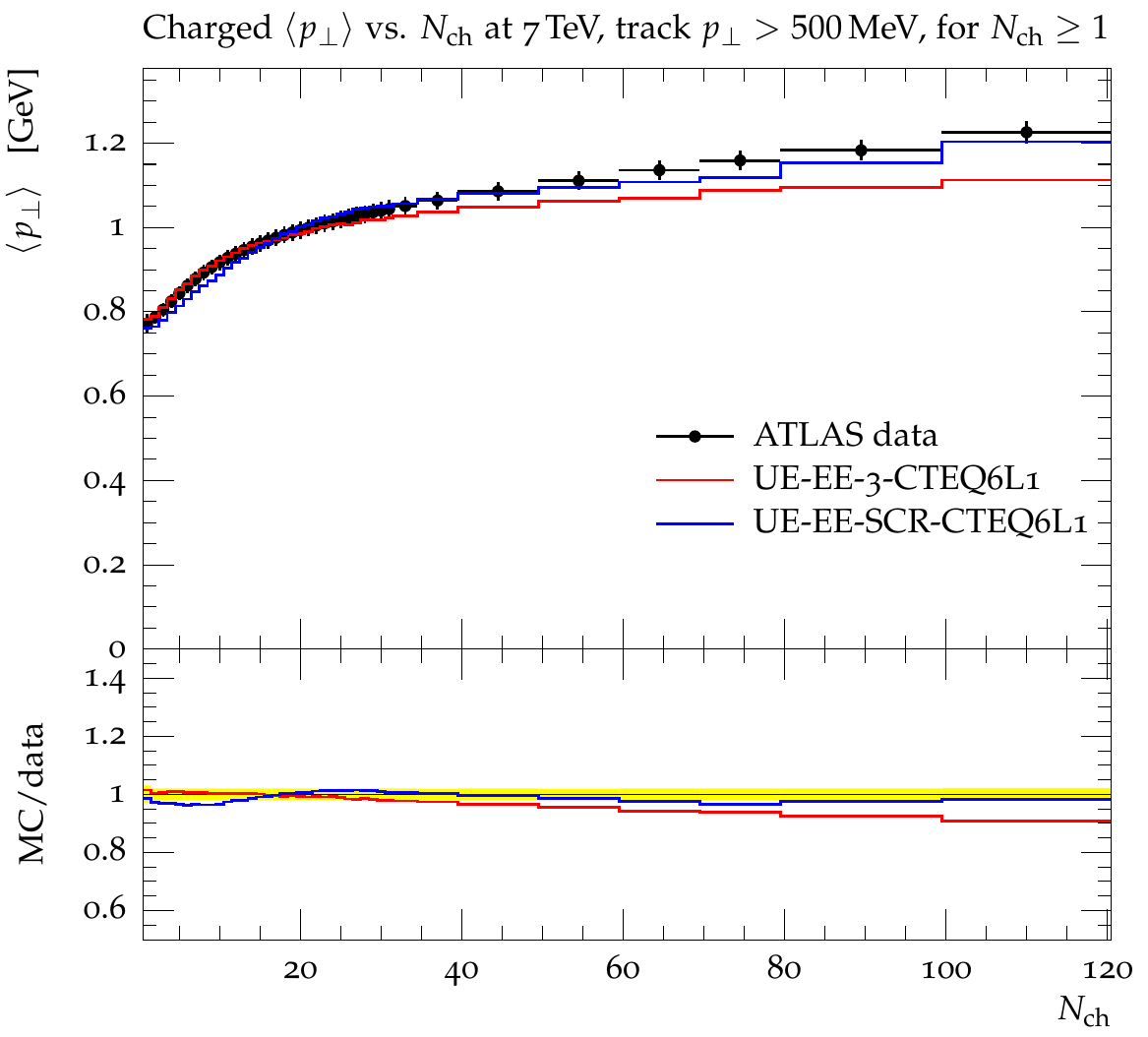}
  \end{minipage}
  \caption{ Comparison of the \EEiiiCTEQ and \EESCRCTEQ tunes to ATLAS
  minimum-bias distributions at $\sqrt{s}=\unit{7}{\TeV}$, with $N_{\mathrm{ch}}
  \ge 6$, $p_{\perp} > \unit{500}{\MeV}$ and $|\eta| < 2.5$. }
  \label{fig:ATLAS_7000_Nch6}
\end{figure*}

\begin{figure*}[f]
  \begin{minipage}[t]{0.44\textwidth}
    \includegraphics[width=\textwidth,keepaspectratio]{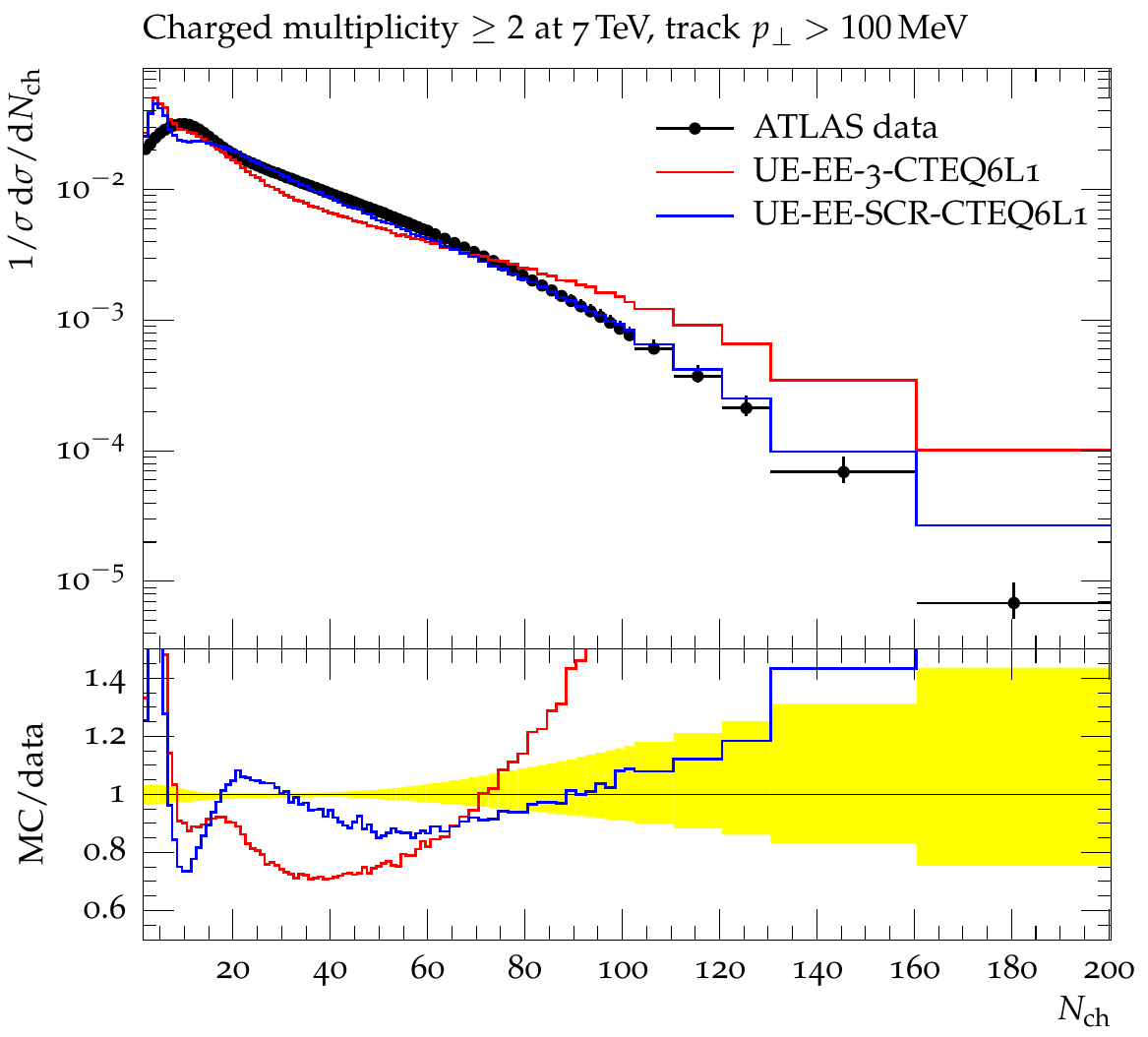}
  \end{minipage}
  \hfill%
  \begin{minipage}[t]{0.44\textwidth}
    \includegraphics[width=\textwidth,keepaspectratio]{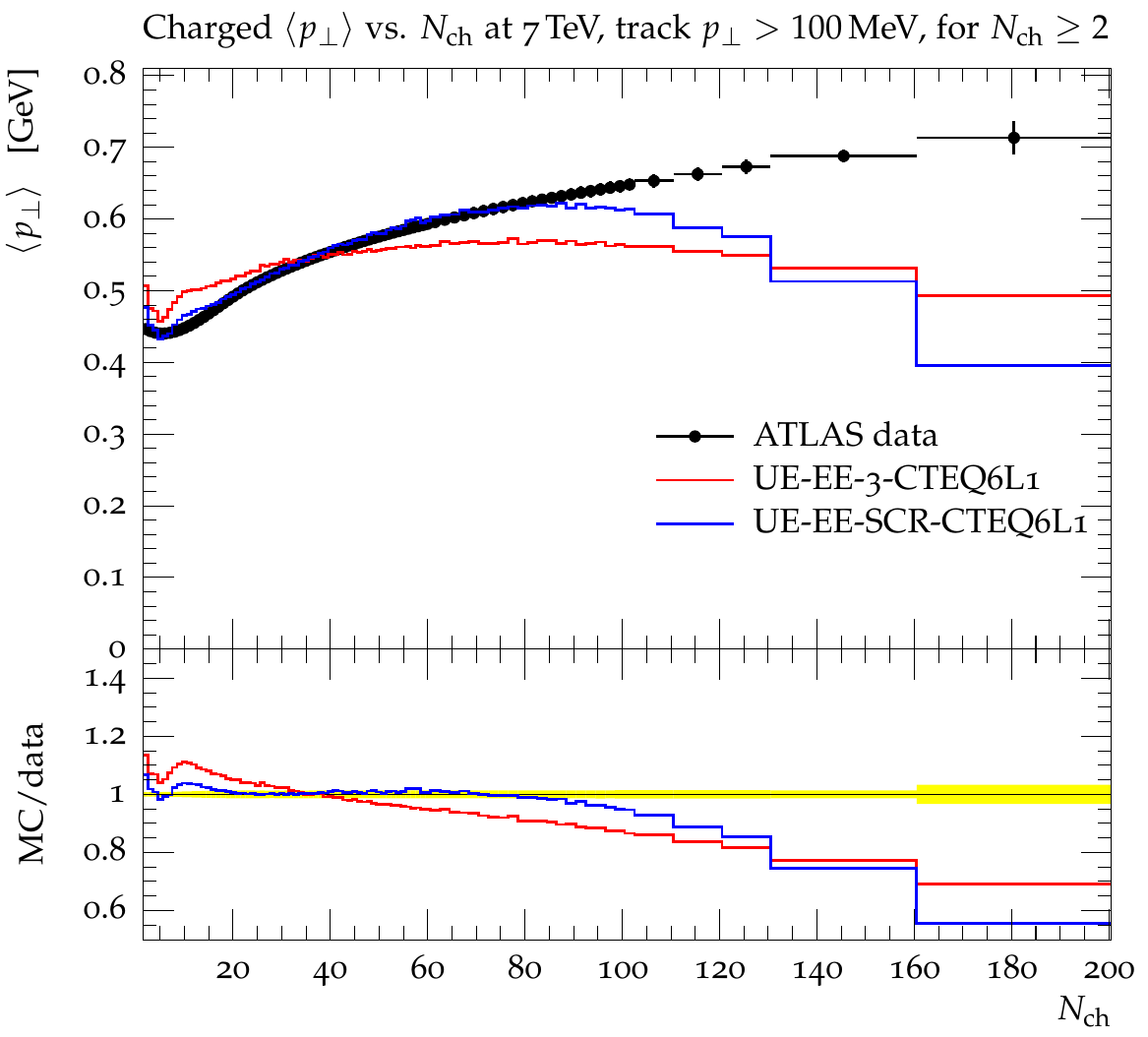}
  \end{minipage} 
  \caption{ Comparison of underlying-event tunes to presumably
  diffraction-enhanced MB observables, measured by ATLAS at
  $\sqrt{s}=\unit{7}{\TeV}$, with $N_{\mathrm{ch}} \ge 2$, $p_{\perp} >
  \unit{100}{\MeV}$ and $|\eta| < 2.5$. }
  \label{fig:ATLAS_7000_Nch2}
\end{figure*}

For the preparation of the energy-extrapolated tunes we did not use any MB
observables. Nevertheless, we show a comparison of the \EEiiiCTEQ and \EESCRCTEQ
tunes to the diffraction-reduced ATLAS MB data at \unit{7}{\TeV} (with $N_{\rm
ch} \ge 6$) in Fig.~\ref{fig:ATLAS_7000_Nch6}.  We see that the data is
described slightly better by the \scr{} than by the \pcr{} tune. Moreover,
although these data sets were not taken into account in both tunes, the results
are close to the experimental data.

In the future, we plan to study the energy scaling of the model
parameters using diffraction-reduced minimum-bias data, and then, in
more detail, the possibility of achieving a common description of the UE
and MB data, cf.~\cite{Schulz:2011qy}.  As can be seen in
Fig.~\ref{fig:ATLAS_7000_Nch2}, the UE tunes fail to reproduce the ATLAS
MB data at \unit{7}{\TeV} with a less tight cut on the number of charged
particles, $N_{\rm ch} \ge 2$, and where all charged particles with
$p_{\perp} > \unit{100}{\MeV}$ are taken into account.  This is not
surprising, however, since \herwig{} lacks a model for soft diffractive
physics so far.  That explains the poor description of both the charged
multiplicity and the average transverse momentum in the low-multiplicity
bins.  On the other hand, the unsatisfactory description of the shown
observables in the high multiplicity tail may indicate missing physics
in the model. It might, however, as well be resolved by a dedicated MB
tune. Both possibilities are left for future work.  In particular, we point out
the lack of an explicit model for diffractive events.  A more complete
description of the MB data should also include a modelling of
these.

\section{Conclusions}

We have introduced two different models for non-perturbative colour
reconnections in \herwig{}.  The models are of slightly different
computational complexity but give very similar results.  The tuning
results have shown that the \scr{} is preferred to have parameters that
force a quick `cooling' of the system and therefore results in a very
similar model evolution as in the simpler \pcr{} model.  We therefore
consider the \pcr{} as a special case of the \scr{} model for quick
cooling and keep the \scr{} as the more flexible model for future
versions of \herwig{}++.  As a consequence, we understand that the data
demands a final state that does not obey a perfectly minimized colour
length.  We interpret this as a model limitation.  At some point the
picture of colour lines breaks down.  Colour lines themselves are only a
valid prescription up to leading order in the $N_C\to \infty$ limit.
Furthermore, the mechanism addresses the non-perturbative regime where
the picture of the colour triplet charges themselves is already a model
by itself and possibly completely washed out.  

\afterpage{\clearpage}

We have studied the mechanism of colour reconnection in detail and found
that in fact the non-perturbative parts of the simulation demand the
colour reconnection mechanism in order to repair the lack of information
on the colour flow.  The intuitive picture we have based our model on
could be verified.  The idea of colour preconfinement is meaningful in
the context of the hadronization model and has to be rectified when a
model of multiple partonic interactions is applied without further
information on the colour structure in between the multiple scatters.  

Furthermore, we have shown that by tuning the MPI model with CR we can obtain a
proper description of non-diffractive MB ATLAS observables.  We present the
energy-extrapolated tune \EEiii, which is an important step towards the
understanding of the energy dependence of the model.  Finally, we have unified
the different tunes of the MPI model in \herwig{}++ into a simple
parametrization of the $\ptmin$ dependence in a way that allows us to describe
data at different energies with only one set of parameters. News concerning
\herwig{} tunes are available on the tune wiki page \cite{tune_wiki}.

\begin{acknowledgements} 
We are grateful to the other members of the \herwig{} collaboration for critical
discussions and support.  We acknowledge financial support from the Helmholtz
Alliance ``Physics at the Terascale''. This work was funded in part (AS) by the
Lancaster-Manchester-Sheffield Consortium for Fundamental Physics under STFC
grant ST/J000418/1.
\end{acknowledgements} 

\bibliographystyle{h-physrev}
\bibliography{references}

\end{document}